\documentclass[manuscript]{aastex}
\newcommand{\gsim}{\;\lower.6ex\hbox{$\sim$}\kern-7.75pt\raise.65ex\hbox{$>$}\;}
\newcommand{\lsim}{\;\lower.6ex\hbox{$\sim$}\kern-7.75pt\raise.65ex\hbox{$<$}\;}

\slugcomment{to appear ...}

\shorttitle{}
\shortauthors{}

\begin{document}


\title{Chemical abundances and properties of the ionized gas in NGC~1705}


\author{F. Annibali \altaffilmark{2}, M. Tosi \altaffilmark{2}, A. Pasquali \altaffilmark{4}, A. Aloisi \altaffilmark{3}, M. Mignoli\altaffilmark{2}, 
D. Romano \altaffilmark{2}}

\altaffiltext{2}{INAF-Osservatorio Astronomico di Bologna, 
Via Ranzani 1, I-40127 Bologna, Italy; francesca.annibali@oabo.inaf.it}

\altaffiltext{3}{Space Telescope Science Institute, 3700 San Martin Drive, 
Baltimore, MD 21218, USA}

\altaffiltext{4}{Astronomisches Rechen-Institut, Zentrum fuer Astronomie der Universitaet Heidelberg, Moenchhofstr. 12 - 14,
            69120 Heidelberg, Germany}



\begin{abstract}

We obtained [O III] narrow-band imaging and multi-slit MXU spectroscopy of the blue compact dwarf (BCD) galaxy NGC~1705 with FORS2@VLT to derive chemical abundances of PNe and H II regions and, more
in general, to characterize the properties of the ionized gas. 
The auroral [O III]$\lambda4363$ line was detected in all but one of the eleven analyzed regions, allowing for a direct estimate of their electron temperature. 
The only object for which the [O III]$\lambda4363$ line was not detected is a possible low-ionization PN, the only one detected in our data.
For all the other regions, we derived the abundances of  Nitrogen, Oxygen, Neon, Sulfur and Argon out to $\sim$1 kpc from the galaxy center. We detect for the first time in NGC~1705 a negative radial gradient in the oxygen metallicity of $-0.24 \pm 0.08$ dex kpc$^{-1}$. 
The element abundances are all consistent  with values reported in the literature for other samples of dwarf irregular and blue compact dwarf galaxies. 
However, the average (central) oxygen abundance, $12 + \log(O/H)=7.96 \pm 0.04$, is $\sim$0.26 dex lower than previous literature estimates for NGC~1705 based on the [O III]$\lambda4363$ line. 
From classical emission-line diagnostic diagrams, we exclude a major contribution from shock excitation. On the other hand, the radial behavior of the emission line ratios is consistent with the progressive dilution of radiation with increasing distance from the center of NGC~1705.   
This suggests that the strongest starburst located within the central $\sim$150 pc is responsible for the ionization of the gas out to at least $\sim$1 kpc. 
The gradual dilution of the radiation with increasing distance from the center reflects the  gradual and continuous transition from the highly ionized H II regions in the proximity of the major starburst into the diffuse ionized gas. 

\end{abstract}



\section{Introduction}

Dwarf galaxies play a key role in our understanding of galaxy formation and evolution. Within the framework of hierarchical formation, they are considered 
the first systems to collapse, supplying the building blocks for the formation of more massive galaxies through merging and accretion  \citep[e.g.][]{kau93}.
Moreover, late-type dwarf galaxies (dwarf irregular (dIrr) and blue compact dwarf (BCD) galaxies), with their low metallicity, high gas content, and blue colors, resemble the properties of primordial galaxies \citep[e.g.][]{it99}, and are the best laboratories where to study star formation processes similar to those occurring in the early Universe.

 Because of their poorly evolved nature, dIrrs and BCDs  where initially suggested to be young galaxies at their first bursts of star formation. However, studies of the star formation histories (SFHs) based on color-magnitude diagrams (CMDs) of the resolved stars in a large number of dwarf galaxies within the Local Group and beyond \citep[see][for a review]{tht09} showed that these systems started forming stars as long ago as the look-back time set by the available photometry. In practice, all of them were already active at least $\sim 1-2$ Gyr ago, and, when the photometry was deep enough to reach the horizontal branch or the most ancient main-sequence turnoffs, or when RR Lyrae stars were found, at epochs as old as $\sim$ a Hubble time \citep[e.g.][]{dolphin00,clementini03,mcconnachie06}. This poses a 
 major challenge for chemical evolution models of dIrrs and BCDs, which have to reconcile the low observed metallicity with the old ages, roughly continuous SFHs, 
 and  fairly high star formation rates of the most metal-poor systems.

Since three decades, metal enriched winds seem to be the most viable mechanism to
reconcile the high rates of star formation observed in starburst dwarf galaxies with
their low metal abundances \citep[e.g.][]{mt85,pilyugin93,marconi94,carigi95,romano06}.
 Outflows driven by supernova explosions and stellar winds  are in fact expected to play an especially important role in these systems, whose relatively shallow potential wells make them susceptible to wind-driven loss of gas and newly created metals \citep[e.g.][]{dekel86,martin99}.
In practice, even the accretion of large amounts of metal-free gas is usually insufficient to explain the low chemical abundances 
observed in dwarf galaxies with very intense episodes of star formation, where the metal production is quite high.
Alternative models have sometimes been proposed, but never succeded in reproducing
all the observed properties of starburst dwarfs. For instance, recently, a scenario
that considers infall of primordial external gas, but does not include any outflow
or galactic wind, has been proposed by \cite{gavilan13}  to explain the observed
properties of gas-rich dwarf galaxies. However, these models have proven to be
successful in reproducing the gas fractions, chemical abundances and photometric
properties only for ``quiet'' field dwarf galaxies, with recent star formation
activity but not strong intensity episodes (i.e. dIrr galaxies).  Moreover, NGC1705,
as well as a few other starburst dwarfs, does show observational evidence of
galactic winds \citep{meurer92,papaderos94,marlowe95}.

Element abundance estimates in late-type dwarf galaxies are fundamental to constrain chemical evolution models. 
While H II regions define the present-day galaxy properties, chemical abundances for individual stars or planetary nebulae (PNe) offer a view of the galaxy  chemical properties at earlier epochs, but are challenging outside the Local Group, where the most actively star-forming BCDs are found. 
In addition to the evolution of elements in time, the spatial behaviour of the chemical abundances plays a key role in constraining dwarf galaxy models. 
The majority of literature studies indicate that, within the observational uncertainties, dIrrs and BCDs show nearly spatially constant chemical abundances \citep[e.g.,][]{kobul97,croxall09,lagos13,haurberg13}. This could indicate that the ejecta from stellar winds and supernovae are dispersed and 
mixed across the interstellar medium (ISM) on timescales of $<10^7$ yr, but an alternative possibility is that the freshly synthesized elements remain unmixed with the surrounding interstellar medium and reside in a hot $10^6$ K phase or a cold, dusty, molecular phase \citep{kobul97}.

NGC~1705 is a late-type dwarf galaxy classified as BCD, famous for its high recent star formation (SF) activity and for the unambiguous evidence of galactic winds
\citep{meurer92,heckman01}. Several H II regions have been found in NGC~1705 \citep{melnick85,meurer92}, and most recently studied with EFOSC2 by \cite{lee04}, hereafter LS04.  Among their 16 analyzed H II regions, LS04 were able to measure the [O  III]$\lambda$4363 line in 5 central 
(distances $\lesssim$300 pc from the galaxy center) regions, from which they inferred a spatially homogeneous oxygen abundance of 
$12+\log(O/H)=8.21 \pm0.05$ (i.e. Z$\sim$0.005), $\log(N/O)=-1.63\pm0.07$ and $\log(Ar/O)=-2.31\pm0.11$; they noticed large differences with the corresponding abundances estimated by \cite{heckman01} for the neutral gas observed with the FUSE satellite, where N, O and Ar are all significantly lower than in the H II regions  \footnote{Notice however that the lines of N, O, and Ar sampled by FUSE have issues (e.g., saturation) that likely give an underestimate of the real column density of these elements in the
neutral gas.}. 
The chemistry of the neutral gas in NGC~1705 was confirmed with further FUSE data by \cite{aloisi05}. We observed NGC~1705 in UBVIJH with the HST/WFPC2 and resolved its individual stars down to V$=$28 mag, $\sim$2 mag below the RGB tip \citep{tosi01}. The HST-derived CMDs led to a safe determination of the galaxy distance (5.1 Mpc, i.e. $DM=(m-M)_0=28.54$, \cite{tosi01}) from the luminosity of its RGB tip, and to a detailed description of its SF history \citep[][hereafter, A03, A09]{anni03,anni09}. 
One-zone chemical evolution models for NGC~1705 based on the SFH of A03, and assuming metal-enriched galactic winds,  
were able to reproduce the observed H II region metallicity in NGC~1705, but could not explain the low N/O ratio as derived by LS04 \citep{romano06}.  

We acquired FORS2@VLT data  (programme ID 084.D-0248 and 086.D-0761, PI Tosi) with the aim of deriving chemical abundances in PNe and H II  regions and, more in general, to characterize the properties of the warm gas in NGC~1705.  Pre-imaging and spectroscopy are described in Sections~2 and 3, respectively. The emission line fluxes, corrected for underlying Balmer absorption and reddening, are given in Section~4. In Section~5 we provide the derived chemical abundances. In Section~6 we discuss the comparison with  classical emission-line diagnostic diagrams. In Section 7, we discuss the comparison with literature data for NGC~1705 and for other samples of late-type dwarf galaxies, 
discuss relative element abundances, and provide evidence for a metallicity gradient. Our conclusions are in Section~8.

\section{Pre-imaging}

FORS2 \citep{fors} pre-imaging exposures were taken in order to identify the PNe and H II regions selected to build the mask for our multi-object spectroscopy. 
We imaged NGC~1705 on March 25, November 15 and December 1, 2010 for a total of $\sim$5 h. Seeing ranged between 0.63 arcsec in March, 0.97 in November 
and 0.81 in December.  Observations were performed with the standard resolution collimator in the narrow-band filters  
FILT500$-$5$+$85 (sampling the [OIII]$\lambda$5007 emission line) and OIII/6000$+$52 (centered at $\lambda=5105$ \AA, and sampling the continuum 
emission adjacent to the [OIII] line). The total exposure time was $\sim$2.5 h for each filter, split into three subexposures. The images have a 
pixel scale of 0.25''/pixel and a field of view of 6.8' $\times$ 6.8'.  

\begin{figure*}
\epsscale{1}
\plotone{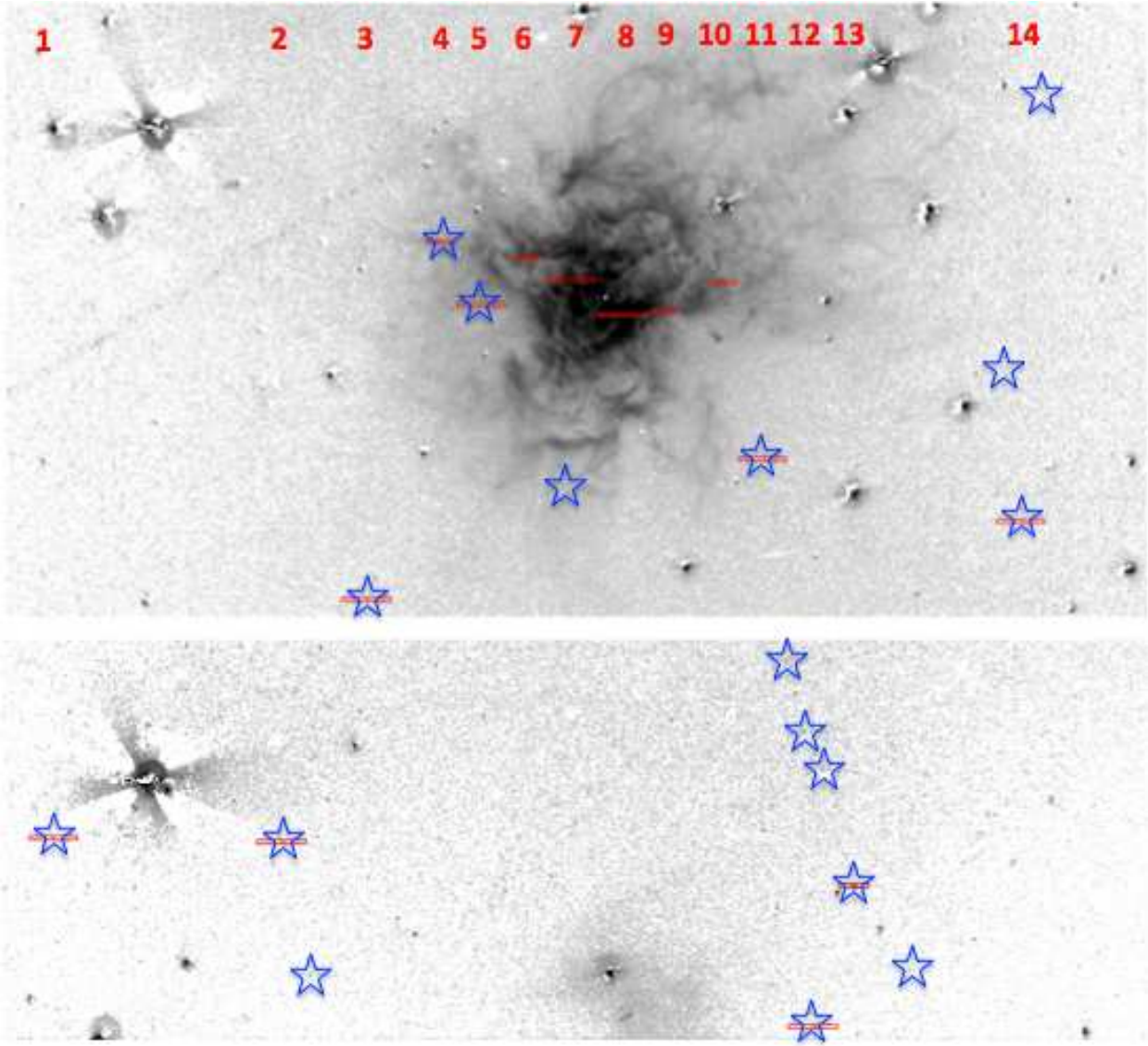}
\caption{VLT FORS2 continuum-subtracted [OIII] image of NGC~1705. The blue stars indicate the PN candidates. The mask slits were positioned 
on PN candidates and on extended [OIII] emission regions. The slit identification numbers are given at the top of the image in correspondence with the slit positions. 
The displayed field of view is $\approx6.5'\times5.8'$.  North is up and East is left. 
\label{o3}}
\end{figure*}

After bias subtraction and flat field calibration, we combined the individual images into a single frame for each filter using the DRIZZLE task \citep{drizzle} 
in the IRAF environment\footnote{ IRAF is distributed by the National Optical Astronomy Observatory, which is operated by the Association of Universities for Research in Astronomy, Inc., under cooperative agreement with the National Science Foundation.}. We normalized the images to account for the different filter widths, and applied a Gaussian smoothing to bring them to the same resolution;  then we constructed a [O III] continuum-subtracted image, which is shown in Figure~\ref{o3}.

 The morphology of the [O III] image resembles that in H$\alpha$ already discussed by \cite{meurer92}. The ionized gas presents a bipolar morphology extending out to
$\sim$100 '' ($\sim$2.5 kpc) from the galaxy center,  and it is organized into thin filaments, loops, and arcs. The presence of  ``cellular'' structures  or cavities is striking, making the gas appear as ``fragmented'' at a global look. The cells have sizes from $\sim$1'' to $\sim$4'' ($\sim$25 pc to $\sim$100 pc), with a median value of $\sim$2'' 
($\sim$50 pc). 
 
\begin{figure*}
\epsscale{1}
\plotone{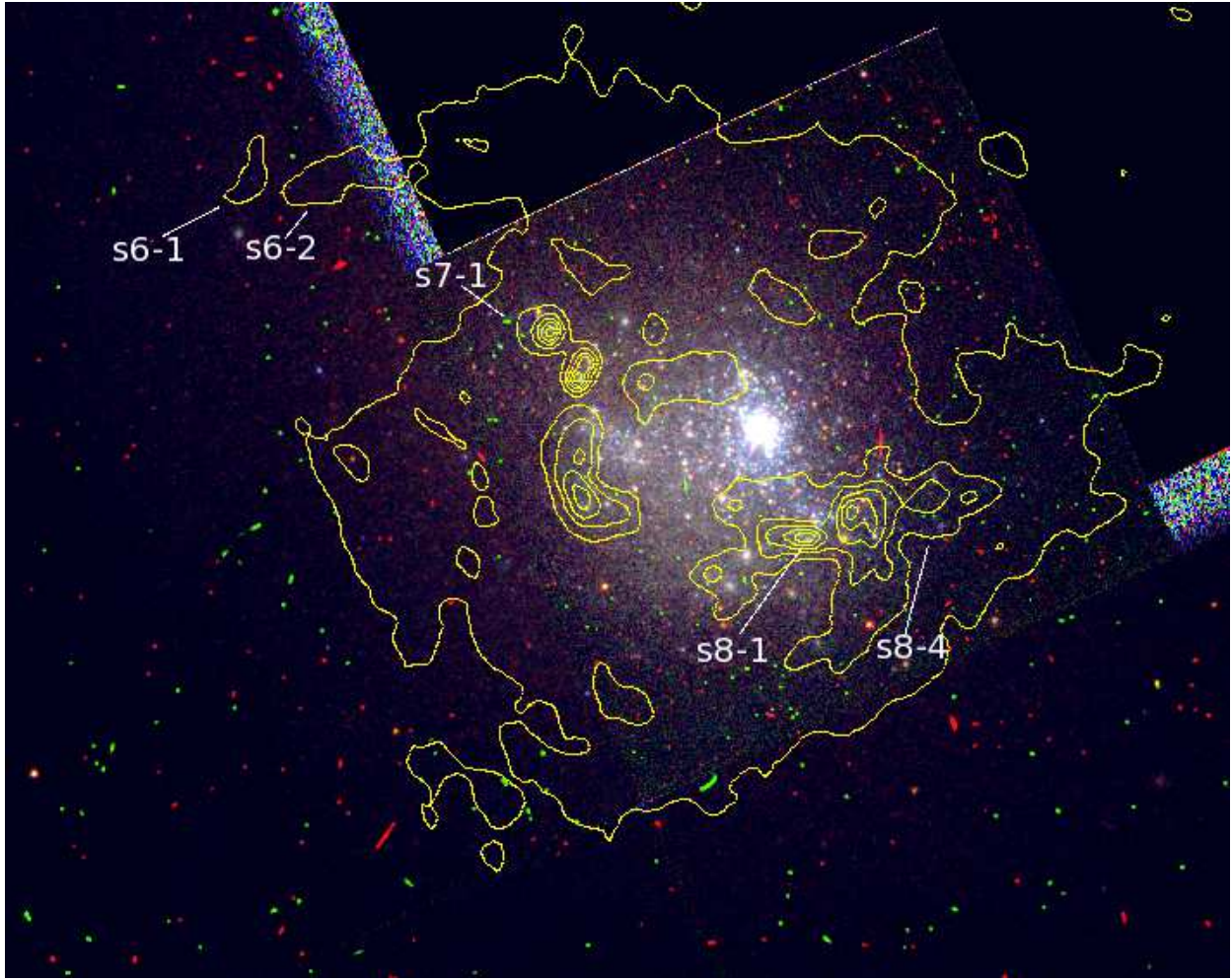}
\caption{HST WFPC2 color-combined image of NGC~1705 with superimposed contour plots of the continuum-subtracted [O III] image. 
The F380W ($\sim$U), F555W ($\sim$V), and F814W ($\sim$I) WFPC2 images correspond to the blue, green, and red channels, respectively.  Some of the regions targeted  for FORS2 spectroscopy are indicated. The field of view is $\sim 70''\times 60''$.  North is up and East is left. 
\label{rgb}}
\end{figure*}

The ionized gas extends much further away than the optical image of the galaxy. This is shown in Figure~\ref{rgb}, where we overplot isocontours for a portion of the  [O III] continuum-subtracted image over a color image of NGC~1705 obtained combing WFPC2 images in the F380W ($\sim$U), F555W ($\sim$V),  and F814W ($\sim$I) filters.  
The field of view is $\sim 70''\times 60''$, i.e. just $\sim$3\% of the total field of view in Figure~\ref{o3}. Indicated are also some of the regions that we targeted for spectroscopy (see next section). The figure shows a strong concentration of ionized gas organized in a sort of shell around the high surface brightness optical component of the galaxy. The central region of NGC~1705, with the highest density of young star formation, corresponds to a zone of significantly lower [O III] surface brightness compared to the surrounding shell. A few blue stars are observed in correspondence of  some of the [O III] knots: this is the case for  s7-1, for the double-peaked knot just nearby it, and for s8-1. 
However, we will show in Section~\ref{dd} that, despite the presence of local association of young stars, the main ionization source seems to come from the central starburst in NGC~1705.

The  [O III] continuum-subtracted image was used to identify PN candidates and H II regions.  
At the distance of NGC~1705 \citep[$5.1\pm0.6$ Mpc or $DM=28.54\pm0.26$,][]{tosi01}, PNe, with typical diameters smaller than 1 pc, should appear as point-like objects emitting in the [OIII] line. Thus we selected point-like sources in the [OIII] continuum-subtracted image to identify PN candidates, although we could not exclude that these sources were compact H II regions instead. 
PN central stars typically have absolute visual magnitudes of 
$M_V \sim-2.0$ \citep{mendez92}, corresponding to an apparent magnitude of $m_V\sim26.5$ in NGC~1705, too faint to be detected in our continuum image.  
Therefore, when selecting PN candidates, we also checked that no continuum emission was associated to the point-like sources in the [OIII]-continuum subtracted image.  
Following these criteria,  we identified 17 PN candidates in the frame, whose positions are indicated in Figure~\ref{o3}.

\section{Spectroscopy}

Spectroscopy was obtained with FORS2 in the MXU configuration from January 2011 until January 2012  in service mode. 
The slit mask (see Figure~\ref{o3}), was constructed to observe the largest possible number of PN candidates, giving priority to the brightest ones. We were able to target 9 out of
 the 17 identified candidates (slit number s1, s2, s3, s4, s5, s11, s12, s13, s14). Two slits (s7 and s8) were positioned on H II regions observed by \cite{lee04}, while three slits (s6, s9, and s10) were positioned on extended [O III] emission regions and gas filaments. We also positioned six slits on regions where no diffuse gas emission was observed, to serve as sky templates. These were used to subtract the background in those slits totally occupied by the emission.  To avoid confusion, the sky slits are not shown in the mask of Figure~\ref{o3}.
  
\begin{figure*}
\epsscale{1}
\plotone{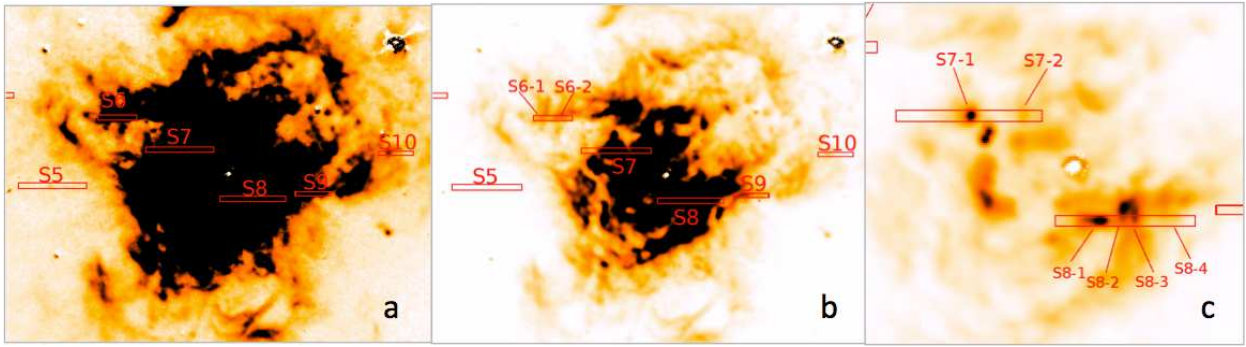}
\caption{Continuum-subtracted OIII image portions of NGC~1705 with the FORS2 MXU slits superimposed. Indicated are the positions of the eleven extracted 
regions (s5, s6-1, s6-2, s7-1, s7-2, s8-1, s8-2, s8-3, s8-4, s9 and s10). 
The field of view is $\approx100''\times80''$ for panels a) and b), and $\approx42''\times38''$ for panel c). 
Different cuts are used to highlight fainter and brighter structures. 
\label{o3bis}}
\end{figure*}

To cover the spectral range from the [O III]$\lambda$3727 to the [S II]$\lambda\lambda$6716,31 doublet 
the grism GRIS$-$300V  was employed, alone in the blue, and coupled with the GG435 filter in the red.
NGC~1705 was observed in the blue with 7$\times$2700 s exposures, for a total of 5.2 h,  and in the red with 20$\times$2700 s exposures, for a total of $\sim$15 h. 
The seeing varied from $\sim$0.7'' to $\sim$1.3'' (median value of 1.1'') for the blue observations, and from $\sim$0.6'' to 1.5''  (median value of  0.9'') for the red observations,
higher than our requirement of a seeing $\leq$0.8'' .
The pixel scale and the dispersion are 0.126 arcsec/pixel and $\sim$2 \AA/pixel, respectively.  
The effective resolution is $\lambda/FWHM \sim780$ at $\lambda \sim5500$\  \AA.  Wavelength calibration was based on four arc lamps (a He, a HgCd, and two Ar arc lamps) to provide a sufficient number of lines over the whole spectral range. 
Data were reduced with the standard procedure using the {\it onedspec} and {\it twodspec} packages inside IRAF.  The science and arc frames were bias subtracted. 
For each science exposure, the individual 2d spectra were extracted from the frame. The exact same extraction apertures used for the science frames were used for the arc frames. 
Then, the 2d spectra were wavelength calibrated using the LONGSLIT {\it identify},  {\it fitcoords} and  {\it transform}  tasks. After that we subtracted the background. 
This step deserves a particular explanation: in the case of point-like sources (PN candidates or compact H II regions), 
we defined the background using two windows at the opposite sides of the central source, and  
performed the subtraction with the {\it{background}} task in IRAF. Typically, the background windows were chosen at a distance larger than $\sim$ 2 times the FWHM of the spatial profile from the source peak.
This procedure removes both the sky emission and the background emission from NGC~1705. 
However, in presence of extended emission occupying the entire slit, it was not possible to define the background windows; then, we subtracted the sky templates from the object spectra. Notice that this procedure allows to subtract the sky emission but does not remove the galaxy background.  
We applied the sky template subtraction to slits number 6, 8, 9 and 10. For all the other slits  we used the iraf ``background'' task.
After background subtraction, the individual exposures were combined into a median 2d spectrum, for the blue and for the red observations separately. 
Then 1d spectra were extracted using the {\it apall} task.  
In the case of PN candidates, we centered the extraction aperture on the point-like source, while in the case of extended emission, multiple apertures were selected within one slit (e.g., apertures s6-1, s6-2, s7-1, s7-2, s8-1, s8-2, s8-3, s8-4, see Figure~\ref{o3bis}). 

The 1d spectra were flux calibrated using the spectrophotometric  standard stars 
Feige~56, GD~108, LTT~3218,  LTT~4816, and Feige~110, EG~21, LTT~1020, LTT~2415,  LTT~3218, respectively for the blue and for the red observations.
The standards were observed with a 5 arcsec wide slit positioned at the center of the MOS instrument. After bias subtraction, wavelength calibration, and background 
removal, a 1d spectrum was extracted for each standard. To obtain the sensitivity function we ran the iraf tasks {\it standard} and {\it sensfunc}.
The  {\it standard} task provides the standard star observed counts, along with the associated calibration fluxes, over some pre-defined bandpasses. 
Then the {\it sensfunc} uses the {\it standard} output to  determine the system sensitivity as a function of wavelength.  In the process of deriving the sensitivity curve, 
we assumed the atmospheric optical extinction curve derived by \cite{patat11} for Cerro Paranal.

\begin{figure*}
\epsscale{1}
\plotone{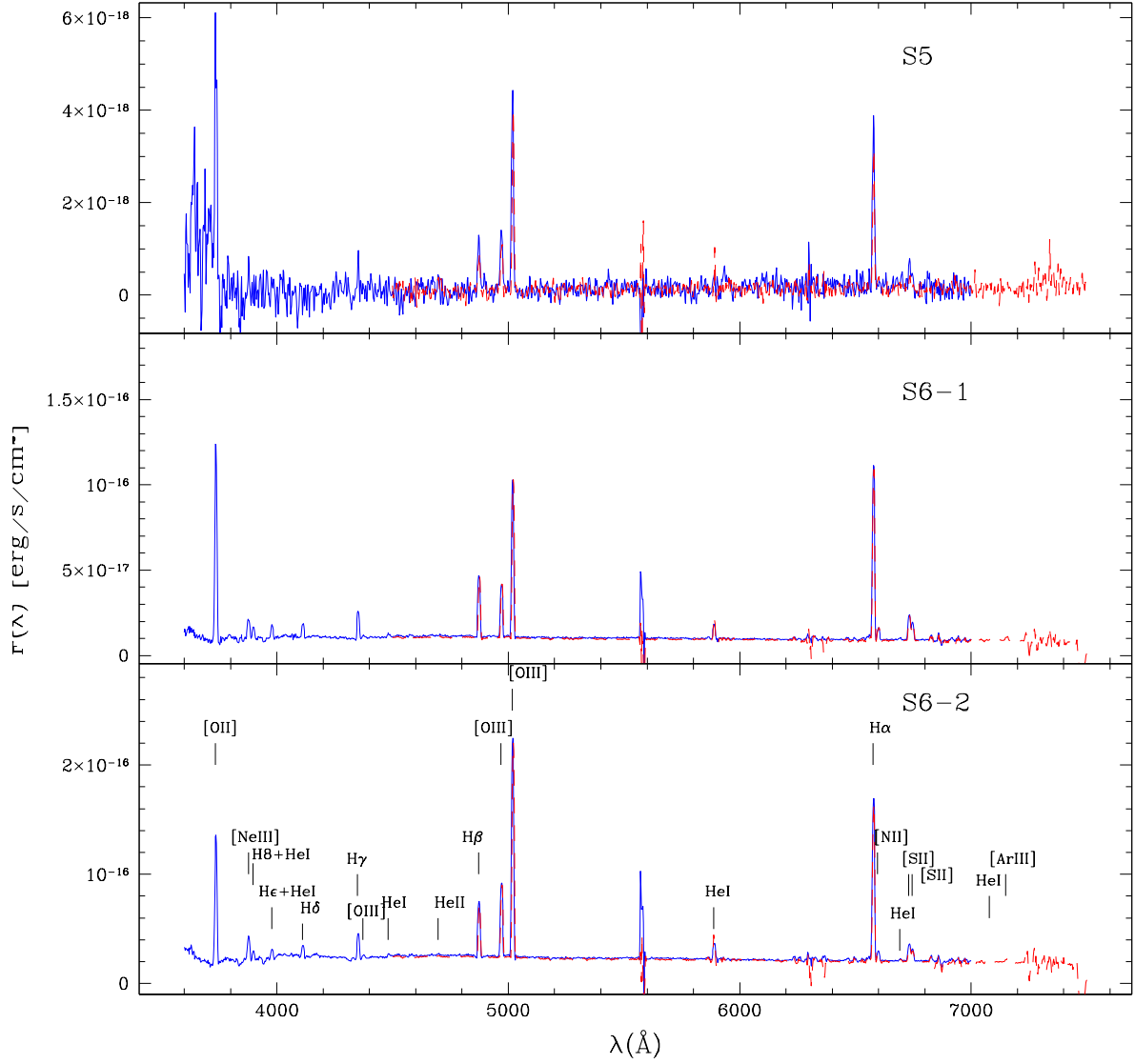}
\caption{Flux-calibrated spectra obtained with the blue configuration (solid line) and with the red configuration (dashed line) for regions 
s5, s6-1 and s6-2.
\label{spectra1}}
\end{figure*}

\begin{figure*}
\epsscale{1}
\plotone{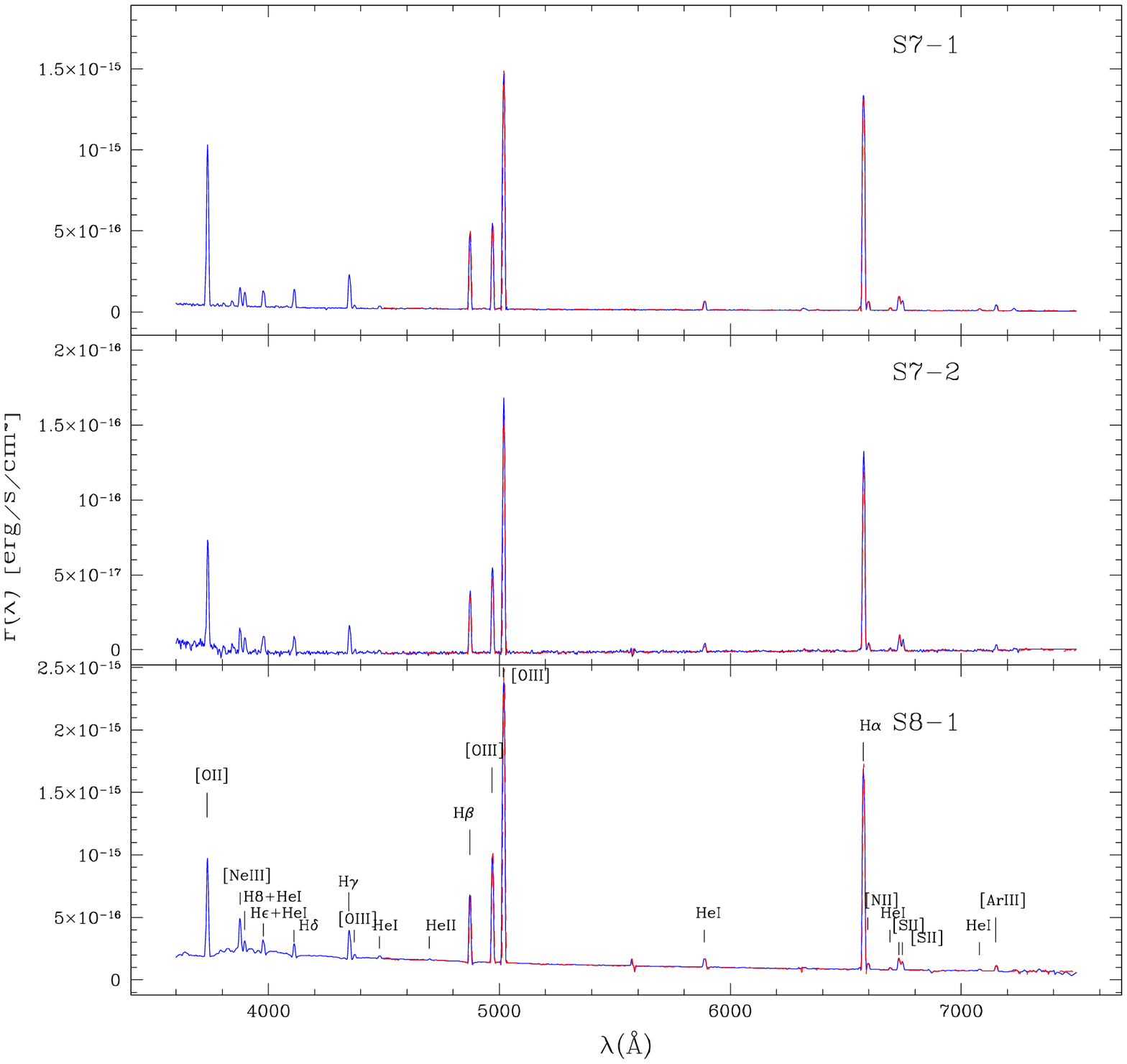}
\caption{Flux-calibrated spectra obtained with the blue configuration (solid blue line) and with the red configuration (dashed red line) for regions 
s7-1, s7-2 and s8-1.
\label{spectra2}}
\end{figure*}

\begin{figure*}
\epsscale{1}
\plotone{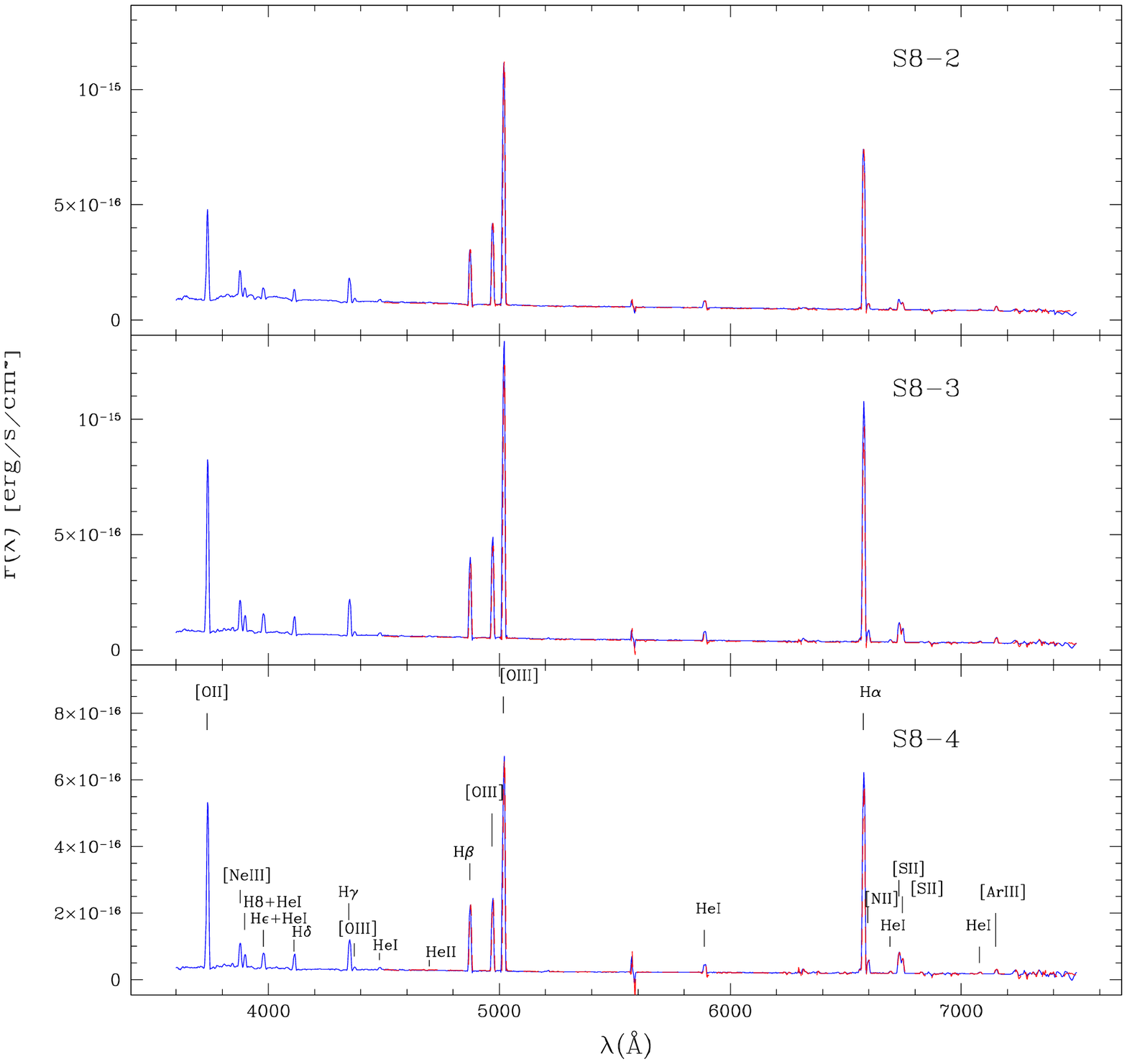}
\caption{Flux-calibrated spectra obtained with the blue configuration (solid blue line) and with the red configuration (dashed red line) for regions 
s8-2, s8-3 and s8-4.
\label{spectra3}}
\end{figure*}

\begin{figure*}
\epsscale{1}
\plotone{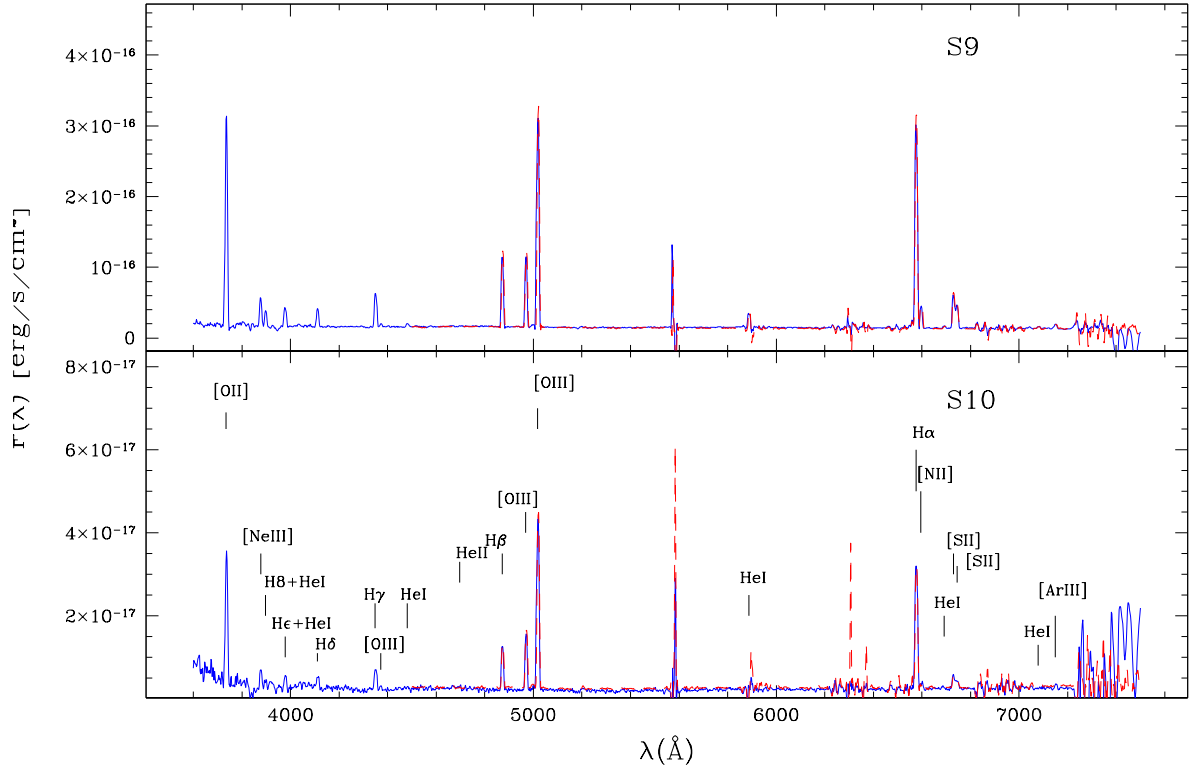}
\caption{Flux-calibrated spectra obtained with the blue configuration (solid blue line) and with the red configuration (dashed red line) for regions 
s9 and s10. The lines at $\sim$6300 \AA \ and   $\sim$6400 \AA\ in the red spectrum of s10 are residuals of sky subtraction.
\label{spectra4}}
\end{figure*}

The {\it calibrate} task was run to obtain flux calibrated spectra. 
The spectra obtained with the blue (GRIS$-$300V) and red (GRIS$-$300V$+$GG435) configurations are shown in Figures~\ref{spectra1}, \ref{spectra2}, \ref{spectra3} and \ref{spectra4}.
We do not show the regions below $\sim$3600 \AA \ and  below $\sim$4500 \AA \ for the blue and red spectra, respectively, because the 
instrumental sensitivity curves significantly decrease below these wavelengths and are very uncertain. Also, below $\sim$3600 \AA, the atmospheric transmission is low. 
Figs.~\ref{spectra1} to \ref{spectra4} show that the blue spectra extend up to $\sim$7000 \AA \  and, in some cases, even redder. 
The red spectra always cover wavelengths redder than 7000 \AA. The agreement in flux between the red and blue spectra is excellent, 
with the only exception of the region below $\sim$5000 \AA, due to the higher uncertainty of the sensitivity curve in the red configuration below these wavelengths. 

With the calibrated spectra in hands, we were able to asses the nature of our PN candidates. Unfortunately, almost the totality of them turned out to be 
background objects. In particular, the targets in slit s1, s2, s11, s12 and s14 show just one prominent emission line around $\sim$5000 \AA, and no (or very weak) stellar 
continuum, and are likely to be Ly$\alpha$ emitting galaxies at $z\sim3$. The target in slit s14 also shows another prominent 
line at $\lambda\sim6157$ \AA, identified as the C IV $\lambda$1539 line at $z\sim3$. The targets in s3 and s13 are two background galaxies at $z\sim0.3$. The object in s4 is likely a QSO.

At this point we are left only with the PN candidate in s5. Some specific lines and line ratios can be used to discriminate between compact H II regions and PNe. 
For instance, an He II $\lambda$4686 emission exceeding a few percent of H$\beta$ unquestionably indicates a  PN; also, [O III]$\lambda$5007/H$\beta$ ratios larger than $\sim$ 4 are found only in PNe \citep[e.g.][]{pena07,magrini09}. From the blue spectrum of s5,  the [O III]$\lambda$5007/H$\beta$ ratio is $\sim$3.4, lower than what typically observed in PNe, but still consistent with a low-ionization PN classification \citep[e.g.][]{depew11,gorny14}. 
On the other hand, due to the low signal-to-noise, the detection of the  He II $\lambda$4686 line in s5 is challenging. For this reason, we performed  
some statistical analysis on the bidimensional spectrum in order to asses the presence of the line, and in case, the significance of the detection.
More specifically, we derived the average counts in 6 $\times$ 7 px$^2$ boxes centered on the line and on different regions of the background. The significance of the detection was computed as $(<C>_{line} - <C>_{bg})/STD_{bg}$ , where  $<C>_{line}$  and $<C>_{bg}$ are the average counts for the line and for the background, respectively, and  $STD_{bg}$ is the standard deviation of the background computed from the different measurements. 
We obtain that the He II $\lambda$4686 line is detected at $\sim$3 sigma. 
From the 1D blue and red spectra (see next section), we measure He II $\lambda$4686/H$\beta$ ratios of 0.28$\sim$0.11 and 0.35$\sim$0.14, respectively. Concluding, there is a significant possibility that s5 is a low-excitation PN.

\section{Emission line measurement}

Emission line fluxes were derived with the {\it deblend} option available in the  {\it splot} iraf task. 
We used this function to fit single lines, group of lines, or blended lines. 
Lines were fitted with Gaussian profiles, treating the centroids and the widths as free parameters. In presence of  blended lines we forced the lines  to have 
all the same widths.  
The continuum was defined choosing two continuum windows to the left and to the right of the line or line complex, and fitted with a linear regression. The final emission fluxes were obtained 
repeating the measurement several times with slightly different continuum choices, and averaging the results. 
The errors were derived from the standard deviation of the different measurements. 
The emission line measurements for the blue and red spectra are given in Tables~1 and 2. The errors, which in the case of bright lines can be very small (even lower than 
$\sim$1\%), do not account for flux calibration errors or for other systematic uncertainties. A comparison between the fluxes measured for the lines in common between the blue and red spectra provides more realistic errors: the differences are typically within $\sim$4\% for the brightest lines, and within $\sim$10\% for the faintest lines. 
 
\subsection{Balmer absorption}

Particular attention was paid to the correction of the emission lines for Balmer absorption due to the underlying NGC~1705' s  stellar population. 
 Balmer absorption lines  are visible in our spectra as absorption wings in correspondence of the H$\delta$, H$\gamma$, and H$\beta$ 
emission lines, while no absorption wings are visible in  the H$\alpha$ region due to the blend with the [N II]$\lambda$6548,84 emission lines. 
We attempted a fit of the absorption wings in order to recover the full strength of the Balmer absorption lines, but these were not deep enough  
to allow for a ``stable'' solution:  the resulting absorption strengths were highly dependent on the particular choice of 
 the input parameters (for instance, the windows for the continuum evaluation). Thus we adopted a different approach than a direct spectral fit to all the regions.  
 More specifically,  we selected a specific region in NGC~1705 with the minimum possible contribution of the emission with respect to the underlying stellar absorption (and with the deepest absorption wings), determined the corrections directly from its spectrum, and then extended the derived corrections to other regions of NGC~1705 using population synthesis models. In the following subsections we describe in details the three different steps of our semi-empirical approach : a) Derivation of the absorption lines 
from an empirical spectrum in a region of NGC~1705;  b) Comparison with predictions from population synthesis models; c) Extension of the correction to all regions in NGC~1705 through population synthesis models.

\subsubsection{a) Absorption lines from  empirical spectral data}

\begin{figure*}
\epsscale{1}
\plotone{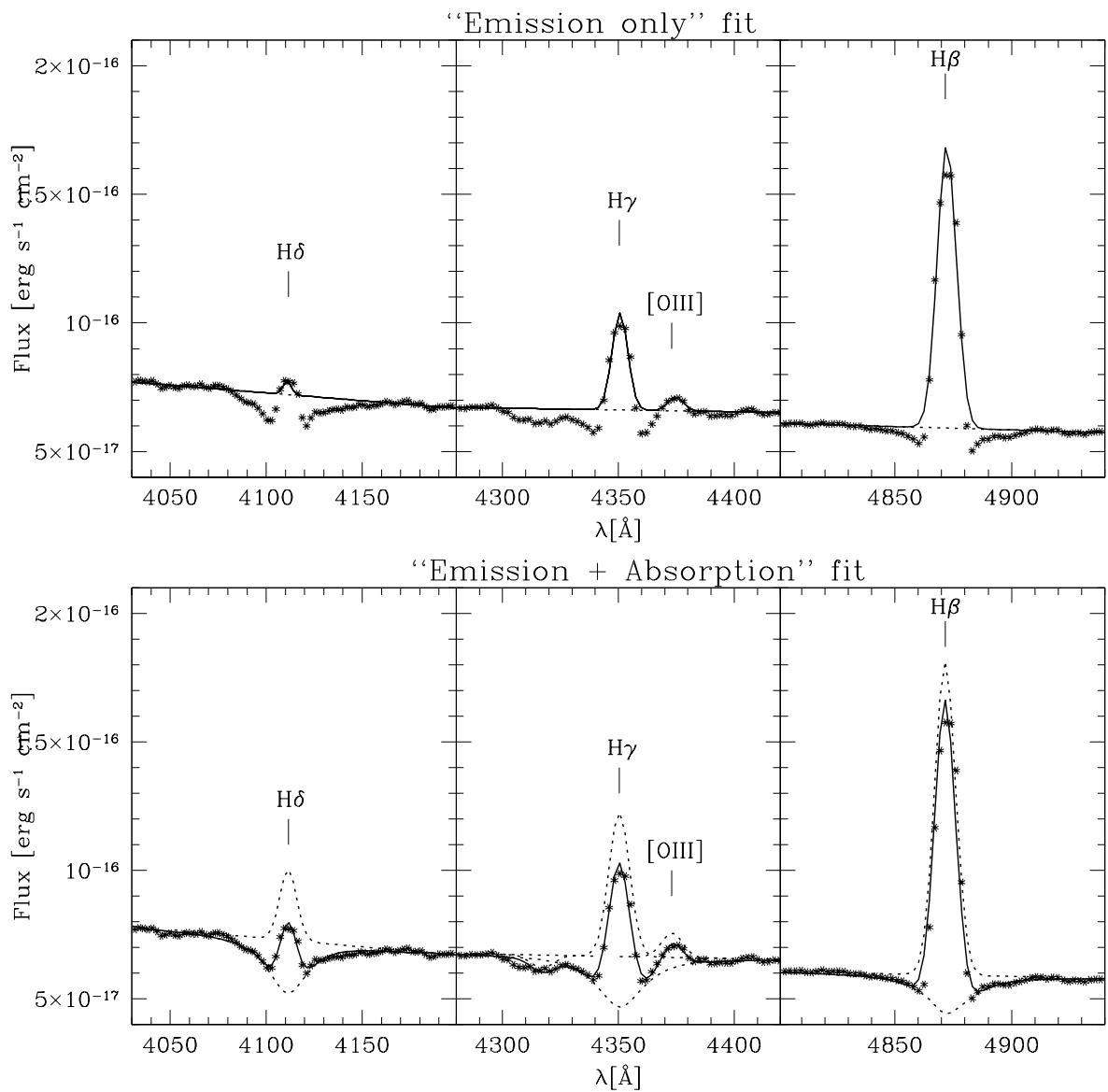}
\caption{Top panels: fit to the H$\delta$, H$\gamma$, and H$\beta$ regions of spectrum s7-a assuming only emission components.
Bottom panels: multi-component fit to the same regions assuming components both in emission and in absorption. 
The asterisks are the observed spectrum, while the continuous line is the best fit. The individual components of the fit are plotted with a dashed line: 
linear continuum, Gaussian profiles for the emission lines, and Voigt profiles for the Balmer absorption lines. 
\label{balmer}}
\end{figure*}

Unfortunately, it was not possible from our data to extract an emission-free spectrum with a signal to noise high enough to allow for a reliable characterization of the Balmer absorption lines in NGC~1705: 
 in fact,  all the slits positioned near the galaxy center, where the stellar surface brightness is higher,  are severely 
contaminated by the strong emission from the ionized gas, while the galaxy stellar surface brightness is too low in more external regions. 
Given the impossibility of working with an emission-free spectrum, we looked for a region with the highest possible strength of the absorption lines in comparison with the emission lines, to allow for a good fit of the absorption wings. This region is located between the two emission peaks at  s7-1 and s7-2  (see Figure~\ref{o3bis}, panel c), and we will call it region s7-a hereafter.  The strong absorption wings around H$\delta$, H$\gamma$ and H$\beta$ in region s7-a are well visible in  Figure~\ref{balmer}. 

To compute the corrections, we adopted the same procedure adopted by  \cite{lee03} for a sample of dwarf irregular galaxies: this consists in fitting the Balmer regions first with a model constructed with both absorption and emission lines, and then with a model that assumes only emission lines
(see Figure~\ref{balmer}). The correction is then computed as the difference between the emission strengths obtained with the ``emission only'' and with the ``emission $+$ absorption'' fits. 
In our specific case, we used the  IRAF {\it splot} task to fit the Balmer absorption features with Voigt profiles (resulting from the convolution of a Gaussian and a Lorentzian profile), and the emission lines with Gaussians,  while the linear continuum was estimated from two windows located on both sides of the emission lines, away from the Balmer absorption wings. 
 For H$\delta$ and H$\beta$, we assumed a line in absorption plus a line in emission, with the peaks and the FWHMs  treated as free-parameters. Instead, the more complex H$\gamma$ region was fitted assuming two absorption profiles (a Gaussian for the G band around $\sim$ 4300 \AA \  and a Voigt profile for the H$\gamma$ in absorption) plus two Gaussians for the H$\gamma$ and [O III]$\lambda$4363 emission lines.
Finally, the H$\alpha$ region (not shown in Figure~\ref{balmer}) was fitted adopting three Gaussians in emission for the H$\alpha$ $+$ [N II]$\lambda\lambda6548,84$ blend, plus a Voigt profile for the H$\alpha$ in absorption. Due to the absence of visible absorption wings, the parameters of the H$\alpha$ absorption line were fixed: the absorption equivalent width was taken from 
population synthesis models (see point {\it b)} in Section 4.1.2), and the (Lorentzian and Gaussian) FWHMs were fixed to the values obtained from the fits to the  H$\delta$, H$\gamma$, and H$\beta$ lines.

 The ``emission $+$ absorption'' fits are shown in the bottom panels of Figure~\ref{balmer}, and the results  are  given in columns 2 and 3 of Table~3. The values were obtained repeating the fitting procedure 
 several times, and computing average values and standard deviations. 
The absorption strengths for the H$\delta$, H$\gamma$ and H$\beta$ lines are  $9.9 \pm 0.4$ \AA, $10.2 \pm 0.6$ \AA, and $8.9 \pm 0.5$ \AA, respectively. 
The ``emission only'' fits are instead shown in the top panels of Figure~\ref{balmer}, and the results are given in column 4 of Table~3.
For each line, the  final correction  (column 5) was computed as the difference between the emission equivalent widths obtained with the ``emission only'' and ``emission $+$ absorption'' fits. 

We notice that the corrections are lower than the full strength of the Balmer absorption lines; this is because 
the absorption lines are wider than the emission ones, with the consequence that only part of the absorption affects the emission line (see Figure~\ref{balmer}).  
We obtain corrections of 4.0 $\pm$ 0.2 \AA, 4.5 $\pm$ 0.3 \AA, 3.1 $\pm$ 0.4 \AA, and 1.5 $\pm$ 0.6 \AA \ to the equivalent widths in H$\delta$, 
H$\gamma$, H$\beta$, and H$\alpha$, respectively. Our values are higher than the average correction of 1.59$\pm$0.56 \AA \ to H$\beta$ derived 
by  \cite{lee03} for their sample of dwarf irregular galaxies, and also higher than the  constant 2 \AA\ correction usually applied in the literature to all Balmer lines \citep[e.g.,][]{mccall85,kennicutt96,lee03,lee04}. Furthermore, our correction to the H$\beta$ 
line is roughly twice as high as that in H$\alpha$. Thus, applying the same correction to H$\beta$ and H$\alpha$ may lead to significant uncertainties in reddening estimates, 
in particular if the emission lines are weak. 
We also notice from Table~3 that there are non negligible 
corrections to the [O III]$\lambda$4363, [N II]$\lambda6548$, and  [N II]$\lambda6584$ lines (0.9 $\pm$ 0.1  \AA, 0.5 $\pm$ 0.1  \AA, and 0.5 $\pm$ 0.1  \AA, respectively), whereas such lines are usually not corrected in the literature.

\subsubsection{b) Comparison with population synthesis models}

Since our results are based on  fits to the absorption wings rather than on direct measurements of the Balmer absorption lines, we performed 
a consistency-check with population synthesis models.  To this purpose, we used the results by \cite{anni03}, hereafter A03,  and \cite{anni09}, hereafter A09, who derived the star formation history (SFH) in 
different regions of NGC~1705 (regions 7 to 0 from the center outwards) from the CMDs of the resolved stars \citep{tosi01}. 
From a comparison of our FORS2-MXU slit mask in Figure~\ref{o3bis} with WFPC2 images of NGC~1705 (see Figure~2 in A09),   
we find that region s7-a defined in Section~4.1.1 is located within region~5 of A09. We thus compare our fits to the empirical spectrum in s7-a (see Section~4.1.1) with population synthesis models based on the SFH from A03 and A09 for region~5. 

In order to construct the synthetic spectrum, simple stellar populations (SSPs) from the Padova group for a metallicity of $Z=0.004$, based on the tracks of \cite{marigo08} \citep[see also][]{chavez09}, were combined according to the SFH of region 5 in NGC~1705. 
The adopted SSPs are based on the MILES empirical spectral library \citep{sanchez06}, which consists of 2.3 \AA \ FWHM optical spectra of $\sim$1000 stars spanning a large range of atmospheric parameters. Balmer absorption lines were then measured on the synthetic spectrum using, as for the real data,  the {\it splot} task in IRAF.  The results are given in  Column~2 of Table~4. The synthetic absorption equivalent widths \footnote{Absorption strengths obtained with the Starburst~99 \citep{sb99} SSPs turned out to be consistent with those obtained using the Padova models. }, $H\delta=8.8 \pm 0.2$ \AA, $H\gamma=8.7 \pm 0.2$ \AA, and  $H\beta=8.7 \pm 0.3$ \AA \ are consistent, within the errors, with the values of $9.9 \pm 0.4$ \AA, $10.2 \pm 0.6$ \AA, and $8.9 \pm 0.5$ \AA \  derived from the empirical spectrum in s7-a. This indicates that our fit to the absorption wings provided sensible values for the Balmer lines, and that our corrections are reliable.  
From the synthetic spectrum of region~5, we also measured an absorption EW in H$\alpha$ of 5.9 $\pm$ 0.2 \AA, which we used in Section~4.1.1 to compute the emission corrections.

\subsubsection{c) Extension of the correction to other regions}

Our corrections, which are given in terms of equivalent widths, do not depend on the continuum absolute flux, but depend on the (luminosity-weighted) age and metallicity of the  underlying stellar population.
Thus, given the spatial variation of the SFH within NGC~1705, we expect different amounts of Balmer absorption over the galaxy field of view. 
To extend the corrections computed for region~5 to other regions, we constructed synthetic spectra for regions 6 to 0 in NGC~1705 using the 
SFHs derived by A03 and A09, and combining the SSPs as described in Section~4.1.2. We want to stress out that the corrections were not  directly derived from the synthetic spectra: what we did was instead to scale the corrections derived from the empirical spectrum in s7-a (Table~3) by a factor predicted from population synthesis models. 
In doing this, we used {\it{ratios}} of absorption lines as predicted by the models and {\it not absolute values}; this minimizes some of the uncertainties 
 related to the adopted models and to the assumed input parameters (e.g., the IMF).

More in details, for each $X$ region (6, 4-3, or 2-1-0), the new corrections were obtained by multiplying the values in col. 5 of Table~3 by the quantity  $EW_{reg X}/EW_{reg5}$, i.e.  by the ratio between the absorption strengths from the synthetic spectra. The results are given in Table~4. The absorption lines from the synthetic spectra, the $EW_{reg X}/EW_{reg5}$ ratios, and the 
final corrections are given in  columns~2, 3 and 4, respectively. 
The comparison of the FORS2-MXU slit mask  with the WFPC2 images of NGC~1705 provides the following spatial association:  s8-1 with region~6; s6-2, s7-1, s7-2, s8-2, s8-3, and s8-4 with region~5; s6-1 and s9 with region 4-3; s5 and s10 with region~2-1-0. Notice that no correction needs to be applied to spectra s5, s7-1 and  s7-2, where both the sky and the galaxy background were subtracted, as described in Section~3.

\subsection{Reddening correction}

For each region, the reddening was estimated from the observed H$\alpha$/H$\beta$ ratio using  the Cardelli et al. (1989) extinction law according to the formula: 

\begin{equation}
E(B-V)= \frac {log R_o/ R_t  }{0.4\times R_V \times [A_{H\beta}/A_V - A_{H\alpha}/A_V ]}
\end{equation}

where $R_o$ and $R_t$ are the observed and theoretical H$\alpha$/H$\beta$ ratios, respectively; 
the magnitude attenuation ratio $A_{\lambda}/A_V$ was taken from the Cardelli' s law adopting $R_V = 3.05$.  
To compute $R_o$, the H$\alpha$ and H$\beta$ emission fluxes were corrected for underlying Balmer absorption as described in Section~4.1. 
We assumed a theoretical Balmer decrement of  $R_t=2.8$ from Table~4.4 of \cite{oster89},  assuming $n_e=100 \ cm^{-3}$ and $T_e=15,000$ K. 
In fact, the measured [S II] $\lambda$6716 / $\lambda$6731 ratios imply densities lower than this ($n_e \lesssim 30 \ cm^{-3}$, Tables~7 and 8)
but from models \citep{panuzzo03} we verified that $R_t$ does not significantly depend on density below $n_e\sim 100  \ cm^{-3}$.

The $E(B-V)$ values derived from the blue and the red spectra are given in Tables~7 and~8. When the H$\alpha$/H$\beta$ ratio resulted into a negative reddening, we 
adopted $E(B-V)=0$. For many regions (the most external ones), the derived reddenings  are consistent with zero or very low levels of extinction. 
However, along some, more central,  lines of sight, we do detect significant extinction. For example, the extinction derived for s7-1 and s7-2 is as high as 
$E(B-V)=0.10$ and 0.16, respectively, which is significantly higher than the Galactic reddening of $E(B-V)=0.045$ from \cite{meurer92}, or  the newer  
$E(B-V)\sim0.007$ from \cite{schlegel} and \cite{schlafly}.
 We notice that our red spectra tend to provide systematically lower reddening values than the blue ones. 

The intrinsic emission line fluxes for the blue and red spectra were computed according to the formula :

\begin{equation}
F_c = F_o \times 10^{0.4\times A_{\lambda}},
\end{equation}

where $F_o$ is the measured line flux, corrected only for the effect of underlying Balmer absorption, and $F_c$ is the extinction-corrected flux; $A_{\lambda}$, the magnitude attenuation at the 
line wavelength, is derived from the  Cardelli et al. (1989) extinction law. We provide in Tables~5 and 6 the emission line fluxes corrected for both 
Balmer underlying absorption and dust extinction. The H$\gamma$/H$\delta$ ratios from the corrected fluxes result into     
1.80 $\pm$ 0.06, 1.77 $\pm$ 0.27, 1.79 $\pm$ 0.08, 1.57 $\pm$ 0.08, 2.03 $\pm$ 0.21, 1.92 $\pm$ 0.15, 1.86 $\pm$ 0.08, 1.87 $\pm$ 0.08, 1.81 $\pm$ 0.10, 1.73 $\pm$ 0.40 
for regions s6-1, s6-2, s7-1, s7-2, s8-1, s8-2, s8-3, s8-4, s9, and s10, respectively. With the exception of region s7-2, they are all 
consistent, within the errors, with the theoretical ratio of H$\gamma/H\delta \sim 1.8$ predicted for case B  recombination in \cite{oster89}.

\section{Chemical abundances}

Temperatures, densities, and chemical abundances were derived using two different tools, the {\it abund} task in IRAF \citep{abund} and the 
PyNeb tool \citep{pyneb}, both based upon the FIVEL program  developed  by  \cite{fivel}. 
The electron density  $n_e$ was computed using the density-sensitive [S II] $\lambda$6716/ $\lambda$6731 diagnostic line  ratio, which provides the density 
of the low-ionization ions. This density was assumed to be constant all over the nebula. 
 For all the regions but s5, where the [O III]$\lambda$4363 was not detected, the [O III]($\lambda$4959$+$5007)/$\lambda$4363 diagnostic line ratio, measured in the blue spectra, was used instead to derive the temperature of the $O^{+2}$ ions
(the so-called ``direct'' $T_e$ method). 
Both the IRAF {\it abund} task and the {\it getCrossTemDen} option in PyNeb adopt an iterative process to derive temperatures and densities:
the quantity derived with one line ratio is inputted into the other, and the process is iterated; at the end of the iteration process, the two 
temperature-sensitive and density-sensitive diagnostics give self-consistent results. 
The high [S II]$\lambda\lambda$6716,31 line ratios measured in NGC~1705 imply very low densities. In those, numerous, cases where the [S II] line ratio exceeded the theoretical 
value for $n_e\rightarrow0$, we provided an upper limit for the density at the  $-1 \sigma$ value of the measured [S II] line ratios. 
Typically, the derived densities are  $\lesssim 30 \ cm^{-3}$. 
In those cases where a solution could not be found at the $-1 \sigma$ level, we assumed a density of $n_e = 30 \ cm^{-3}$ to compute the temperature. We checked that 
assuming $n_e = 10 \ cm^{-3}$ or  $n_e = 100 \ cm^{-3}$ had no effect on the temperatures. 
Direct temperatures could be derived for all regions in NGC~1705 but s5, where the [O III]$\lambda$4363 line was not detected. The derived values are in the range $\sim 12 000 - 15000$ K.  
We provide in Tables~7 and~8 the temperatures and densities obtained with PyNeb. These results are consistent, within the errors, with the values obtained 
 with the  IRAF {\it{abund}} task, but we notice a systematic shift  of $\sim +300$ K for the temperatures  derived with  {\it abund} with respect to those obtained with 
PyNeb.  
 
 Starting from the temperature of the $O^{+2}$ ions directly obtained from the [O III]($\lambda$4959$+$5007)/$\lambda$4363 ratio, we derived the temperatures 
 of the other ions assuming that  $T_e(Ne^{+2})=T_e(O^{+2})$, and $T_e(N^{+})=T_e(S^{+})=T_e(O^{0})=T_e(O^{+})$. In the literature, the relation between 
 $T_e(O^{+2})$ and $T_e(O^+)$ has been widely discussed \citep[e.g.][]{campbell86,perez03,izotov06,pilyugin07}. In this work, we used the formula in equation (14) of \cite{izotov06}  for the intermediate metallicity regime, i.e.  $t_e (O^+)= -0.744 + t_e (O^{+2}) \times (2.338 - 0.610\times t_e (O^{+2}))$, where $t_e = T_e/10^4$ K.  
 The temperature of the $Ar^{+2}$ and $S^{+2}$ ions was assumed to be intermediate between that of $O^+$ and $O^{+2}$, and from  \cite{izotov06}, eq. (15),  we adopted 
$t_e (Ar^{+2})=t_e (S^{+2})=  -1.276 + t_e (O^{+2}) \times (2.645 - 0.546\times t_e (O^{+2}))$.
We used the [N II] $\lambda$6548, 6584 lines for the determination of the $N^+$ abundance, and the [O II] $\lambda$3727 and  [O III] $\lambda$4959, 5007 lines for the 
$O^+$ and  $O^{+2}$ abundances, respectively; when detected (regions  s7-1, s8-1, s8-2, s8-3, and s8-4), the  [O I] $\lambda$ 6300 line provided the $O^{0}$ abundance.
We used the [Ne III] $\lambda$ 3869 line for the determination of the $Ne^{+2}$ abundance; the [S II] $\lambda$ 6716, 6731 doublet and, when detected, the   
[S III] $\lambda$ 6312 line, for the determination of the $S^+$ and $S^{+2}$ abundances, respectively;  from the [Ar III] $\lambda$ 7135 we derived the $Ar^{+2}$ abundance.   
The fluxes used for the abundance computations are those listed in Tables~5 and 6, i.e. corrected for the effect of underlying Balmer absorption and dust extinction.    

 For all regions but s5, the ion abundances obtained with the {\it getIonAbundance} option in PyNeb are given in Tables~7 and 8. Due to the wavelength coverage, the abundance of the $O^+$ and $Ne^{+2}$ ions  
was derived only from the blue spectra, while for all the other ions we derived two different values respectively from the blue and from the red spectra (although in both cases the temperature was that 
derived from the blue spectra). The ion abundances derived with PyNeb and with {\it abund} are generally consistent within the errors, nevertheless we notice systematic trends 
for some elements:  the $O^{+2}$, $Ne^{+2}$, $N^{+}$, and $S^{+}$ abundances derived with PyNeb are systematically higher than those obtained with  {\it abund}  by $\sim$0.03, 0.04, 0.03, and 0.04 dex, respectively;  on the other hand, the abundances of the $O^{+}$ and $S^{+2}$ ions are systematically lower by $\sim$0.02 and 0.01 dex, respectively.

The total element abundances can be derived from the abundances  of ions seen in the optical spectra using ionization correction factors (ICFs). 
In the case of oxygen, the most abundant ions $O^+$ and $O^{+2}$ are seen in our blue spectra, allowing for an immediate derivation of the total abundance as
 $O/H= O^+/H + O^{+2}/H$. For regions s71, s81, s82, s83, and s84, where we measured the [O I] $\lambda$ 6300 line, we added the  contribution from the 
 $O^0$ ion to the computation of the total oxygen abundance. This contribution however is very small, amounting to $\lesssim$0.02 dex. 
Concerning the $O^{+3}$ ion,  \cite{izotov06} estimated a contribution larger than $\sim$1\% to the total oxygen abundance only in the highest-excitation 
H II regions with $O^+ /(O^+ + O^{+2}) \lesssim 0.1$. From our Tables~7 and 8, we notice that this condition does never occur 
in our regions, implying that the contribution from the $O^{+3}$ ions can be neglected. 
The oxygen abundances derived for regions s6-1 to s10 in NGC~1705 are in the range   $7.75 \lesssim 12 + \log(O/H) \lesssim 8.05$ (see Tables~7 and 8), 
with an average  value and standard deviation of $7.91 \pm 0.08$ dex.  The tabulated values were computed from the blue spectra, due to the lack of the [O II] $\lambda$3727 line 
in the red spectra. 

To compute the abundances of the other elements, we adopted the following ICFs from  \cite{izotov06}: 
 
 \begin{equation}
 ICF(N^+)= -0.809 v + 0.712 + 0.852/v, 
  \end{equation}
  
  \begin{equation} 
 ICF(Ne^{+2}) = -0.405 w + 1.382 + 0.021/w, 
  \end{equation}
  
 \begin{equation}
 ICF(S^+ + S^{+2})=0.155 v + 0.849 + 0.062/ v, 
 \end{equation}
 
  \begin{equation}
 ICF(Ar^{+2})=0.285 v + 0.833 + 0.051/v, 
 \end{equation}
 
 where
 
 \begin{equation}
 v = O^+/(O^+ + O^{+2}),  w = O^{+2}/(O^+ + O^{+2}).
  \end{equation}

The ICFs were computed from the $O^+$ and $O^{+2}$ abundances derived from the blue spectra, but were used to correct the ions both for the blue and the red spectra. 
The total sulfur abundance could be computed only for those regions where we measured the [S III] $\lambda$6314 line, providing the 
$S^{+2}$ abundance.  Our results for the total abundances of N, O, Ne, S, and Ar  in regions s6-1 to s10 are given in Tables~7 and 8. 
The abundances of N, S, and Ar obtained from the blue and red spectra are consistent within the errors. Hereafter, we will use the abundances based on the blue spectra.

\section{Line ratio diagnostic diagrams \label{dd}}

The \cite{bpt} and \cite{vo87} diagnostic diagrams are a powerful tool to classify ionized regions according to the main excitation mechanism.
In the most commonly used $\log([O III]\lambda5007/H\beta)$ versus $\log([N II]\lambda\lambda6548,84/H\alpha)$, $\log([S II]\lambda\lambda6716,31/H\alpha)$, 
or  $\log([O I]\lambda6300/H\alpha)$ diagrams, starbursts fall onto the 
left-hand region, while Seyfert galaxies are located in the upper right, and LINERs lie in the lower right zone. 
\cite{kewley01}, hereafter Ke01, produced starburst grids on the optical diagnostic diagrams combing stellar population synthesis models with the MAPPINGS III photoionization 
and shock code \citep[][and references therein]{dopita00} for a wide range of metallicities ($Z = 0.05 -3$ times solar) and ionization parameters 
($q=5\times10^6 - 3\times10^8 \ cm \ s^{-1}$). Using these grids, Ke01 defined a theoretical upper limit for starburst models on the optical diagnostic diagrams, called the ``maximum starburst line'' (MSL): objects lying above and to the right of the MSL can not be explained with pure starburst models, but require an additional contribution from a harder ionizing source such as an active galactic nucleus or shocks.  On the other hand, objects located below and to the left of the MSL may have a non negligible contribution  
(up to $\sim30\%$, Ke01) from excitation mechanisms different than stellar photoionization. 

\begin{figure*}
\epsscale{0.9}
\plotone{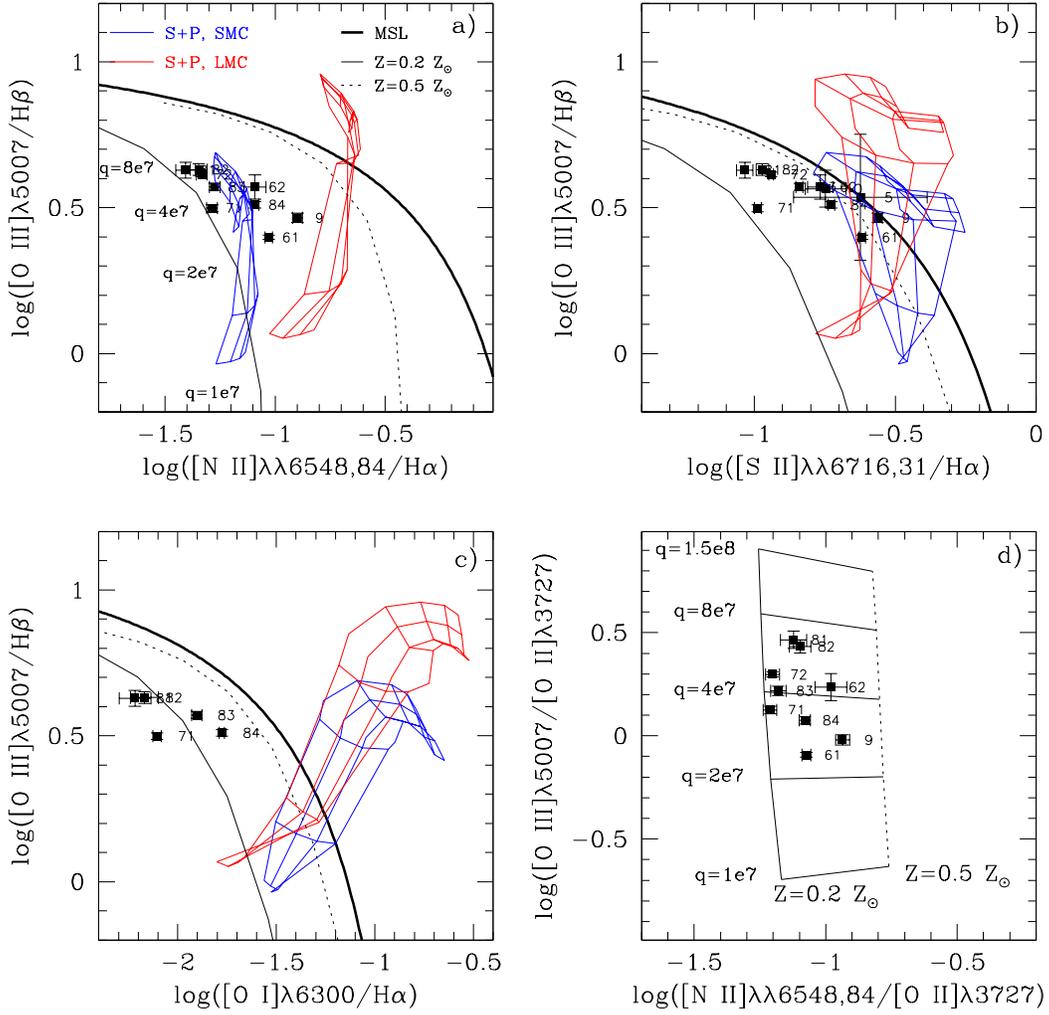}
\caption{Diagnostic diagrams 
 for regions s5 to s10 in NGC~1705 compared with models. In panels a) to d), the solid and dotted curves are the starburst models of \cite{kewley01} (Ke01) obtained with the PEGASE v2.0 code for 
a continuous star formation, density $n=10\  cm^{-3}$, and for metallicities of 0.2 and 0.5 times solar, respectively. 
Panels a), b), and c): the models run from bottom right to top left in the diagrams with increasing 
ionization parameter q. The thick solid curve is the maximum starburst line (MSL) defined by Ke01. 
The grids are the shock $+$ precursor (S$+$P) models of \cite{allen08} for the SMC (blue) and LMC (red) chemical compositions. The shock models span velocities in the range $125-800$  km  $s^{-1}$ from bottom to top, and magnetic fields in the range $0.5 - 10 \ \mu G$ from left to right. 
Panel d): in the Ke01 models, q decreases from top to bottom, while the metallicity increases from left to right. 
 \label{bpt}}
\end{figure*}

In order to investigate if excitation mechanisms other than pure stellar photoionization are present in NGC~1705, we compared the line ratios observed in regions 
s5 to s10 against models.  
In Figure~\ref{bpt} we plotted the Ke01 models based on the PEGASE v2.0 population synthesis code \citep{pegase}, utilizing the Padova group tracks \citep{bressan93}, for a continuous star formation, density $n=10\ cm^{-3}$, and metallicities of 0.2 and 0.5 times solar. 
The Ke01 models suggest a higher metallicity for NGC~1705 than that derived by the direct method, but large differences are observed in the starburst grids depending on the adopted population synthesis code (STARBURST~99 or PEGASE v2.0), stellar evolution tracks, continuous or instantaneous burst, and on other input parameters. 
In the diagrams of panels a), b), and c), the models run from top left to bottom right with decreasing ionization parameter q, defined as the ratio between the ionizing photon flux through a unit area 
and the local number density of hydrogen atoms. We found that the Ke01 models based on the STARBURST99 \citep{sb99} code utilizing the Geneva group tracks \citep{geneva} provide a worst agreement with the NGC~1705 data. The thick solid line in Figure~\ref{bpt} defines the maximum starburst sequence by Ke01. 

In panels a), b), and c) of Figure~\ref{bpt}, we also plotted the fast shock$+$precursor models of \cite{allen08} for the SMC and LMC chemical compositions obtained with the MAPPINGS III code. In these models, the ionizing radiation generated by the cooling of hot gas behind the shock front generates a strong radiation field of extreme ultraviolet and soft X-ray photons, which leads to significant photoionizing effects. At shock velocities above a certain limit ($v\sim170$  km $s^{-1}$), the ionization front velocity exceeds that of the shock, and expands to form a precursor H II region ahead of the shock. For a given metallicity and pre-shock gas density, the emission line ratios are mainly determined by the shock velocity and by the magnetic field, which acts to limit the compression through the shock. 
For the shock models in Figure~\ref{bpt}, the $[O III]/H\beta$ ratio increases with shock velocity (with plotted velocities in the range $125-800$  km $s^{-1}$ ), while the $[N II]/H\alpha$, $[S II]/H\alpha$,
and $[O I]/H\alpha$ ratios increase with magnetic field (with plotted magnetic fields in the range $0.5 - 10 \  \mu G$).  The models assume a density of $n=1 \ cm^{-3}$ (the only value available for non-solar abundance models). 
The diagrams in panels a), b), and c) clearly show that shocks are better separated from stellar photoionization in the $[S II]/H\alpha$ and $[O I]/H\alpha$ ratios, while 
 photo-ionized and shock excited components are strongly degenerate in the  $[O III]/H\beta$ vs $[N II]/H\alpha$ diagram \citep[e.g.][]{mat72,dopita97}. 

From the comparison of the NGC~1705 data with models in Figure~\ref{bpt} we can draw the following conclusions. 
First,  the emission line ratios measured in regions s5 to s10 are below the MSL in all the a), b), and c) diagnostic diagrams, implying that the emission lines  
can in principle be explained with pure stellar photoionization. Furthermore, a major contribution from shock excitation can be excluded on the basis of diagrams b) and c), 
where the data show few or no overlap with the shock models. Indeed, the data completely avoid the shock grids in the  $[O III]/H\beta$ versus $[O I]/H\alpha$ diagrams, 
where stellar photoionization and shock excitation are best separated  (but notice that the most external regions  s5, s6-1, s6-2, s9, and s10 are missing in this diagram). 
Of course, we can not exclude a minor contribution from shocks projected on relatively stronger stellar photoionized regions. 
Furthermore, it is possible that our data are biased toward stellar photoionization, 
and that shock-dominated zones are actually present in NGC~1705; in fact, non-radiative ionization is typically confined to regions of low H$\alpha$ surface brightness, as observed by many authors who studied the occurrence of shocks in star forming galaxies \citep{ferguson96,martin97,calzetti04}, while our slits were positioned in regions of relatively high [O III]  (and H$\alpha$) surface brightness in order to obtain spectra with a good signal-to-noise. Indeed, the presence of  a kiloparsec-scale  supershell of ionized gas centered on the nucleus of NGC~1705 and expanding at a velocity of $(53\pm10)$ km $s^{-1}$  \citep{meurer92,sahu97,heckman01} suggests that shock excitation is a possible mechanism.

The second consideration from an inspection of the diagrams in Figure~\ref{bpt} is that the data tend to move from top left to bottom right in diagrams a), b) and c) (decreasing $[O III]/H\beta$ values and increasing $[N II]/H\alpha$, $[S II]/H\alpha$,  and $[O I]/H\alpha$ ratios) with increasing distance from the galaxy center. 
We can exclude that this effect is due to dust extinction, given the very weak effect of dust absorption on the BPT diagrams and the low $E(B-V)$ values 
derived for NGC~1705 (see Tables~7 and 8); we can also exclude a density effect, as that shown in \cite{califa} for a sample of $\sim$ 5000 H II regions in the CALIFA survey, given the homogeneously low $n_e$ derived for regions s6 to s10.  
The behaviour that we see in Figure~\ref{bpt} has already been observed for other galaxies,  and it is commonly explained with the gradual and continuous transition from the core of the H II regions  into the diffuse ionized gas (DIG) 
\citep[e.g.,][]{ferguson96,martin97}. 
In fact, the trend of the data from top left to bottom right follows that of the  starburst models of Ke01 with decreasing ionization parameter q. 
\cite{califa} have shown that there is a strong correlation between 
position in the BPT diagram, metallicity and ionization parameter: more metal-rich H II regions have the lowest ionization parameters, and are located in the bottom right part of the 
star-forming sequence, while low-metallicity H II regions display the highest ionization parameters and are located top left. However, in the case of NGC~1705, the observed trend 
from top left to bottom right in the BPT diagram  with increasing galactocentric distance goes in the opposite direction, being associated to a decrease in metallicity as derived by the direct $T_e$ method. 
In this context, the observed gradual decrease in the ionization parameter could be explained with the dilution of radiation from a centralized source with increasing distance from the main star forming regions.

\begin{figure*}
\epsscale{1}
\plotone{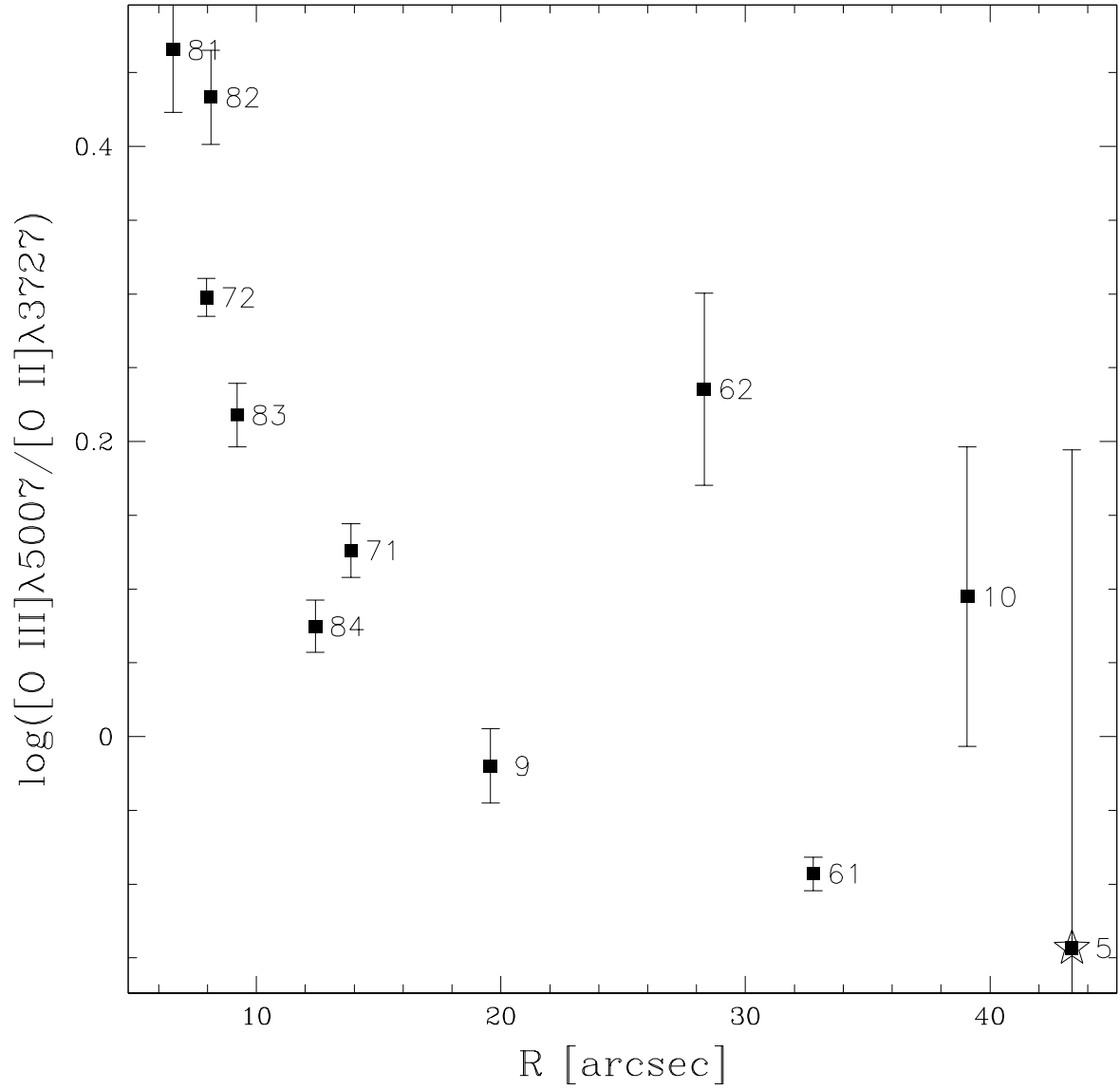}
\caption{
$\log([O III]\lambda5007/[O II]\lambda3727)$ line ratio versus projected distance R from the galaxy center for regions s5 to s10 in NGC~1705. 
The possible low-ionization PN is indicated with a star.
 \label{ion}}
\end{figure*}

 \cite{dopita00} showed that the $[O III]\lambda5007/[O II]\lambda3727$ versus  $[N II]\lambda\lambda6548,84/[O II]\lambda3727$ diagram provides an optimal separation between the ionization parameter and the chemical abundance. This diagram is shown in panel d) of Figure~\ref{bpt}, where the NGC~1705 data have been plotted against the Ke01 starburst models for metallicities of 0.2 and 0.5 solar. In this metallicity range, the $[O III]\lambda5007/[O II]\lambda3727$ ratio is almost independent of metallicity and decreases with decreasing ionization parameter q. For the plotted models, the data span ionization parameters from $q=2\times10^7  \ cm \ s^{-1}$ to  $q=8\times10^7  \ cm \ s^{-1}$, with the most internal regions (s8-1, s8-2) having the largest q values, and the most external ones (s6-1, s9) corresponding to the lowest q values. 
To better show this radial trend, we have plotted in Figure~\ref{ion}  the $[O III]\lambda5007/[O II]\lambda3727$ ratio against the projected distance of regions s5 to s10 from the galaxy center. This plot suggests a well defined decreasing trend of the ionization parameter with increasing distance from the galaxy center, with regions s6-2 and s10  as outliers. 
This trend indicates that the main ionizing source is located within the central $\sim$6.6 arcsec ($\sim$160 pc), which is the distance of region s8-1 from the galaxy center, and that the diluted radiation from this centralized source is able to ionize the gas out to at least $\sim$32.8 arcsec ($\sim$800 pc, the distance of region s6-1). 
Indeed, Figure~\ref{rgb} shows that the high surface brightness optical component of NGC~1705, where the young starburst is concentrated, is well within our innermost region s8-1. Young blue stars are found in correspondence of two of our targeted  H II regions, i.e. regions s7-1 and s8-1. Through a comparison with the map of stars 
younger than $\sim$5 Myr derived from the CMDs in \cite{anni09}, Figure~14,  we ascertained that these stars are indeed young enough to be able to ionize the gas. 
Nevertheless, the trend of the ionization parameter with distance from the galaxy center suggests that 
the main ionizing source is the most central starburst, or perhaps the super star cluster (SSC), in NGC~1705, and that local association of young stars likely contribute to a minor extent.  
Regions  s6-2 and s10, which deviate from the global decreasing trend  (notice however the particularly large errors in their measured line ratios), could be affected by different excitation mechanisms than pure stellar photoionization in a larger amount.

\section{Results and Discussion}

\subsection{Comparison with literature abundances}

We compared our results with the work of  \cite{lee04} who derived chemical abundances for H II regions in NGC~1705. 
Lee et al. obtained long-slit spectra with the EFOSC2 instrument on the 3.6 m telescope at ESO La Silla Observatory for 16 H II regions, and detected the [O III]$\lambda$4363 line in five of them. There are a few H II regions in common with our sample, namely region s8-1 (their A1) and region s7-1 (their region C1); our regions s8-2 and s8-3 partially overlap with their regions A2 and A3, respectively. 
Among the regions in common, the [O III]$\lambda$4363 is detected only in region A3, while for regions A1 and C1 Lee et al. provide an upper limit for the flux. 
For region A3, our oxygen abundance is $12 + \log(O/H)=7.97\pm0.02$, while they obtain $8.21^{+0.11}_{-0.16}$, which is consistent with our results within 3$\sigma$ errors.  
For regions s8-1 and s7-1 our oxygen abundances are $12 + \log(O/H)=7.95\pm0.04$ and $7.93\pm0.04$, respectively, while for A1 and C1 they provide lower limits of
$12 + \log(O/H)>7.57$ and $>8.16$, respectively. Thus, while our result for region s8-1 is consistent with their abundance, our value for region s7-1 turns out significantly lower than their lower limit. 
For the five regions with detected [O III]$\lambda$4363 line, the average oxygen abundance by \cite{lee04} is  $12 + \log(O/H)=8.22\pm0.05$, while our average oxygen abundance is $12 + \log(O/H)=7.91\pm0.08$ (see Table~9).
However, we notice that all the  five H II regions in Lee et al. are relatively close to the galaxy center, while some of our slits are located at much higher distances; 
due to the evidence for a negative oxygen metallicity gradient in NGC~1705 (see Section~7.3),  
we re-computed the average abundance using only the most central regions s7-1, s7-2, s8-1, s8-2, s8-3, and s8-4, 
and obtained $\log(O/H)=7.96\pm0.04$, which is still $\sim0.26$ dex lower than the 
Lee et al.'s abundance. 
We identified a few reasons for the discrepancy between our abundances on those by Lee et al. : 1) First of all, our temperatures are systematically higher than those derived by them (their average $T_e(O^{+2})$ is  $\sim11700 \  K$, while our average   $T_e(O^{+2})$ is $\sim13000 \ K$ for the regions in slits 7 and 8); our higher temperatures  imply 
$\sim0.15$  dex lower abundances.  Although both temperature estimates are based on the direct detection of the [OIII]$\lambda$4363 line, we were able to detect this feature with a better signal-to-noise than Lee et al., and thus we tend to consider our temperature estimates more reliable;   
2) In the second instance, we notice that the [O II]$\lambda3727$/H$\beta$ reddening-corrected fluxes published by Lee et al. are systematically higher than ours (almost twice as high for regions C1, A1, and A2 in common with us), while the [O III] fluxes are more in agreement with our values.  
Their higher [O II] fluxes imply $\sim$ 0.1 dex higher (total) oxygen abundances. Possible reasons for the discrepancy between the [O II] fluxes are the drop 
in the instrumental sensitivity below $\sim$4000 \AA\  and the higher atmospheric extinction toward bluer wavelengths: systematic uncertainties in the determination 
of the sensitivity curve or in the atmospheric extinction function in this wavelength range could lead to significant differences in the estimated emission fluxes;  
3) Finally, we notice that our $t(O^{+})$ temperatures are based on the \cite{izotov06} relation, while Lee et al. adopted \cite{campbell86}; the Izotov et al. formula provides $\sim$ 500 k higher $O^{+}$ temperatures, implying  $\sim 0.01- 0.03$ dex lower total oxygen abundances.

\begin{figure*}
\epsscale{0.9}
\plotone{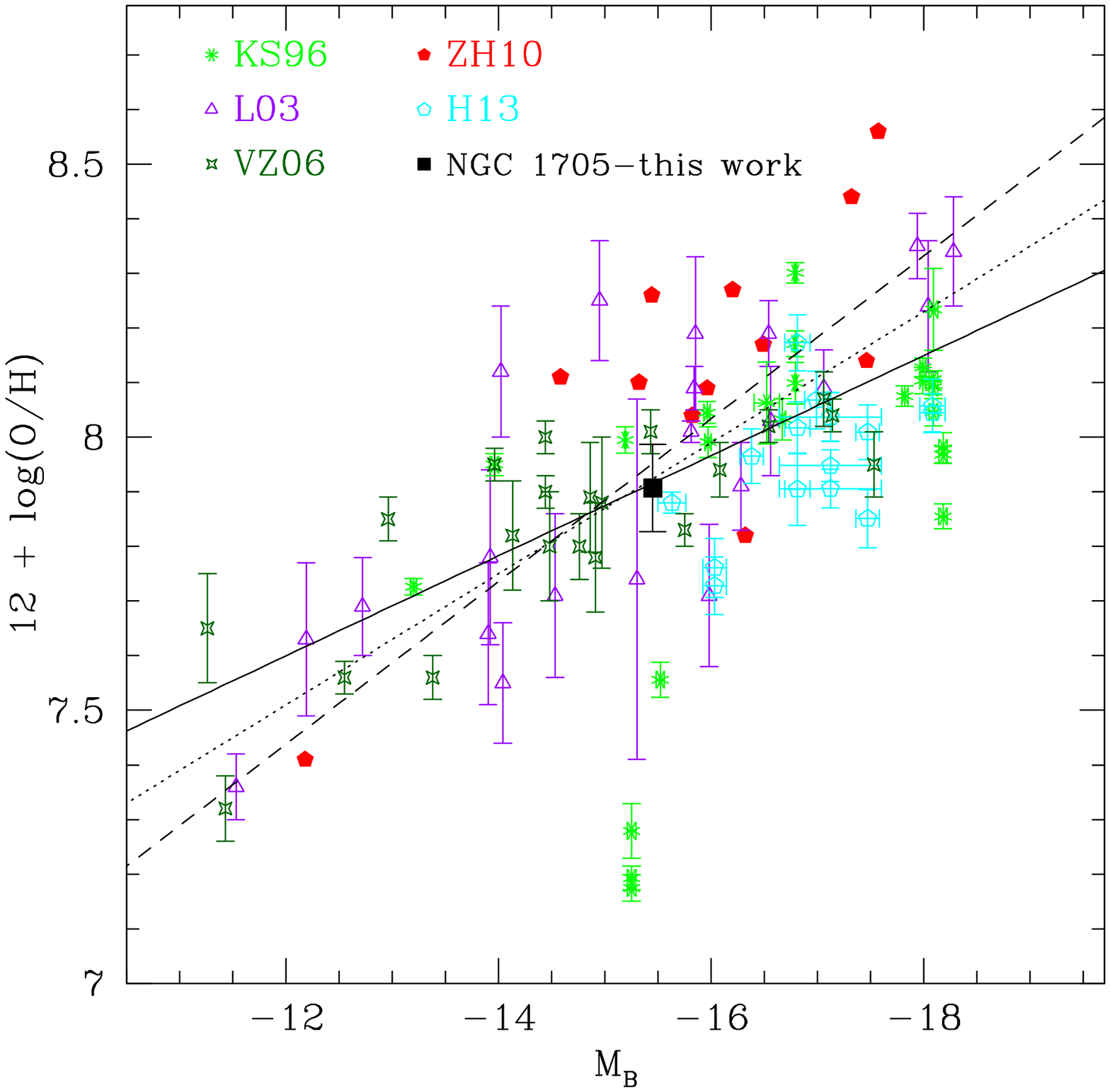}
\caption{Total oxygen metallicity versus galaxy absolute B magnitude for different samples of star-forming dwarf galaxies  in the literature.
 KS96: metal-poor galaxies from \cite{ks96}; L03: dwarf irregular galaxies in the field from \cite{lee03a}; VZ06: isolated dwarf irregular 
 galaxies from \cite{vanzee06}; ZH10: blue compact dwarf galaxies from \cite{zhao10};   
H13:  dwarf irregular galaxies from \cite{haurberg13}; black filled square: average abundance in  NGC~1705 from this work. 
 The solid line is the 3-$\sigma$ rejection linear least square fit to the data points: $12 + \log(O/H)= (-0.092 \pm 0.010) \times M_B + (6.50 \pm 0.17)$, $rms=0.17$. 
 The dashed and dotted lines are respectively the relations from \cite{vanzee06} and \cite{haurberg13}, equation (2).  
 \label{metallicity}}
\end{figure*}

We investigated how our results for NGC~1705 compare with the luminosity-metallicity relationship for other samples of dwarf galaxies \citep{ks96,vanzee06,zhao10,haurberg13}. 
Our results are illustrated in Figure~\ref{metallicity}, where we plotted the absolute B magnitude ($M_B=-15.45$ from $B=13.09$, corrected for Galactic extinction \citep{paz03}, and ($m-m_0)=28.54$ \citep{tosi01}) versus our average oxygen abundance  ($12 + \log(O/H)=7.91\pm0.08$), together with abundance estimates for samples of dIrrs and BCDs in the literature.  The solid line in Figure~\ref{metallicity} is the 3 $\sigma$-rejection linear least squares fit to literature data, excluding NGC~1705
($12 + \log(O/H)= (-0.092 \pm 0.010) \times M_B + (6.50 \pm 0.17)$, $rms=0.17$).  This is flatter than the relations obtained by  \cite{vanzee06}, and by  \cite{haurberg13} (equation 2).  Our average abundance estimate for NGC~1705 is in agreement with all the three metallicity-luminosity relations.

\cite{lee04} derived also abundances for nitrogen, neon, and argon. From their Table~8, we computed average abundances and standard deviations of $12 + \log(N/H)=6.51\pm0.06$ ($\log(N/O)=-1.69\pm0.10$),   
$12 + \log(Ne/H)=7.81\pm0.07$ ($\log(Ne/O)=-0.41\pm0.05$), and  $12 + \log(Ar/H)=5.93\pm0.05$ ($\log(Ar/O)=-2.30\pm0.07$) (see Table~9). 
Averaging our regions s6-1 to s10, we derive instead mean abundances and standard deviations of $12 + \log(N/H)=6.63\pm0.07$ ($\log(N/O)=-1.27\pm0.11$),   
$12 + \log(Ne/H)=7.19\pm0.12$ ($\log(Ne/O)=-0.71\pm0.06$),  and  $12 + \log(Ar/H)=5.62\pm0.06$ ($\log(Ar/O)=-2.31\pm0.03$). We also derive a sulfur average abundance 
of $12 + \log(S/H)=6.34\pm0.04$ ($\log(S/O)=-1.61\pm0.04$). 
If we use only regions s7-1, s7-2, s8-1, s8-2, s8-3, and s8-4 to compute the average abundances, we obtain $12 + \log(N/H)=6.62\pm0.05$ ($\log(N/O)=-1.34\pm0.06$),   
$12 + \log(Ne/H)=7.26\pm0.09$ ($\log(Ne/O)=-0.70\pm0.07$),  and  $12 + \log(Ar/H)=5.65\pm0.05$ ($\log(Ar/O)=-2.32\pm0.03$).
Thus, while our average central $N/H$ abundance is marginally consistent within the errors with that of Lee et al., 
(the effect of their lower temperatures is partly compensated by their slightly lower [N II]$\lambda$6584 fluxes),  our $\log(N/O)$ ratio is $\sim$0.35 dex higher, as a consequence of our lower oxygen abundance. There is significant discrepancy also for Ne and Ar
, since the Lee et al. abundances  are $\sim0.55$ and $\sim$0.28 dex higher than ours. 
For Ne, the particularly large difference is due to the combined effect of their lower temperatures and their $\sim$ 50\% higher [Ne III]$\lambda$3869  fluxes. The authors themselves noticed the anomalously high Ne abundances  derived in their study.  
Figures~\ref{elements} and ~\ref{nsuo}, where we compared our abundances with literature values for different samples of late-type dwarf galaxies \citep{ks96,it99,vanzee06,guseva11,haurberg13,esteban14,nicholls14},  show that our abundances for NGC~1705 are in very good agreement with literature trends.

\subsection{Relative abundances of N and $\alpha$-process elements}

\begin{figure*}
\epsscale{1}
\plotone{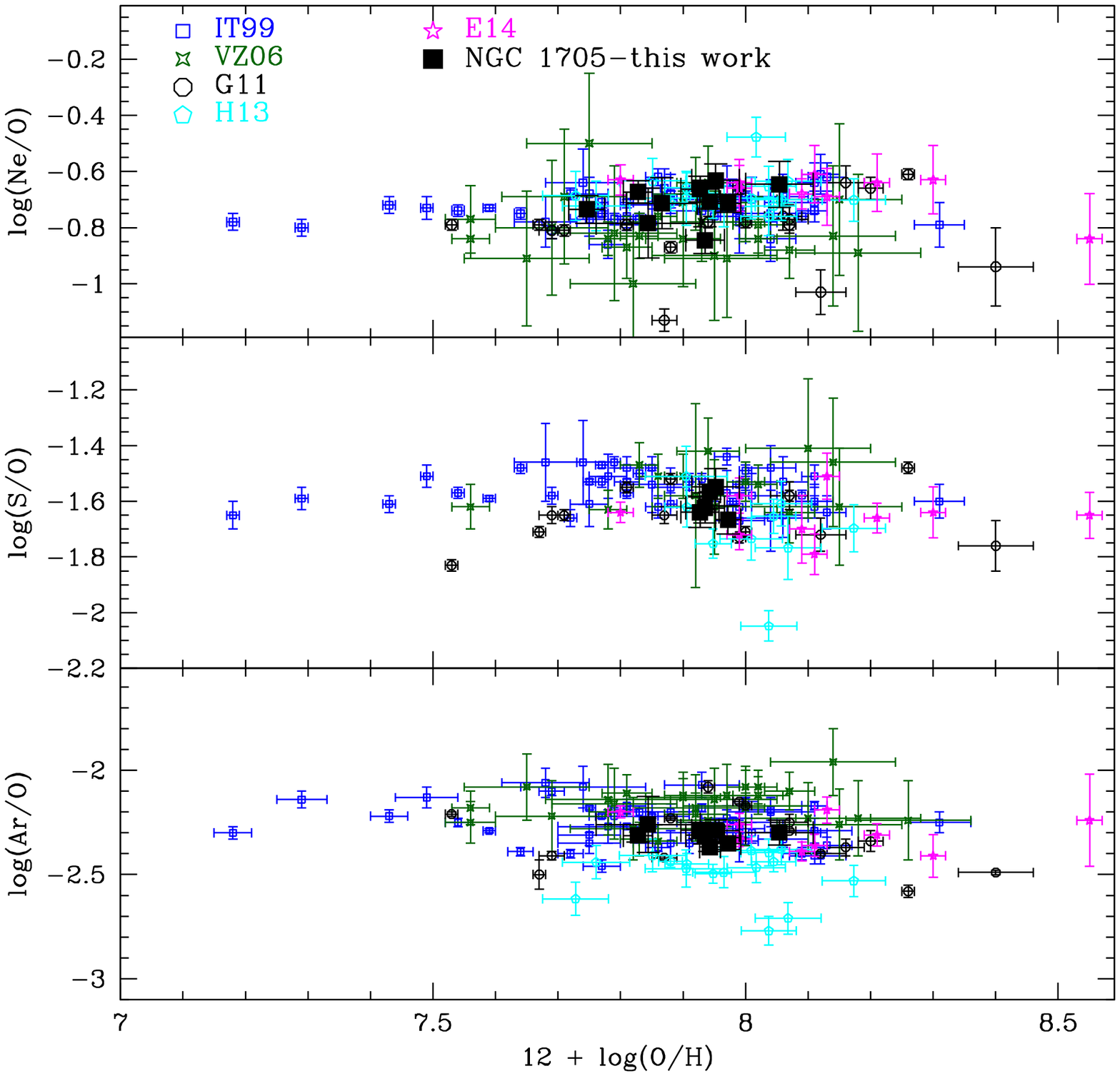}
\caption{Ne, S, and Ar relative abundances versus $12+\log(O/H)$ for different samples of star-forming dwarf galaxies  in the literature.
IT99:  low-metallicity blue compact galaxies from \cite{it99}; VZ06: isolated dwarf irregular galaxies from \cite{vanzee06}; G11: low-metallicity emission-line galaxies from \cite{guseva11}; 
H13:  dwarf irregular galaxies from \cite{haurberg13}; E14: low-metallicity star-forming galaxies from \cite{esteban14}; 
black filled squares: abundances in H II regions of NGC~1705 from this work. 
 \label{elements}}
\end{figure*}

\begin{figure*}
\epsscale{0.9}
\plotone{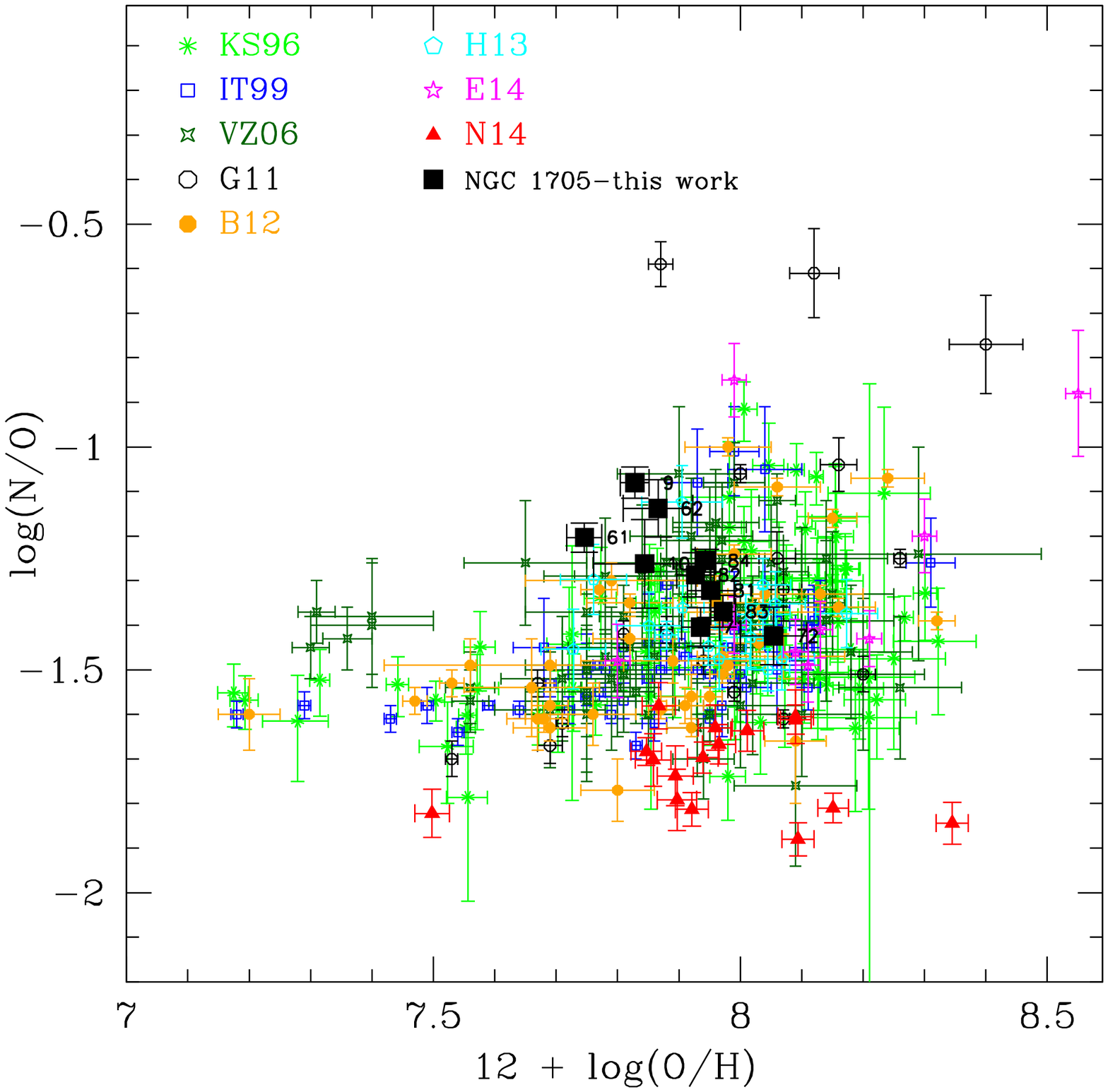}
\caption{$\log(N/O)$ versus $12+log(O/H)$ for different samples of star-forming dwarf galaxies  in the literature.  KS96: metal-poor galaxies from \cite{ks96}; 
IT99:  low-metallicity blue compact galaxies from \cite{it99}; VZ06: isolated dwarf irregular galaxies from \cite{vanzee06}; G11: low-metallicity emission-line galaxies from \cite{guseva11}; 
B12: low-luminosity galaxies in the Spitzer Local Volume Legacy survey from \cite{berg12}; H13:  dwarf irregular galaxies from 
\cite{haurberg13}; E14: low-metallicity star-forming galaxies from \cite{esteban14}; N14: isolated gas-rich irregular dwarf galaxies 
from \cite{nicholls14}; black filled squares: abundances in H II  regions of NGC~1705 from this work. 
 \label{nsuo}}
\end{figure*}

Since the $\alpha$-elements (such as O, Ne, S, and Ar) are all synthesized by massive stars on similar timescales, the relative abundance of these elements is expected to remain constant, independent of the metallicity of the H II regions. This has been observationally demonstrated by many authors \citep[e.g.,][]{vanzee97,it99,vanzee06}, and it is also shown in  Figure~\ref{elements}, where we plotted the Ne/O, S/O, and Ar/O ratios as a function of total oxygen abundance for different samples of late-type dwarf galaxies. 
 Figure~\ref{elements} shows that this trend is also valid within NGC~1705  for the different H II regions: within the $\sim$0.3 dex oxygen abundance range spanned by our data, the $\alpha$-element ratios remain roughly constant with total oxygen abundance.   
 
The case of nitrogen is more complex since it may have a ``primary'' and a ``secondary'' origin. Nitrogen which is produced only out of the original hydrogen in a star is referred to as ``primary'' nitrogen. The major contributors of primary nitrogen are believed to be intermediate mass stars in the range of 4 to 8 $M_{\odot}$ \citep{rv81}. It has also been suggested that primary nitrogen is produced in low metallicity massive stars \citep[e.g.,][]{ww95,meynet02}. ``Secondary'' nitrogen refers instead to nitrogen which has been produced from C and O originally incorporated into the star when it formed. In presence of a substantial  primary nitrogen component, the N/O ratio is expected to be independent of the oxygen abundance. On the other hand, one expects secondary nitrogen to become more and more important as oxygen abundance increases.  Indeed, this is what is observed: for very low-metallicity galaxies the N/O ratio is roughly constant at $\log(N/O)\sim-1.6$, suggesting a substantial primary nitrogen component. Then, as oxygen increases,  the N/O ratio starts to increase with O, suggesting a major secondary nitrogen component \citep{vanzee98}. 
We see in Figure~\ref{nsuo} that our H II regions are in the regime where nitrogen has  both a primary and a secondary component. 
The N/O ratios exhibit a significant scatter, with the more external regions  in NGC~1705 (see also Figure~\ref{radial}) populating the upper envelope of the distribution of late-type dwarf galaxies.

\subsection{Abundance radial trends}

\begin{figure*}
\epsscale{1}
\plotone{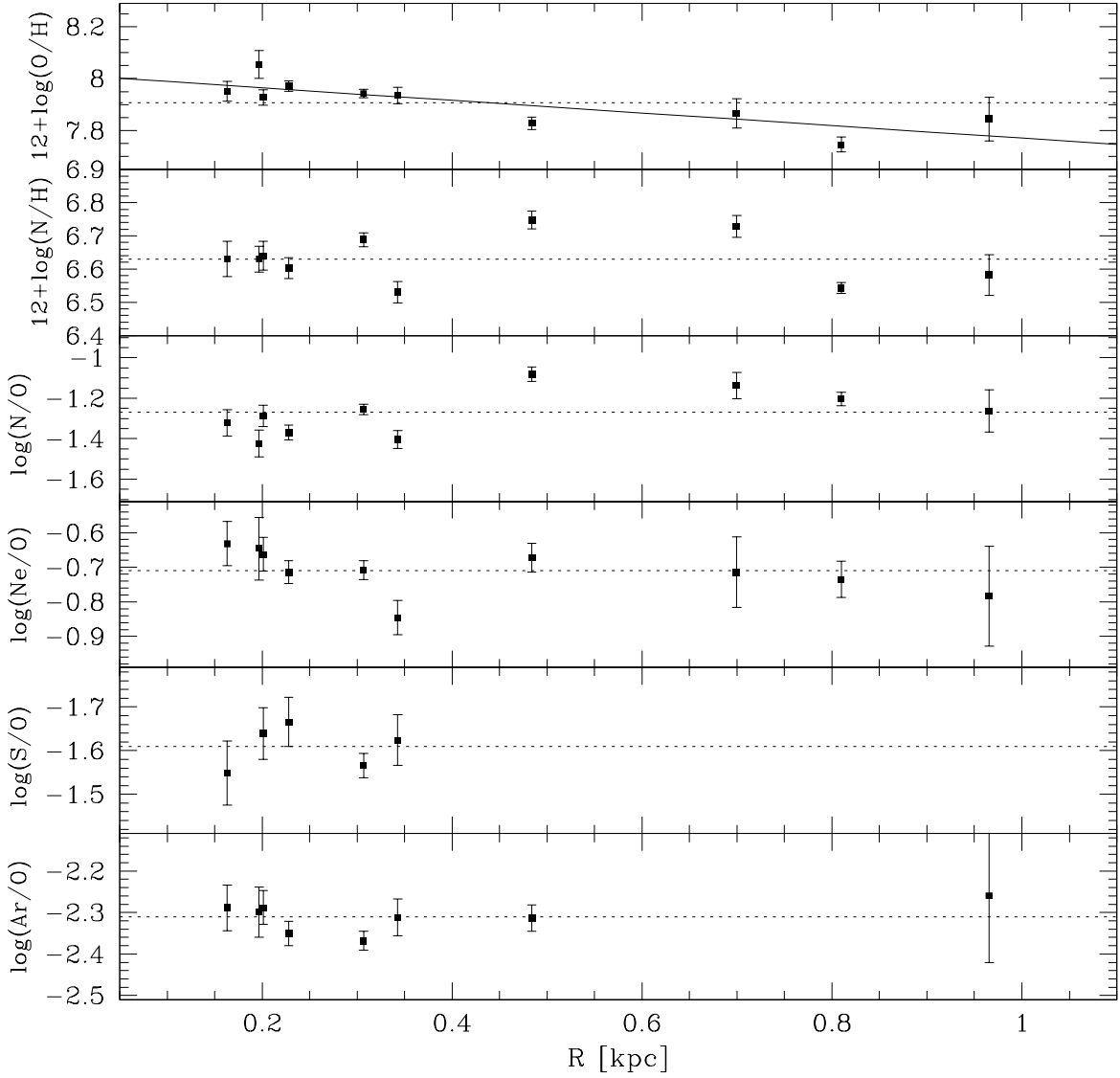}
\caption{Total oxygen  and nitrogen abundances, and N/O, Ne/O, S/O, and Ar/O abundance ratios as a function of distance R from NGC~1705's  center.
In each panel, the dotted line is the average abundance. In the top panel, the solid line is the linear least squares fit to the data points.
 \label{radial}}
\end{figure*}

 We illustrate in Figure~\ref{radial} the behaviour of the total oxygen and  nitrogen abundances, and of the N/O, Ne/O, S/O, and Ar/O ratios, with distance from the galaxy center. 
 While no gradients in the relative abundances of the $\alpha$-elements are detected, the plot suggests a decreasing trend of the total oxygen abundance 
 with increasing distance from the center of NGC~1705. A linear least squares fit to the data provides the following relation:  
  $12 + log(O/H) = (-0.24\pm0.06) \times R \ kpc^{-1} + (8.01\pm0.03)$, which implies a radial gradient of $d \log(O/H)/d r = -0.24 \pm 0.08$ dex \ $kpc^{-1}$.   This is steeper than what usually found in spiral galaxies and in the Milky Way \citep[see][for a review]{hw99}.
The presence of metallicity gradients in late-type dwarf galaxies has been widely investigated and discussed in the literature. The majority of studies indicate that,  
 within the observational uncertainties, dIrr and BCD galaxies  show nearly spatially constant chemical abundances \citep[e.g.,][]{kobul97,croxall09,lagos13,haurberg13}. 
Two possible explanations for  the absence of significant metallicity gradients in these systems are that a) the 
ejecta from stellar winds and supernovae are dispersed and mixed across the ISM on timescales of $<10^7$ yr , or that b) freshly synthesized elements remain unmixed with the surrounding interstellar medium and reside in a hot $10^6$ K phase or a cold, dusty, molecular phase \citep{kobul97}.
Detections of negative metallicity gradients from stars and H II regions have been reported in the literature for the dIrr NGC~6822  \citep{venn04,lee06}, with an oxygen abundance gradient of 
$-0.16 \pm 0.05$ dex \ $kpc^{-1}$. More recent studies based on spectroscopy of individual RGB stars have detected slightly negative gradients in [Fe/H]  for the 
 SMC, the LMC and the dIrr WLM \citep[][and references therein]{leaman14}. A very recent study by \cite{pilyugin15} finds a correlation between H~II region abundance gradients and surface brightness profiles in irregular galaxies:  galaxies with a flat inner profile show shallow (if any) radial abundance gradients,  while irregular galaxies with a steep inner profile show rather steep radial abundance gradients, in agreement with our result.

\begin{figure*}
\epsscale{1}
\plotone{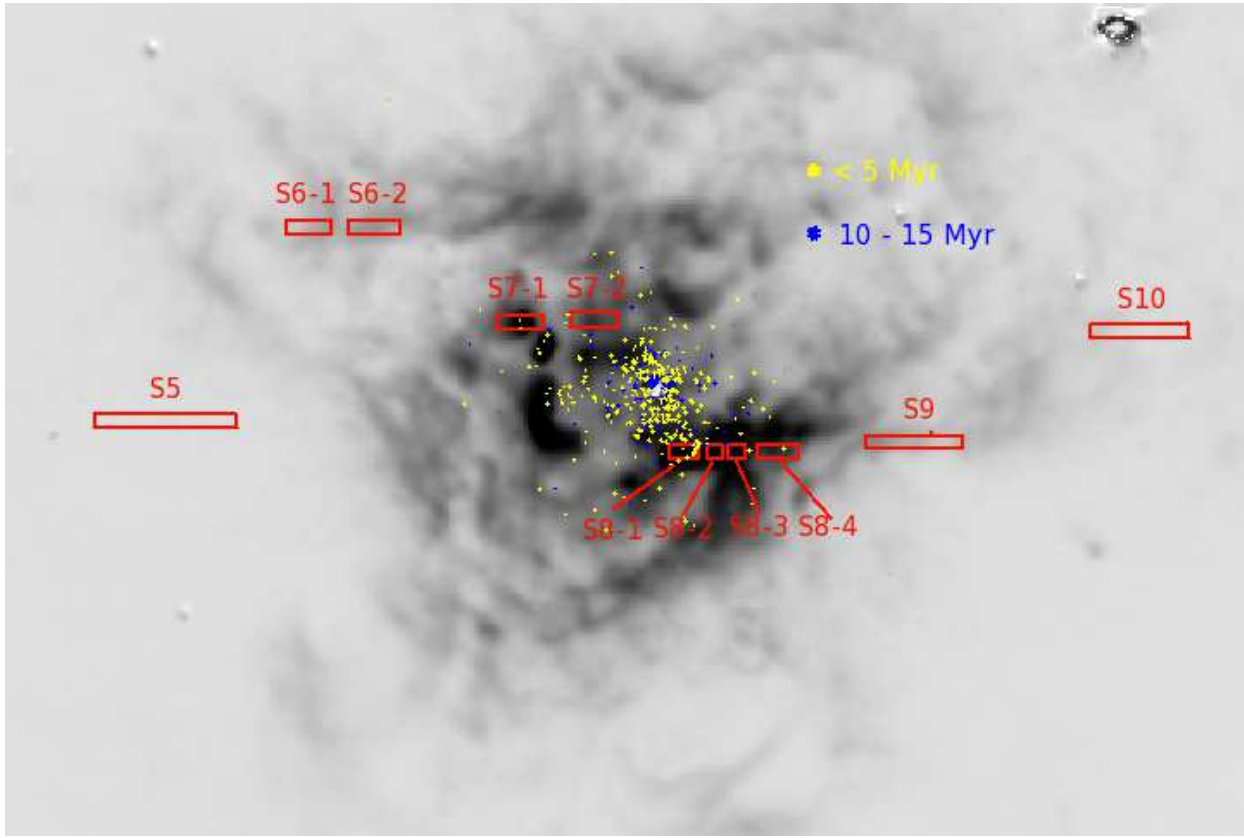}
\caption{Continuum-subtracted [OIII] image of NGC~1705 with the FORS2-MXU slits. Superimposed are the stars younger than $\sim$ 5 Myr (yellow points) and 
with ages in the range 10-15 Myr (blue points) as derived by \cite{anni09}. 
 \label{bursts}}
\end{figure*}

Finally, in Figure~\ref{radial} there is some evidence for larger N/O ratios in the external regions of NGC~1705 ($R>$ 0.45~kpc), in correspondence of which no young stars (age $<$ 15 Myr) are found \citep[][and Figure~\ref{bursts} in this paper]{anni09}.  The N/O behaviour is a consequence of the decreasing radial oxygen trend, since the total nitrogen abundance does not show any definite trend, and it is consistent with a large scatter around a flat distribution.
The innermost regions share common values (at a 2$\sigma$ level) of 
$\langle$log(N/O)$\rangle_{\mathrm{int}}$~= $-$1.34$\pm$0.06 and 
$\langle$12$+$log(O/H)$\rangle_{\mathrm{int}}$~= 7.96$\pm$0.04, while the 
external regions appear characterized by lower oxygen abundances and higher N/O ratios, 
$\langle$log(N/O)$\rangle_{\mathrm{ext}}$~= $-$1.17$\pm$0.07 and 
$\langle$12$+$log(O/H)$\rangle_{\mathrm{ext}}$~= 7.82$\pm$0.04.  
A higher oxygen abundance and lower N/O toward the galaxy center may be the natural
consequence of a burst in the inner regions sufficiently recent to have polluted the
medium in oxygen but not yet in nitrogen \citep[see, e.g. ][]{pilyugin92}. This scenario is consistent 
with the location of the most central  ($R<$ 0.45~kpc) s7-1, s7-2, and s8-1 to s8-4 H II regions,  
which are at the periphery of a region that has been forming stars during the last $\sim$15 Myr (see Figure~\ref{bursts}), 
and that has consequently been enriched in oxygen. 
We will test this hypothesis quantitatively when the LEGUS HST Survey \citep{legus} will provide more precise information on the recent star formation activity of different regions in NGC~1705. This information will be used as an input to multi-zone chemical evolution models, thus improving upon the one-zone model approach adopted in \cite{romano06}. 

We also want to point out
that the history of nitrogen enrichment in our Galaxy is still poorly constrained by N
measurements in stars, especially in the halo \citep[see e.g.,][their figure~3, and references and discussion therein]{romano10}. 
Therefore, the determination of N abundances in the gaseous component of low-metallicity  
external systems is of the highest importance to improve our 
understanding of N stellar nucleosynthesis.

\section{Conclusions}

We obtained [O III] narrow-band imaging and multi-slit optical spectroscopy of NGC~1705 with FORS2@VLT  to infer chemical abundances for planetary nebulae (PNe) and H II regions and, more in general, to characterize the properties of the ionized gas. 

From the [O III] pre-imaging, the ionized gas presents a bipolar morphology extending out to  $\sim$2.5 kpc  from the galaxy center, and it is organized into thin filaments, loops, and arcs, in agreement with previous studies in the literature based on H$\alpha$ narrow-band imaging \citep[e.g.][]{meurer92}. 
It also exhibits ``cellular'' structures or cavities, with a typical cell size of $\sim$50 pc. 

Spectra in the wavelength range $\sim3600-7200$  \AA \ were obtained in the MXU configuration for some selected regions. 
We positioned the slits on planetary nebulae (PN) candidates, extended [O III] emission regions, and gas filaments. 
Nine out of the seventeen PN candidates identified from the [O III] pre-imaging were targeted for spectroscopy. However,  almost the totality of them ultimately  turned out to be 
background objects, with the exception of one, which is a possible low-ionization PN.  
We were then left with five MXU slits positioned on the ionized gas in NGC~1705 out to $\sim$1 kpc from the galaxy center. 
From those, we extracted spectra for ten apertures (regions),  plus the low-ionization PN. 
The auroral [O III]$\lambda4363$ line was detected in all regions but the PN,  allowing for a
direct estimate of the electron temperatures. We derived the abundances of  nitrogen, oxygen, neon, sulfur and argon out to $\sim$1 kpc from the center of NGC~1705. 
The lack of detection of the  [O III]$\lambda4363$  line in the PN, coupled with the low signal-to-noise of its spectra, prevented us from deriving any 
reliable characterization of the chemical properties for this object.

We detected for the first time in NGC~1705 a negative radial gradient for the oxygen abundance of $-0.24 \pm 0.08$ dex \ $kpc^{-1}$. 
 On the other hand, nitrogen exhibits a large dispersion around a roughly flat spatial distribution. As a consequence,   
the external regions ($R>$ 0.45~kpc) of NGC~1705 present larger N/O ratios than the central ones. These trends may be the natural consequence of oxygen 
enrichment in the central regions of NGC~1705 from the starburst 
activity occurred over the last $\sim$15 Myr, as derived from the CMDs of the resolved stars.

The average abundances and standard deviations derived for O, N, Ne, S, and Ar are:   
$12 + \log(O/H)=7.91 \pm 0.08$,  $12 + \log(N/H)=6.63\pm0.07$ ($\log(N/O)=-1.27\pm0.11$),   
$12 + \log(Ne/H)=7.19\pm0.12$ ($\log(Ne/O)=-0.71\pm0.06$),  $12 + \log(S/H)=6.34\pm0.04$ ($\log(S/O)=-1.61\pm0.04$), 
and  $12 + \log(Ar/H)=5.62\pm0.06$ ($\log(Ar/O)=-2.31\pm0.03$), in agreement with values reported in the literature for other samples of dwarf irregular and blue compact dwarf galaxies. On the other hand, our average oxygen 
abundance is $\sim$ 0.31 dex lower than previous literature estimates for NGC~1705 based on the [O III]$\lambda4363$ line, while the  $\log(N/O)$ ratio is $\sim$0.42 dex 
higher. Part of the discrepancy with other estimates of NGC~1705 oxygen abundance is due to the presence of the negative radial gradient. However, if we recompute the average O abundance using only the most central regions (comparable in distance from the center 
to regions used in the literature), we obtain  $12 + \log(O/H)=7.96 \pm 0.04$, which is still $\sim$0.26 dex lower than the average abundance reported in the literature.

Using classical emission-line diagnostic diagrams, we investigated the possible contribution from components different than pure stellar photoionization in NGC~1705. A major contribution from shock excitation can be excluded in our analyzed regions. Furthermore, the radial behavior of the emission line ratios is consistent with the progressive dilution of radiation with increasing distance from the center of NGC~1705.   
This suggests that the strongest starburst located within the central $\sim$150 pc is the major ionizing source out to distances of at least $\sim$1 kpc, and 
possibly even more. The gradual dilution of the radiation (decrease of the ionization parameter) with increasing distance from the galaxy center reflects the  gradual and continuous transition from the highly ionized H II regions in the proximity of the major starburst into the diffuse ionized gas.

In order to explain the observed element abundances in NGC~1705, we plan to run in the future new multi-zone chemical evolution models based both on the SFH already derived from WFPC2 data, and on the more precise information on the recent star formation activity that will come from the LEGUS HST Survey. In particular, we will quantitatively test the hypothesis that the presence of a negative radial gradient in the oxygen abundance, coupled with a flat nitrogen distribution,  
can be explained by the active star formation derived from the CMDs of the resolved stars in the most central regions of  NGC~1705 over the last $\sim$15 Myr.

\acknowledgments

F. A. and this work have been supported by PRIN MIUR through grant 2010LY5N2T\_006.
D. R. and M. T. have also been supported by the same grant. 
We  acknowledge support  from the ESO telescope operator I. Condor.
Based on observations made with ESO Telescopes at the La Silla Paranal Observatory under programme ID 084.D-0248 and 086.D-0761. 




\clearpage

%\begin{deluxetable}{lccc}
%\tablecolumns{4} 
%\tablewidth{0pc} 
%\tablecaption{Fits to the emission lines in s7-abs with and without underlying stellar absorption.}
%\tablehead{
%& \multicolumn{2}{c}{Fit with emission $+$ absorption} & Fit with emission only \\ 
%\colhead{Line} &  \colhead{Absorption EW from fit} &  \colhead{Absorption EW from SB99} &  \colhead{Correction EW} \\
%\colhead{} &  \colhead{[\AA]} &   \colhead{[\AA]} &  \colhead{[\AA]} \\
%}
%\startdata
%H$\delta$ & 10.1 $\pm$ 0.3 &  10.3 $\pm$ 0.1  & 3.6 $\pm$ 0.1  \\
%H$\gamma$ & 10.6 $\pm$ 0.4 &  10.4 $\pm$ 0.7  & 3.9 $\pm$ 0.1  \\
%{[O III]} $\lambda$4363    &  &   & 0.8 $\pm$ 0.1  \\ 
%H$\beta$ & 8.7 $\pm$ 0.2 &  10.0 $\pm$ 0.2  & 3.7 $\pm$ 0.1  \\
%{[NII]} $\lambda$6548 &  &   & 0.5 $\pm$ 0.1  \\
%H$\alpha$ &  &  5.5 $\pm$ 0.1  & 1.9 $\pm$ 0.2  \\
%{[NII]} $\lambda$6584 &  &   & 0.6 $\pm$ 0.1  \\
%\enddata
%\tablecomments{}
%\end{deluxetable}

\begin{deluxetable}{lcccccc}
\tabletypesize{\scriptsize}
%rotate
\tablecaption{Emission line fluxes for regions s5 $-$ s8-1 in NGC~1705}
%q\tablewidth{0pt}
\tablehead{
\colhead{Line} & \colhead{s5} &  \colhead{s6-1}  &  \colhead{s6-2}  &  \colhead{s7-1} &  \colhead{s7-2} &  \colhead{s8-1} 
}
\startdata
{[O II]} $\lambda$3727           & 471.0 $\pm$ 62.0  &  334.4 $\pm$ 5.9  &  240.3 $\pm$ 4.5 & 210.7 $\pm$ 2.0 &   173.1 $\pm$ 0.9   & 142.3 $\pm$ 0.7  \\  

{[Ne III]} $\lambda$3869        & $-$ &  35.8 $\pm$ 1.0   &  42.1 $\pm$  1.3  & 22.7 $\pm$ 1.1 &    37.9 $\pm$ 0.6  & 45.9 $\pm$ 0.5  \\

H8$+$He I $\lambda$3889  & $-$ & 20.2 $\pm$ 0.8  &  14.4 $\pm$ 0.7  & 17.2 $\pm$ 0.8 &    23.5 $\pm$ 0.3  & 12.2 $\pm$ 0.2  \\

H$\epsilon$ $+$ He I $+$[Ne III]  $\lambda$3970 & $-$  & 20.7 $\pm$ 1.0  &  15.6 $\pm$ 1.6 & 20.6 $\pm$ 0.1  &   30.8 $\pm$ 3.0 & 16.5 $\pm$ 0.8  \\ 

H$\delta$ $\lambda$4101      & $-$ & 19.4 $\pm$ 0.4  &  15.8 $\pm$ 0.2 & 23.0 $\pm$ 0.5 &    27.0 $\pm$ 0.1 & 10.9 $\pm$ 0.3  \\

H$\gamma$ $\lambda$4340 & $-$ & 42.0 $\pm$ 0.7  & 40.5 $\pm$ 0.9 & 42.4 $\pm$ 0.4 &    44.2 $\pm$ 1.8   & 36.0 $\pm$ 0.2    \\

{[O III]} $\lambda$4363            & $-$ & 3.3 $\pm$ 0.3 &  3.4 $\pm$ 0.5  & 4.2 $\pm$ 0.3 & 4.6 $\pm$ 0.5  &  3.7 $\pm$ 0.2 \\

He I $\lambda$4471                & $-$ & 5.1 $\pm$ 0.8 &  $-$ & 3.2 $\pm$ 0.2 &     $-$ &  3.3 $\pm$ 0.1 \\

He II  $\lambda$4686              & 28.0 $\pm$ 11.1 \tablenotemark{b}  & $-$  & $-$ & 0.8 $\pm$0.1   \tablenotemark{b} &     $-$ &  1.5 $\pm$ 0.2 \tablenotemark{b}   \\ 
\nodata         & 35.0 $\pm$ 14.0 \tablenotemark{r}  &$-$ & $-$ & $-$ & $-$& $-$ \\

H$\beta$ $\lambda$4861       &  100.0 $\pm$ 10.4 \tablenotemark{b}  & 100.0 $\pm$ 1.4 \tablenotemark{b}  &   100.0 $\pm$ 1.4  \tablenotemark{b} & 100.0 $\pm$ 0.5  \tablenotemark{b}  &   100.0 $\pm$ 0.4 \tablenotemark{b}   & 100.0 $\pm$ 0.2 \tablenotemark{b}   \\
\nodata                                       &  100.0 $\pm$ 5.6 \tablenotemark{r}  & 100.0 $\pm$ 0.8 \tablenotemark{r} & 100.0 $\pm$ 0.7 \tablenotemark{r}   & 100.0 $\pm$ 0.3 \tablenotemark{r}  &   100.0 $\pm$ 0.2  \tablenotemark{r}   & 100.0 $\pm$ 0.2  \tablenotemark{r}  \\

{[O III]} $\lambda$4959            &  115.0 $\pm$ 14.9  \tablenotemark{b}  & 90.6 $\pm$ 1.5 \tablenotemark{b}  &  142.9 $\pm$ 2.4  \tablenotemark{b}   & 112.5 $\pm$ 1.8 \tablenotemark{b} &   144.0 $\pm$ 0.7  \tablenotemark{b}  & 163.8 $\pm$ 3.3 \tablenotemark{b}  \\
\nodata                                        &  141.8 $\pm$ 10.8  \tablenotemark{r}  & 89.8 $\pm$ 1.0 \tablenotemark{r}  &  139.5 $\pm$ 1.0 \tablenotemark{r} & 111.1 $\pm$ 1.0  \tablenotemark{r}  &   132.8 $\pm$ 0.4   \tablenotemark{r}  & 157.9 $\pm$ 0.5   \tablenotemark{r}  \\

{[O III]} $\lambda$5007            &  343.9 $\pm$ 37.6 \tablenotemark{b}  & 270.0 $\pm$ 4.0 \tablenotemark{b} & 423.2 $\pm$ 6.1  \tablenotemark{b}   & 319.4 $\pm$ 2.2  \tablenotemark{b} &   420.2 $\pm$ 1.8 \tablenotemark{b}  & 457.8 $\pm$ 3.6 \tablenotemark{b}  \\
\nodata                                       &  492.2 $\pm$ 29.5 \tablenotemark{r}  & 264.3 $\pm$ 2.4 \tablenotemark{r} &  412.9 $\pm$ 2.9  \tablenotemark{r}  & 312.7 $\pm$ 1.5  \tablenotemark{r}   &   411.9 $\pm$ 0.9   \tablenotemark{r}  & 443.2 $\pm$ 1.2   \tablenotemark{r}   \\

He I $\lambda$5876                &  $-$ & 22.7 $\pm$ 0.9 \tablenotemark{b} &  30.1 $\pm$ 0.7   \tablenotemark{b}  & 11.7 $\pm$ 0.2 \tablenotemark{b}  &   12.5 $\pm$ 0.2 \tablenotemark{b}   & 13.3 $\pm$ 0.5  \tablenotemark{b}  \\
\nodata                                       &  44.6  $\pm$ 2.6  \tablenotemark{r}  & 29.1 $\pm$ 0.3 \tablenotemark{r}   & 35.0 $\pm$ 0.4  \tablenotemark{r} & 10.9 $\pm$ 0.1  \tablenotemark{r}  &   10.9 $\pm$ 0.1   \tablenotemark{r}    & 12.8 $\pm$ 0.1  \tablenotemark{r}  \\

{[OI]}  $\lambda$6302              & $-$  & $-$  & $-$  & 2.4 $\pm$ 0.1 \tablenotemark{b} &     $-$  &  1.9 $\pm$ 0.3  \tablenotemark{b}  \\
\nodata                                        & $-$  & $-$  & $-$  & 2.7 $\pm$ 0.1 \tablenotemark{r}  &     $-$ &  $-$  \\

{[S III]} $\lambda$6314            & $-$  & $-$  & $-$  & 1.7 $\pm$ 0.1 \tablenotemark{b}  &     $-$ &  2.2 $\pm$ 0.1 \tablenotemark{b}  \\
\nodata                                       & $-$  & $-$  & $-$  & 1.8 $\pm$ 0.1 \tablenotemark{r}   &     $-$  &  $-$ \\

{[OI]} $\lambda$6365               &  $-$ & $-$  & $-$  & 1.0 $\pm$ 0.1\tablenotemark{b}  &     $-$  &  $-$ \\   
\nodata                                       &  $-$ & $-$  & $-$  & 0.8 $\pm$ 0.1\tablenotemark{r}  &     $-$  &  $-$  \\   

{[NII]} $\lambda$6548              & $-$ & 7.5 $\pm$ 0.6 \tablenotemark{b}  &  7.2 $\pm$ 0.7  \tablenotemark{b} &  3.2 $\pm$ 0.5 \tablenotemark{b}  &   3.1 $\pm$ 0.4  \tablenotemark{b} &  2.2 $\pm$ 0.5   \tablenotemark{b}  \\
\nodata                                        & $-$ & 3.7 $\pm$ 0.2 \tablenotemark{r} & $-$ & $-$  &   $-$  &  $-$ \\ 

H$\alpha$ $\lambda$6563     & 283.6 $\pm$ 30.1 \tablenotemark{b} & 298.1 $\pm$ 4.2 \tablenotemark{b}  &  318.1 $\pm$ 4.4   \tablenotemark{b} &   310.3 $\pm$ 1.7  \tablenotemark{b} &  330.0 $\pm$ 1.3 \tablenotemark{b}  & 320.7 $\pm$ 3.2 \tablenotemark{b}   \\
\nodata                                        & 355.3 $\pm$ 22.8 \tablenotemark{r}  & 278.2 $\pm$ 2.4 \tablenotemark{r}   &  300.4 $\pm$ 2.1   \tablenotemark{r}  &   306.3 $\pm$ 0.5 \tablenotemark{r}   &  328.6 $\pm$ 0.9  \tablenotemark{r}  & 308.0 $\pm$ 0.8   \tablenotemark{r}   \\

{[N II]} $\lambda$6584             &   $-$     & 18.5 $\pm$ 0.7 \tablenotemark{b} &  15.6 $\pm$ 0.9  \tablenotemark{b} & 12.8 $\pm$ 0.7 \tablenotemark{b}  &  12.3 $\pm$ 0.6  \tablenotemark{b}   &  9.1 $\pm$ 0.7   \tablenotemark{b}  \\
\nodata                                        &  32.0 $\pm$ 8.2  \tablenotemark{r}  & 20.0 $\pm$ 0.4 \tablenotemark{r} & 16.2 $\pm$ 0.3   \tablenotemark{r}  & 11.6 $\pm$ 0.3 \tablenotemark{r}   &  12.4 $\pm$ 0.6  \tablenotemark{r}   &  9.1 $\pm$ 0.1   \tablenotemark{r}   \\

He I $\lambda$6678                & $-$ & 3.6 $\pm$ 0.1 \tablenotemark{b}  &  3.9 $\pm$ 0.1   \tablenotemark{b}  & 3.4 $\pm$ 0.2 \tablenotemark{b}  &   3.9 $\pm$ 0.2 \tablenotemark{b}  &   3.4 $\pm$ 0.1  \tablenotemark{b}  \\

\nodata                                       & $-$ &  3.8 $\pm$ 0.1 \tablenotemark{r} &  4.1 $\pm$ 0.1 \tablenotemark{r}  & 3.5 $\pm$ 0.1 \tablenotemark{r}  &   3.5 $\pm$ 0.1  \tablenotemark{r}   &  3.8 $\pm$ 0.1   \tablenotemark{r}  \\

{[S II]} $\lambda$6716             &  42.0 $\pm$ 5.1  \tablenotemark{b} & 43.9 $\pm$ 0.9 \tablenotemark{b}  &  33.3 $\pm$ 0.6 \tablenotemark{b}  & 19.0 $\pm$ 0.2  \tablenotemark{b}  &  23.0 $\pm$ 0.1 \tablenotemark{b}   &  17.5 $\pm$  0.1 \tablenotemark{b}   \\
\nodata                                       & $-$ & 41.7 $\pm$ 0.4 \tablenotemark{r} &  32.3 $\pm$ 0.3  \tablenotemark{r}  & 19.5 $\pm$ 0.1 \tablenotemark{r}  &  25.1 $\pm$ 0.1  \tablenotemark{r}  &  17.9 $\pm$  0.2   \tablenotemark{r}  \\

{[S II]} $\lambda$6731             &  25.4 $\pm$ 4.3 \tablenotemark{b}  & 28.9 $\pm$ 1.1 \tablenotemark{b}  &  22.3 $\pm$ 0.6   \tablenotemark{b} & 13.1 $\pm$ 0.1 \tablenotemark{b} &  15.4 $\pm$ 0.1 \tablenotemark{b}   & 12.5 $\pm$ 0.1 \tablenotemark{b}   \\
\nodata                                       & $-$ & 28.1 $\pm$ 0.3 \tablenotemark{r}  &  21.4 $\pm$ 0.2 \tablenotemark{r}  & 13.6 $\pm$ 0.1  \tablenotemark{r}  &  17.2 $\pm$ 0.1  \tablenotemark{r}  & 12.7 $\pm$ 0.1  \tablenotemark{r}  \\

He I $\lambda$7065                & $-$ & $-$ & $-$ &  2.4 $\pm$ 0.1\tablenotemark{b}  &    $-$    & 2.4 $\pm$ 0.1 \tablenotemark{b}   \\
\nodata                                       & $-$ & $-$ & $-$ &   2.4 $\pm$ 0.1  \tablenotemark{r}  &  2.0 $\pm$ 0.1   \tablenotemark{r}   & 2.4 $\pm$ 0.2  \tablenotemark{r}   \\

{[Ar III]} $\lambda$7136           & $-$ & $-$ & $-$ &  7.7 $\pm$ 0.1 \tablenotemark{b}  &   9.4 $\pm$ 0.1 \tablenotemark{b}  & 8.6 $\pm$ 0.2  \tablenotemark{b}   \\
\nodata                                        & $-$ & 7.3 $\pm$ 0.3 \tablenotemark{r} & 7.8 $\pm$ 0.3  \tablenotemark{r}   &  8.3 $\pm$ 0.1 \tablenotemark{r}  &  9.1 $\pm$ 0.2  \tablenotemark{r}   & 8.6 $\pm$ 0.1  \tablenotemark{r}  \\

\enddata
%% Text for table notes should follow after the \enddata but before
%% the \end{deluxetable}. Make sure there is at least one \tablenotemark
%% in the table for each \tablenotetext.
\tablecomments{Fluxes are given on a scale where F(H$\beta$)$=$100. The tabulated values are the raw measures, i..e. with no correction for underlying 
Balmer absorption and reddening.
The values obtained from the blue and red spectra are indicated with the letters $b$ and $r$ respectively.}
\end{deluxetable}

\begin{deluxetable}{lccccc}
\tabletypesize{\scriptsize}
%rotate
\tablecaption{Emission line fluxes for regions s8-2 $-$ s10 in NGC~1705}
%q\tablewidth{0pt}
\tablehead{
\colhead{Line} & \colhead{s8-2} &  \colhead{s8-3}  &  \colhead{s8-4}  &  \colhead{s9} &  \colhead{s10} 
}
\startdata
{[O II]} $\lambda$3727   & 163.1 $\pm$ 1.0  & 222.7 $\pm$ 6.1  &   261.4 $\pm$ 4.1   &  305.1 $\pm$ 6.9  & 300.2 $\pm$ 21.5\\          

{[Ne III]} $\lambda$3869   &  45.9 $\pm$ 0.3 &  37.5 $\pm$ 0.6   &  36.1 $\pm$ 0.8  &  39.8 $\pm$ 0.2  &  36.7 $\pm$ 3.3 \\        

H8$+$He I $\lambda$3889  &  13.8 $\pm$ 0.4 &  18.3 $\pm$ 0.2 &  18.5 $\pm$ 0.5  &  22.3 $\pm$ 0.1 & 14.7 $\pm$ 2.7   \\  

H$\epsilon$ $+$ He I $+$[Ne III]  $\lambda$3970  &  16.2 $\pm$ 0.6 &  23.2 $\pm$ 0.8  &  23.5 $\pm$ 0.6  &  26.7 $\pm$ 0.5 & 23.1 $\pm$ 3.2  \\

H$\delta$ $\lambda$4101 &  13.9 $\pm$ 0.1   &  19.4 $\pm$ 0.4  &  20.2 $\pm$ 0.1 &  23.6 $\pm$ 0.4 &   22.1 $\pm$ 0.7 \\

H$\gamma$ $\lambda$4340  &  39.0 $\pm$ 0.2  &  42.5 $\pm$ 0.1  &  43.5 $\pm$ 0.1   &  46.7 $\pm$ 0.3  &   44.1 $\pm$ 2.0  \\

{[O III]} $\lambda$4363   &   4.3 $\pm$ 0.1    &  3.9 $\pm$ 0.1    &   3.6 $\pm$ 0.1   &  4.2 $\pm$ 0.2 &   6.5 $\pm$ 1.1  \\

He I $\lambda$4471     &   4.3 $\pm$ 0.2   &  3.1 $\pm$ 0.1     &   3.5 $\pm$ 0.1     &  4.6 $\pm$ 0.3 &  $-$     \\

He II  $\lambda$4686      & 1.2 $\pm$ 0.1  &  0.6 $\pm$ 0.1    & $-$    & $-$   & $-$  \\

H$\beta$ $\lambda$4861   &  100.0 $\pm$ 0.2 \tablenotemark{b} & 100.0 $\pm$ 0.1   \tablenotemark{b}  &  100.0 $\pm$ 0.1   \tablenotemark{b} &  100.0 $\pm$ 0.4  \tablenotemark{b}  &   100.0 $\pm$3.1 \tablenotemark{b}  \\
\nodata    &  100.0 $\pm$ 0.3 \tablenotemark{r}  &  100.0 $\pm$ 0.2 \tablenotemark{r}  &  100.0 $\pm$ 0.3   \tablenotemark{r}   &  100.0 $\pm$ 0.6  \tablenotemark{r}  &   100.0 $\pm$ 1.9 \tablenotemark{r}   \\                    

{[O III]} $\lambda$4959     & 156.7 $\pm$ 1.3   \tablenotemark{b} & 130.7 $\pm$ 1.2 \tablenotemark{b}  &  113.1 $\pm$ 1.1  \tablenotemark{b}  &  101.3 $\pm$ 1.3  \tablenotemark{b}  &   125.9 $\pm$ 4.8 \tablenotemark{b}   \\
\nodata    &  149.8 $\pm$ 0.6 \tablenotemark{r}  & 125.9 $\pm$ 0.3  \tablenotemark{r}  &  109.6 $\pm$ 0.5   \tablenotemark{r}    &  96.5 $\pm$ 0.8  \tablenotemark{r}  &   133.3 $\pm$ 3.2 \tablenotemark{r}   \\                  

{[O III]} $\lambda$5007     & 462.6 $\pm$ 1.4   \tablenotemark{b} & 390.4 $\pm$ 2.8   \tablenotemark{b}   &  338.1 $\pm$ 2.3  \tablenotemark{b} &  304.9 $\pm$ 4.3   \tablenotemark{b} &   383.1 $\pm$ 12.4  \tablenotemark{b}  \\
\nodata        & 449.5 $\pm$ 1.5  \tablenotemark{r}  & 377.4 $\pm$ 0.8  \tablenotemark{r}   &  329.0 $\pm$ 1.1   \tablenotemark{r}   &  286.6 $\pm$ 1.8  \tablenotemark{r}   &   400.9 $\pm$ 8.2  \tablenotemark{r}   \\               

He I $\lambda$5876    &  12.8 $\pm$ 0.2   \tablenotemark{b}     &  11.9 $\pm$ 0.3 \tablenotemark{b}    &  12.6 $\pm$ 0.4 \tablenotemark{b}  &  19.3 $\pm$ 1.0   \tablenotemark{b} &   $-$  \\
\nodata     &  13.0 $\pm$ 0.2  \tablenotemark{r}  &  11.9 $\pm$ 0.1 \tablenotemark{r}   &  12.1 $\pm$ 0.1   \tablenotemark{r}   & $-$  &   $-$ \\                  

{[OI]}  $\lambda$6302     &   2.1 $\pm$ 0.4   \tablenotemark{b}   &  3.8 $\pm$ 0.2  \tablenotemark{b}  &  5.2  $\pm$ 0.1 \tablenotemark{b}  & $-$ &   $-$   \\
\nodata    &  $-$ &  $-$ & $-$ & $-$ &   $-$ \\                  

{[S III]} $\lambda$6314   &   1.7 $\pm$ 0.2   \tablenotemark{b}    &  1.4 $\pm$ 0.2 \tablenotemark{b} &  1.9 $\pm$ 0.1  \tablenotemark{b}  & $-$ &   $-$   \\
\nodata        &   $-$  &  $-$ &  $-$ & $-$  &   $-$ \\

{[OI]} $\lambda$6365    &   1.2 $\pm$ 0.1   \tablenotemark{b} &  1.5 $\pm$ 0.1 \tablenotemark{b} &  2.5 $\pm$ 0.2 \tablenotemark{b}  &   2.3 $\pm$ 0.2  \tablenotemark{b} &   $-$   \\       
\nodata        &   $-$  &  $-$ &  $-$   & $-$   &  $-$ \\                      

{[NII]} $\lambda$6548    &   2.4 $\pm$ 0.3    \tablenotemark{b} &  2.4 $\pm$ 0.3   \tablenotemark{b}  &  4.8 $\pm$ 0.4    \tablenotemark{b}  &   7.0 $\pm$ 0.9  \tablenotemark{b}  & $-$ \\    
\nodata         &   $-$ & $-$& 4.4 $\pm$ 0.4\tablenotemark{r}   &   6.9 $\pm$ 0.1  \tablenotemark{r}  &  $-$  \\                     

H$\alpha$ $\lambda$6563 &  310.4 $\pm$ 0.8 \tablenotemark{b}   &  304.7 $\pm$ 0.5   \tablenotemark{b}  & 309.5 $\pm$ 0.6 \tablenotemark{b}  &  300.3 $\pm$ 1.3   \tablenotemark{b} &   295.2 $\pm$ 9.2 \tablenotemark{b}  \\    
\nodata    &  303.4 $\pm$ 1.0 \tablenotemark{r}   &  294.2 $\pm$ 0.7  \tablenotemark{r}  & 294.5 $\pm$ 1.1   \tablenotemark{r} &  290.2 $\pm$ 2.1  \tablenotemark{r}  &   296.4 $\pm$ 6.5 \tablenotemark{r}    \\                         

{[N II]} $\lambda$6584 &   10.0 $\pm$ 0.8  \tablenotemark{b}  &  13.0 $\pm$ 0.7  \tablenotemark{b} &  19.5 $\pm$ 0.7  \tablenotemark{b}  &  30.1 $\pm$ 1.0   \tablenotemark{b}  &    16.8 $\pm$ 0.9  \tablenotemark{b}  \\            
\nodata       &   10.8 $\pm$ 0.2  \tablenotemark{r}  &  13.9 $\pm$ 0.2  \tablenotemark{r}  &  20.2 $\pm$ 0.5  \tablenotemark{r} &  32.8 $\pm$ 0.9  \tablenotemark{r}  &    15.0 $\pm$ 2.4 \tablenotemark{r}   \\                      

He I $\lambda$6678   &   3.4 $\pm$  0.1  \tablenotemark{b}  &  3.3 $\pm$ 0.1 \tablenotemark{b}  &  3.6 $\pm$ 0.1 \tablenotemark{b}  &  3.6 $\pm$ 0.4   \tablenotemark{b}  & $-$ \\         
\nodata     &   3.5 $\pm$  0.1 \tablenotemark{r}   &  3.2 $\pm$ 0.1  \tablenotemark{r}  &  3.6 $\pm$ 0.1  \tablenotemark{r}  &  3.8 $\pm$ 0.1 \tablenotemark{r}  & $-$ \\                        

{[S II]} $\lambda$6716  &  19.6 $\pm$  0.1  \tablenotemark{b}   &  26.6 $\pm$ 0.1  \tablenotemark{b}  &  35.3 $\pm$ 0.2 \tablenotemark{b}   &  48.9 $\pm$ 0.5  \tablenotemark{b}  &   33.5 $\pm$ 2.4 \tablenotemark{b}   \\        
\nodata     &  20.3 $\pm$  0.1  \tablenotemark{r}  &  27.0 $\pm$ 0.1  \tablenotemark{r}  &  34.9 $\pm$ 0.1   \tablenotemark{r} &  49.3 $\pm$ 0.3 \tablenotemark{r}  &   25.4 $\pm$ 1.1  \tablenotemark{r}   \\                        

{[S II]} $\lambda$6731   &  13.9 $\pm$  0.1   \tablenotemark{b}   &  17.8 $\pm$ 0.1 \tablenotemark{b}  &  23.3 $\pm$ 0.2  \tablenotemark{b} &  33.9 $\pm$ 0.5  \tablenotemark{b} &   19.7 $\pm$ 1.7  \tablenotemark{b}   \\      
\nodata   &   14.1 $\pm$  0.1 \tablenotemark{r}   &  18.0 $\pm$ 0.1 \tablenotemark{r}  &  23.5 $\pm$ 0.1  \tablenotemark{r}  &  34.3 $\pm$ 0.2 \tablenotemark{r}  &   18.3 $\pm$ 1.1 \tablenotemark{r}    \\                          

He I $\lambda$7065      &   2.2 $\pm$ 0.1    \tablenotemark{b}     &  1.9 $\pm$ 0.1   \tablenotemark{b} &  2.1 $\pm$ 0.1 \tablenotemark{b}  &  2.4 $\pm$ 0.1   \tablenotemark{b}  &     $-$  \\  
\nodata     &   2.6 $\pm$ 0.1 \tablenotemark{r}  &  2.2 $\pm$ 0.1  \tablenotemark{r}  &  2.3 $\pm$ 0.1  \tablenotemark{r}    & $-$  &     $-$  \\

{[Ar III]} $\lambda$7136      &   8.4 $\pm$ 0.1  \tablenotemark{b}   &  7.5 $\pm$ 0.1   \tablenotemark{b}  &  7.3 $\pm$ 0.2 \tablenotemark{b}   &  7.4 $\pm$ 0.2   \tablenotemark{b} &    9.0 $\pm$ 2.2 \tablenotemark{b}    \\  
\nodata         &   8.7 $\pm$ 0.3 \tablenotemark{r}    &  7.7 $\pm$ 0.1  \tablenotemark{r}  &  8.0  $\pm$ 0.2   \tablenotemark{r}  &  6.7 $\pm$ 0.2  \tablenotemark{r}  &    8.9 $\pm$ 1.6 \tablenotemark{r}    \\

\enddata
%% Text for table notes should follow after the \enddata but before
%% the \end{deluxetable}. Make sure there is at least one \tablenotemark
%% in the table for each \tablenotetext.
\tablecomments{Fluxes are given on a scale where F(H$\beta$)$=$100. The tabulated values are the raw measures, i..e. with no correction for underlying 
Balmer absorption and reddening. The values obtained from the blue and red spectra are indicated with the letters $b$ and $r$ respectively.}
\end{deluxetable}

\begin{deluxetable}{lcccc}
\tablewidth{0pc} 
%rotate
\tablecaption{Fits to the emission lines in spectrum s7-a with and without underlying stellar absorption}
%q\tablewidth{0pt}
\tablehead{
& \multicolumn{2}{c}{Fit with emission $+$ absorption} & Fit with emission only \\ 
\hline \\
\colhead{Line} &  \colhead{EW$_{em}$} &  \colhead{EW$_{abs}$} & \colhead{EW$_{em}$}  &  \colhead{\bf{EW$_{corr}$}} \\
\colhead{} &  \colhead{[\AA]} &   \colhead{[\AA]} &  \colhead{[\AA]}   &  \colhead{[\AA]}  
}
\startdata
%H$\delta$ & -4.2 $\pm$ 0.1 &  10.4 $\pm$ 0.2  & -0.6 $\pm$ 0.1  \\
H$\delta$ & $-4.5 \pm 0.2$                            &  \multicolumn{1}{c|}{$9.9 \pm 0.4$ } & $-0.5 \pm 0.1$ & \multicolumn{1}{|c}{\bf{$4.0 \pm 0.2$}}  \\ 
H$\gamma$ & $-9.7 \pm 0.3$                       &  \multicolumn{1}{c|}{$10.2 \pm 0.6$}   & $-5.2 \pm 0.1$ & \multicolumn{1}{|c}{\bf{$4.5 \pm 0.3$}} \\
{[OIII]}$\lambda$4363 & $-1.6 \pm 0.1$&   & \multicolumn{1}{|c}{$-0.7 \pm 0.1$}  &  \multicolumn{1}{|c}{\bf{$0.9 \pm 0.1$}}  \\
H$\beta$ & $-22.5 \pm 0.3$ &  $8.9 \pm 0.5$  & \multicolumn{1}{|c}{$-19.4 \pm 0.2$} &  \multicolumn{1}{|c}{\bf{$3.1 \pm 0.4$}} \\

{[NII]} $\lambda$6548  & $-1.3 \pm 0.1$ &     & \multicolumn{1}{|c}{$-0.8 \pm 0.1$} &  \multicolumn{1}{|c}{\bf{$0.5 \pm 0.1$}}  \\
H$\alpha$ & $-100.5 \pm 0.4$ &  $5.9 \pm 0.2$ \tablenotemark{a}  & \multicolumn{1}{|c}{$-99.0 \pm 0.4$}  &  \multicolumn{1}{|c}{\bf{$1.5 \pm 0.6$}}  \\
{[NII]} $\lambda$6584  & $-5.1 \pm 0.1$ &     & \multicolumn{1}{|c}{$-4.6\pm 0.1$} &  \multicolumn{1}{|c}{\bf{$0.5 \pm 0.1$}}  \\
%\tablenotetext{b}{}
\enddata
\tablenotetext{a}{Predicted from population synthesis models for Region 5, see Table~4.}
\end{deluxetable}

\begin{deluxetable}{lccc}
\tablewidth{0pc} 
%rotate
\tablecaption{Applied corrections for Balmer absorption in different regions of NGC~1705}
%q\tablewidth{0pt}
\tablehead{
%& \multicolumn{2}{c}{Fit with emission $+$ absorption} & Fit with emission only \\ 
%\hline \\
\colhead{Line} &  \colhead{Model EW\tablenotemark{a}} &  \colhead{Model (EW/EW$_{reg~5}$)\tablenotemark{b} } & \colhead{EW$_{corr}$\tablenotemark{c}} \\
\colhead{} &  \colhead{[\AA]} &   \colhead{} &  \colhead{[\AA]} 
}
\startdata
\multicolumn{4}{c}{s8-1 (region~6 in A09) }  \\
\hline
H$\delta$                           &  8.1 $\pm$ 0.1  & 0.92 $\pm$ 0.02 & 3.7 $\pm$ 0.2  \\
H$\gamma$                      &  8.4 $\pm$ 0.3  & 0.96 $\pm$ 0.04 & 4.3 $\pm$ 0.3  \\
{[O III]} $\lambda$4363   &   & & 0.9 $\pm$ 0.1  \\ 
H$\beta$                            &  7.5 $\pm$ 0.4  & 0.86 $\pm$ 0.05 & 2.7 $\pm$ 0.4  \\
{[NII]} $\lambda$6548     &   & & 0.5 $\pm$ 0.1  \\
H$\alpha$                         &  6.0 $\pm$ 0.1  & 1.02 $\pm$ 0.04 & 1.5 $\pm$ 0.6  \\
{[NII]} $\lambda$6584    &   & & 0.5 $\pm$ 0.1  \\
%\tablenotetext{b}{}
\hline%{1-4} \\
 \multicolumn{4}{c}{s6-2, s8-2, s8-3, s8-4 (region~5 in A09) } \\ 
\hline% \\
H$\delta$                           &  8.8 $\pm$ 0.2  & 1.00 & 4.0 $\pm$ 0.2  \\
H$\gamma$                      &  8.7 $\pm$ 0.2  & 1.00 & 4.5 $\pm$ 0.3  \\
{[O III]} $\lambda$4363   &   & & 0.9 $\pm$ 0.1  \\ 
H$\beta$                            &  8.7 $\pm$ 0.3  & 1.00 & 3.1 $\pm$ 0.4  \\
{[NII]} $\lambda$6548     &   & & 0.5 $\pm$ 0.1  \\
H$\alpha$                         &  5.9 $\pm$ 0.2  & 1.00 & 1.5 $\pm$ 0.6  \\
{[NII]} $\lambda$6584    &   & & 0.5 $\pm$ 0.1  \\
\hline %\\
\multicolumn{4}{c}{s6-1, s9 (regions~4-3 in A09) } \\ 
\hline %\\
H$\delta$                           &  8.7 $\pm$ 0.1  & 0.99 $\pm$ 0.02 & 3.9 $\pm$ 0.2  \\
H$\gamma$                      &  9.1 $\pm$ 0.4  & 1.05 $\pm$ 0.05 & 4.7 $\pm$ 0.4  \\
{[O III]} $\lambda$4363   &   & & 0.9 $\pm$ 0.1  \\ 
H$\beta$                            &  8.2 $\pm$ 0.4  & 0.94 $\pm$ 0.06 & 2.9 $\pm$ 0.4  \\
{[NII]} $\lambda$6548     &   & & 0.5 $\pm$ 0.1  \\
H$\alpha$                         &  5.9 $\pm$ 0.1  & 1.0 $\pm$ 0.04 & 1.5 $\pm$ 0.6  \\
{[NII]} $\lambda$6584    &   & & 0.5 $\pm$ 0.1  \\
\hline %\\
\multicolumn{4}{c}{s10 (regions~2-1-0 in A09) } \\ 
\hline %\\
H$\delta$                           &  5.9 $\pm$ 0.2  & 0.67 $\pm$ 0.03 & 2.7 $\pm$ 0.2  \\
H$\gamma$                      &  5.3 $\pm$ 0.6  & 0.61 $\pm$ 0.07 & 2.7 $\pm$ 0.4  \\
{[O III]} $\lambda$4363   &   & & 0.5 $\pm$ 0.1  \\ 
H$\beta$                            &  7.0 $\pm$ 0.4  & 0.80 $\pm$ 0.05 & 2.5 $\pm$ 0.4  \\
{[NII]} $\lambda$6548     &   & & 0.4 $\pm$ 0.1  \\
H$\alpha$                         &  4.9 $\pm$ 0.1  & 0.83 $\pm$ 0.03 & 1.2 $\pm$ 0.5  \\
{[NII]} $\lambda$6584    &   & & 0.4 $\pm$ 0.1  \\
\enddata
\tablenotetext{a}{Absorption equivalent width from synthetic spectra}
\tablenotetext{b}{Ratio between the absorption equivalent widths in a specific region and in region 5 from synthetic spectra}
\tablenotetext{c}{Final correction for a specific region}

\end{deluxetable}

\begin{deluxetable}{lcccccc}
\tabletypesize{\scriptsize}
%rotate
\tablecaption{Emission line fluxes corrected for Balmer absorption and for dust extinction for regions s5 $-$ s8-1 in NGC~1705}
%q\tablewidth{0pt}
\tablehead{
\colhead{Line} & \colhead{s5} &  \colhead{s6-1}  &  \colhead{s6-2}  &  \colhead{s7-1} &  \colhead{s7-2} &  \colhead{s8-1} 
}
\startdata
{[O II]} $\lambda$3727           & 477.7  $\pm$ 285.5  &  310.2 $\pm$ 6.1  &  216.7 $\pm$ 25.0 & 235.8 $\pm$ 7.6 &   207.2 $\pm$ 4.7   & 145.8 $\pm$ 10.9  \\  

{[Ne III]} $\lambda$3869        & $-$ &  33.2 $\pm$ 1.0   &  37.9 $\pm$ 4.4  & 25.1 $\pm$ 1.5 &    44.6 $\pm$ 1.2  & 46.6 $\pm$ 3.5  \\

H8$+$He I $\lambda$3889  & $-$ & 18.8 $\pm$ 0.7  &  13.0 $\pm$ 2.1  & 19.0 $\pm$ 1.1 &    27.6 $\pm$ 0.7  & 12.4 $\pm$ 0.9  \\

H$\epsilon$ $+$ He I $+$[Ne III]  $\lambda$3970 & $-$  & 19.3 $\pm$ 0.9  &  14.1 $\pm$ 2.1 & 22.6 $\pm$ 0.7  &   35.8 $\pm$ 3.6 & 16.6 $\pm$ 1.4  \\ 

H$\delta$ $\lambda$4101      & $-$ & 27.6 $\pm$ 0.7  &  29.6 $\pm$ 3.3 & 25.0 $\pm$ 0.9 &    30.7 $\pm$ 0.6 & 23.5 $\pm$ 1.8  \\

H$\gamma$ $\lambda$4340 & $-$ & 49.8 $\pm$ 1.3  & 52.4 $\pm$ 5.7 & 44.8 $\pm$ 1.3 &    48.3 $\pm$ 2.2   & 47.7 $\pm$ 3.4    \\

{[O III]} $\lambda$4363            & $-$ & 5.2 $\pm$ 0.4 &  6.3 $\pm$ 0.9  & 4.4 $\pm$ 0.3 & 5.0 $\pm$ 0.6  &  6.2 $\pm$ 0.6 \\

He I $\lambda$4471                & $-$ & 4.8 $\pm$ 0.7 &  $-$ & 3.4 $\pm$ 0.2 &     $-$ &  3.2 $\pm$ 0.2 \\

He II  $\lambda$4686              & 28.1 $\pm$ 18.1  \tablenotemark{b} & $-$  & $-$ & 0.9 $\pm$ 0.1  \tablenotemark{b} &     $-$ &  1.4 $\pm$ 0.2  \tablenotemark{b} \\ 
\nodata           & 36.3 $\pm$ 14.1 \tablenotemark{r}  & $-$  & $-$ & $-$ & $-$ & $-$ \\

H$\beta$ $\lambda$4861       &  100.0 $\pm$ 51.7 \tablenotemark{b}   & 100.0 $\pm$ 2.3  \tablenotemark{b}  &   100.0 $\pm$ 10.0  \tablenotemark{b}  & 100.0 $\pm$ 2.7 \tablenotemark{b}    &   100.0 $\pm$ 1.9 \tablenotemark{b}   & 100.0 $\pm$ 6.4  \tablenotemark{b}  \\
\nodata    &  100.0 $\pm$ 31.0  \tablenotemark{r}    & 100.0 $\pm$ 1.7  \tablenotemark{r}   &   100.0 $\pm$ 2.3  \tablenotemark{r}   & 100.0 $\pm$ 0.8  \tablenotemark{r}    &   100.0 $\pm$ 1.3  \tablenotemark{r}    & 100.0 $\pm$ 4.6  \tablenotemark{r}    \\

{[O III]} $\lambda$4959            &  114.9 $\pm$ 58.1 \tablenotemark{b}    & 84.1 $\pm$ 1.5  \tablenotemark{b} &  126.0 $\pm$ 12.2  \tablenotemark{b}   & 111.5 $\pm$ 3.4 \tablenotemark{b}  &   141.9 $\pm$ 2.7 \tablenotemark{b}   & 152.8 $\pm$ 10.1 \tablenotemark{b}   \\
\nodata        &  138.9 $\pm$ 42.4  \tablenotemark{r}     & 83.6 $\pm$ 1.2  \tablenotemark{r}   &  123.3 $\pm$ 2.0  \tablenotemark{r}     & 110.2 $\pm$ 0.9  \tablenotemark{r}  &   130.9 $\pm$ 1.6  \tablenotemark{r}    & 148.0 $\pm$ 6.6  \tablenotemark{r}   \\

{[O III]} $\lambda$5007            &  343.4 $\pm$ 170.9  \tablenotemark{b}  & 250.4 $\pm$ 4.3  \tablenotemark{b}  & 372.7 $\pm$ 35.8  \tablenotemark{b}   & 315.2 $\pm$ 8.4  \tablenotemark{b} &   411.3 $\pm$ 7.7  \tablenotemark{b} & 425.8 $\pm$ 26.8  \tablenotemark{b} \\
\nodata      &  477.2 $\pm$ 142.9  \tablenotemark{r}   & 246.1 $\pm$ 3.1  \tablenotemark{r}   & 364.8 $\pm$ 5.9   \tablenotemark{r}    & 309.1 $\pm$ 2.5  \tablenotemark{r}   &   403.4 $\pm$ 5.0  \tablenotemark{r}   & 414.6 $\pm$ 18.4   \tablenotemark{r}   \\

He I $\lambda$5876                &  $-$ & 21.0 $\pm$ 0.9  \tablenotemark{b} &  26.2 $\pm$ 2.4    \tablenotemark{b} & 10.9 $\pm$ 0.3  \tablenotemark{b}  &   11.1 $\pm$ 0.2  \tablenotemark{b}  & 11.8 $\pm$ 0.8  \tablenotemark{b}  \\
\nodata            &  37.9 $\pm$ 10.5   \tablenotemark{r}  & 27.1 $\pm$ 0.4  \tablenotemark{r}   &  31.0 $\pm$ 0.6   \tablenotemark{r}     & 10.3 $\pm$ 0.1  \tablenotemark{r}    &   9.8 $\pm$ 0.1   \tablenotemark{r}   & 11.7 $\pm$ 0.5   \tablenotemark{r}   \\

{[OI]}  $\lambda$6302              & $-$  & $-$  & $-$  & 2.2 $\pm$ 0.1 \tablenotemark{b}  &     $-$  &  1.7 $\pm$ 0.3  \tablenotemark{b}  \\
\nodata          & $-$  & $-$  & $-$  & 2.4 $\pm$ 0.1  \tablenotemark{r}   &     $-$  &  $-$  \\

{[S III]} $\lambda$6314            & $-$  & $-$  & $-$  & 1.5 $\pm$ 0.1 \tablenotemark{b}  &     $-$ &  1.9$\pm$ 0.1  \tablenotemark{b} \\
\nodata         & $-$  & $-$  & $-$  & 1.7 $\pm$ 0.1   \tablenotemark{r}  &     $-$ &  $-$ \\

{[OI]} $\lambda$6365               &  $-$ & $-$  & $-$  & 0.9 $\pm$ 0.1 \tablenotemark{b}  &     $-$  &  $-$ \\   
\nodata           &  $-$ & $-$  & $-$  & 0.7 $\pm$ 0.1  \tablenotemark{r}   &     $-$  &  $-$ \\

{[NII]} $\lambda$6548              & $-$ & 8.0 $\pm$ 0.6  \tablenotemark{b}  &  7.7 $\pm$ 0.9  \tablenotemark{b} &  2.9 $\pm$ 0.4  \tablenotemark{b}  &   2.6 $\pm$ 0.4  \tablenotemark{b}  &  2.5 $\pm$ 0.5   \tablenotemark{b}  \\
\nodata             & $-$ & 4.4 $\pm$ 0.3   \tablenotemark{r}  &  $-$ & $-$ & $-$ & $-$    \\

H$\alpha$ $\lambda$6563     & 280.0 $\pm$ 123.0  \tablenotemark{b} & 279.6 $\pm$ 4.8  \tablenotemark{b}  &  280.0 $\pm$ 23.8  \tablenotemark{b}  &   280.0 $\pm$ 6.5  \tablenotemark{b}  &  280.0 $\pm$ 4.6  \tablenotemark{b}  & 280.0 $\pm$ 15.7  \tablenotemark{b}  \\
\nodata    & 280.0 $\pm$ 74.4  \tablenotemark{r}   & 262.0 $\pm$ 3.4   \tablenotemark{r}  &  270.0 $\pm$ 4.8  \tablenotemark{r}    &   280.0 $\pm$ 2.0  \tablenotemark{r}    &  280.0 $\pm$ 3.1  \tablenotemark{r}   & 280.0 $\pm$ 11.0   \tablenotemark{r}   \\

{[N II]} $\lambda$6584             &   $-$     & 18.2 $\pm$ 0.7  \tablenotemark{b} &  15.0 $\pm$ 1.5  \tablenotemark{b}  & 11.6 $\pm$ 0.6  \tablenotemark{b} &  10.4 $\pm$ 0.6  \tablenotemark{b}    &  8.5 $\pm$ 0.8  \tablenotemark{b}   \\
\nodata          &   25.1 $\pm$ 9.1  \tablenotemark{r}      & 19.6 $\pm$ 0.4  \tablenotemark{r}  &  15.9 $\pm$ 0.5  \tablenotemark{r}   & 10.6 $\pm$ 0.3  \tablenotemark{r}   &  10.5 $\pm$ 0.5    \tablenotemark{r}   &  8.8 $\pm$ 0.4    \tablenotemark{r}   \\

He I $\lambda$6678                & $-$ & 3.3 $\pm$ 0.1 \tablenotemark{b}  &  3.4 $\pm$ 0.3   \tablenotemark{b}  & 3.1 $\pm$ 0.2  \tablenotemark{b}  &   3.3 $\pm$ 0.1  \tablenotemark{b} &   3.0 $\pm$ 0.2   \tablenotemark{b} \\
\nodata             & $-$ & 3.5 $\pm$ 0.1  \tablenotemark{r}   &  3.6 $\pm$ 0.1  \tablenotemark{r}     & 3.2 $\pm$ 0.1  \tablenotemark{r}   &   2.9 $\pm$ 0.1   \tablenotemark{r}  &   3.4 $\pm$ 0.2    \tablenotemark{r}  \\

{[S II]} $\lambda$6716             &  41.5 $\pm$ 18.2  \tablenotemark{b}  & 40.7 $\pm$ 0.9 \tablenotemark{b}   &  28.8 $\pm$ 2.4  \tablenotemark{b}  & 17.1 $\pm$ 0.4  \tablenotemark{b}   &  19.2 $\pm$ 0.3   \tablenotemark{b} &  15.1 $\pm$  0.8  \tablenotemark{b}  \\
\nodata            &  $-$  & 38.8 $\pm$ 0.5 \tablenotemark{r}    &  28.5 $\pm$ 0.5  \tablenotemark{r}   & 17.7 $\pm$ 0.1    \tablenotemark{r}  &  21.1 $\pm$ 0.2  \tablenotemark{r}    &  16.1 $\pm$  0.6   \tablenotemark{r}   \\

{[S II]} $\lambda$6731             &  25.1 $\pm$ 11.4 \tablenotemark{b}   & 26.8 $\pm$ 1.0  \tablenotemark{b} &  19.2 $\pm$ 1.6  \tablenotemark{b}   & 11.7 $\pm$ 0.3 \tablenotemark{b}  &  12.9 $\pm$ 0.2     \tablenotemark{b}  & 10.8 $\pm$ 0.6   \tablenotemark{b} \\
\nodata          &  $-$ & 26.1 $\pm$ 0.4 \tablenotemark{r}    &  18.9 $\pm$ 0.3   \tablenotemark{r}   & 12.3 $\pm$ 0.1  \tablenotemark{r}   &  14.5 $\pm$ 0.2   \tablenotemark{r}   & 11.5 $\pm$ 0.5   \tablenotemark{r}   \\

{[S II]} $\lambda$6716/$\lambda$6731     &  1.65 $\pm$ 0.60  \tablenotemark{b}  & 1.52 $\pm$ 0.06  \tablenotemark{b} &  1.49 $\pm$ 0.10  \tablenotemark{b}   & 1.46 $\pm$ 0.03  \tablenotemark{b}  &  1.50 $\pm$ 0.02  \tablenotemark{b}    & 1.40 $\pm$ 0.06  \tablenotemark{b}   \\
\nodata  &  $-$  & 1.48 $\pm$ 0.02   \tablenotemark{r}  &  1.51 $\pm$ 0.01   \tablenotemark{r}   & 1.44 $\pm$ 0.01  \tablenotemark{r}  &  1.46 $\pm$ 0.02 \tablenotemark{r}     & 1.41 $\pm$ 0.05   \tablenotemark{r}   \\

He I $\lambda$7065                & $-$ & $-$ & $-$ &  2.1 $\pm$ 0.1 \tablenotemark{b}  &    $-$    & 2.0 $\pm$ 0.2   \tablenotemark{b} \\
\nodata             & $-$ & $-$ & $-$ &  2.2 $\pm$ 0.1 &    1.7 $\pm$ 0.1  \tablenotemark{r}     & 2.2 $\pm$ 0.2   \tablenotemark{r}   \\

{[Ar III]} $\lambda$7136           & $-$ & $-$ & $-$ &  6.8 $\pm$ 0.1 \tablenotemark{b}  &   7.7 $\pm$ 0.2  \tablenotemark{b}   & 7.3 $\pm$ 0.4  \tablenotemark{b}   \\
\nodata         & $-$ & 6.8 $\pm$ 0.3 \tablenotemark{r}  & 6.9 $\pm$ 0.3 \tablenotemark{r}    &  7.4 $\pm$ 0.1  \tablenotemark{r}   &   7.5 $\pm$ 0.1  \tablenotemark{r}   & 7.7 $\pm$ 0.3  \tablenotemark{r}     \\

\enddata
%% Text for table notes should follow after the \enddata but before
%% the \end{deluxetable}. Make sure there is at least one \tablenotemark
%% in the table for each \tablenotetext.
\tablecomments{Fluxes are given on a scale where F(H$\beta$)$=$100. The values obtained from the blue and red spectra are indicated with the letters $b$ and $r$ respectively.}
\end{deluxetable}

\begin{deluxetable}{lccccc}
\tabletypesize{\scriptsize}
%rotate
\tablecaption{Emission line fluxes corrected for Balmer absorption and for dust extinction for regions s8-2 $-$ s10 in NGC~1705}
%q\tablewidth{0pt}
\tablehead{
\colhead{Line} & \colhead{s8-2} &  \colhead{s8-3}  &  \colhead{s8-4}  &  \colhead{s9} &  \colhead{s10} 
}
\startdata
{[O II]} $\lambda$3727           & 157.6  $\pm$ 8.9  &  225.6 $\pm$ 9.4  &  272.4 $\pm$ 8.9 & 305.6 $\pm$ 13.9 &   295.5 $\pm$ 54.3     \\  

{[Ne III]} $\lambda$3869        & 44.2 $\pm$ 2.4   &  37.8 $\pm$ 1.3  & 37.3 $\pm$ 1.3 &    39.7 $\pm$ 1.6  & 36.1 $\pm$ 6.8  \\

H8$+$He I $\lambda$3889  & 13.3 $\pm$ 0.8  &  18.4 $\pm$ 0.6  & 19.1 $\pm$ 0.7 &    22.2 $\pm$ 0.9  & 14.4 $\pm$ 3.6  \\

H$\epsilon$ $+$ He I $+$[Ne III]  $\lambda$3970 & 15.5 $\pm$ 1.0  &  23.3 $\pm$ 1.1 & 24.1 $\pm$ 0.9  &  26.5 $\pm$ 1.1 & 22.6 $\pm$ 4.9  \\ 

H$\delta$ $\lambda$4101      & 25.8 $\pm$ 1.5  &  26.3 $\pm$ 0.9 & 26.4 $\pm$ 0.8 &    28.7 $\pm$ 1.2 & 27.5 $\pm$ 4.5  \\

H$\gamma$ $\lambda$4340 & 49.5 $\pm$ 2.7  & 48.9 $\pm$ 1.5 & 49.5 $\pm$ 1.4 &    52.0 $\pm$ 2.0   & 47.5 $\pm$ 7.7    \\

{[O III]} $\lambda$4363            &  6.5 $\pm$ 0.4 &  5.2 $\pm$ 0.2  & 4.8 $\pm$ 0.2 & 5.3 $\pm$ 0.3  &  7.1 $\pm$ 1.5 \\

He I $\lambda$4471                &  4.0 $\pm$ 0.3 &  3.1 $\pm$ 0.1 & 3.5 $\pm$ 0.1 &   4.4  $\pm$ 0.3 &  $-$  \\

He II  $\lambda$4686              & 1.2 $\pm$ 0.1  & 0.6 $\pm$ 0.1   & $-$ & $-$  &     $-$  \\ 

H$\beta$ $\lambda$4861       &  100.0 $\pm$ 4.9    \tablenotemark{b}  & 100.0 $\pm$ 2.7    \tablenotemark{b} &   100.0 $\pm$ 2.5    \tablenotemark{b} & 100.0 $\pm$ 3.4   \tablenotemark{b}   &   100.0 $\pm$ 15.0     \tablenotemark{b}  \\
\nodata      &  100.0 $\pm$ 4.9   \tablenotemark{r}    & 100.0 $\pm$ 3.0    \tablenotemark{r}  &   100.0 $\pm$ 3.0   \tablenotemark{r}   & 100.0 $\pm$ 4.2   \tablenotemark{r}    &   100.0 $\pm$ 11.0    \tablenotemark{r}    \\

{[O III]} $\lambda$4959            &  145.0 $\pm$ 7.0   \tablenotemark{b}   &  125.0 $\pm$ 3.5      \tablenotemark{b} & 108.6 $\pm$ 2.8   \tablenotemark{b}  &   97.2 $\pm$ 3.4   \tablenotemark{b}   & 120.9 $\pm$ 17.8   \tablenotemark{b}  \\
\nodata         &  139.0 $\pm$ 6.6    \tablenotemark{r}   &  120.5 $\pm$ 3.5    \tablenotemark{r}    & 105.5 $\pm$ 3.1   \tablenotemark{r}  &   93.1 $\pm$ 3.9    \tablenotemark{r}   & 126.6 $\pm$ 13.5    \tablenotemark{r}  \\

{[O III]} $\lambda$5007            &   427.5 $\pm$ 20.1  \tablenotemark{b}  & 372.7 $\pm$ 10.1   \tablenotemark{b}     & 323.6 $\pm$ 8.0    \tablenotemark{b}  &   291.9 $\pm$ 10.5    \tablenotemark{b}  & 367.6 $\pm$ 53.1   \tablenotemark{b}   \\
\nodata         &   416.8 $\pm$ 19.7   \tablenotemark{r}   & 361.1 $\pm$ 10.3     \tablenotemark{r}   & 316.4 $\pm$ 9.2    \tablenotemark{r}  &   276.4 $\pm$ 11.4   \tablenotemark{r}   & 380.6 $\pm$ 40.2    \tablenotemark{r}  \\

He I $\lambda$5876                &  11.6 $\pm$ 0.5   \tablenotemark{b}  &  11.0 $\pm$ 0.4   \tablenotemark{b}     & 11.6 $\pm$ 0.4    \tablenotemark{b}  &   18.1 $\pm$ 1.1    \tablenotemark{b}   & $-$    \\
\nodata             &  12.0 $\pm$ 0.5   \tablenotemark{r}   &  11.3 $\pm$ 0.3      \tablenotemark{r}   & 11.5 $\pm$ 0.3    \tablenotemark{r}   &   $-$  & $-$    \\

{[OI]}  $\lambda$6302              & 1.9 $\pm$ 0.3    \tablenotemark{b}  & 3.5 $\pm$ 0.2   \tablenotemark{b}    & 4.7 $\pm$ 0.1    \tablenotemark{b}  & $-$ &     $-$    \\
\nodata       & $-$ & $-$ & $-$ & $-$ &    $-$    \\

{[S III]} $\lambda$6314            &  1.5 $\pm$ 0.2    \tablenotemark{b}  & 1.3 $\pm$ 0.2    \tablenotemark{b}  & 1.7 $\pm$ 0.1    \tablenotemark{b}  &     $-$ &  $-$  \\
\nodata       & $-$ & $-$ & $-$ & $-$ &    $-$    \\

{[OI]} $\lambda$6365               &  1.1 $\pm$ 0.1   \tablenotemark{b}  & 1.4 $\pm$ 0.1   \tablenotemark{b}   & 2.2 $\pm$ 0.2   \tablenotemark{b}   &     2.2 $\pm$ 0.2   \tablenotemark{b}   &  $-$ \\   
\nodata       & $-$ & $-$ & $-$ & $-$ &    $-$    \\

{[NII]} $\lambda$6548              & 2.9 $\pm$ 0.3   \tablenotemark{b}   &  2.6 $\pm$ 0.3   \tablenotemark{b}   &  4.7 $\pm$ 0.4    \tablenotemark{b}  &   7.0 $\pm$ 0.9    \tablenotemark{b}  &  $-$     \\
\nodata          & $-$ &  $-$ &  $-$ &   4.6 $\pm$ 0.4    \tablenotemark{r}   &  7.1 $\pm$ 0.3   \tablenotemark{r}     \\

H$\alpha$ $\lambda$6563     & 280.0 $\pm$ 11.7   \tablenotemark{b}  & 280.0 $\pm$ 6.5   \tablenotemark{b}    &   280.0 $\pm$ 6.0    \tablenotemark{b}  &  280.0 $\pm$ 8.3   \tablenotemark{b}   & 280.0 $\pm$ 35.7   \tablenotemark{b}   \\
\nodata    & 280.0 $\pm$ 11.8    \tablenotemark{r}   &   280.0 $\pm$ 7.1    \tablenotemark{r}  &  280.0 $\pm$ 7.2    \tablenotemark{r}  & 280.0 $\pm$ 35.7    \tablenotemark{r}  & 280.0 $\pm$ 26.2    \tablenotemark{r}  \\

{[N II]} $\lambda$6584             &   9.7 $\pm$ 0.9   \tablenotemark{b}  &  12.3 $\pm$ 0.7    \tablenotemark{b}  & 18.0 $\pm$ 0.7   \tablenotemark{b}   &  28.4 $\pm$ 1.2    \tablenotemark{b}    &  16.4 $\pm$ 2.2    \tablenotemark{b}   \\
\nodata          &   10.6 $\pm$ 0.5  \tablenotemark{r}   &  13.6 $\pm$ 0.4    \tablenotemark{r}  & 19.5 $\pm$ 0.7   \tablenotemark{r}   &  31.9 $\pm$ 1.4     \tablenotemark{r}   &  14.7 $\pm$ 2.6    \tablenotemark{r}    \\

He I $\lambda$6678                & 3.0 $\pm$ 0.1    \tablenotemark{b}  &  3.0 $\pm$ 0.3     \tablenotemark{b}   & 3.2 $\pm$ 0.1   \tablenotemark{b}  &   3.3 $\pm$ 0.4    \tablenotemark{b}  &   6.7 $\pm$ 1.5     \tablenotemark{b}  \\
\nodata              & 3.2 $\pm$ 0.1    \tablenotemark{r}   &  3.0 $\pm$ 0.1    \tablenotemark{r}    & 3.4 $\pm$ 0.2   \tablenotemark{r}   &   3.7 $\pm$ 0.1   \tablenotemark{r}   &  $-$ \\

{[S II]} $\lambda$6716             &  17.5 $\pm$ 0.7    \tablenotemark{b}  &  24.2 $\pm$ 0.5    \tablenotemark{b}  & 31.6 $\pm$ 0.7     \tablenotemark{b}  &  45.3 $\pm$ 1.4    \tablenotemark{b}   &  31.5 $\pm$  4.5    \tablenotemark{b}   \\
\nodata         &  18.5 $\pm$ 0.8   \tablenotemark{r}   &  25.5 $\pm$ 0.6    \tablenotemark{r}   & 33.0 $\pm$ 0.8   \tablenotemark{r}    &  47.3 $\pm$ 1.7   \tablenotemark{r}    &  23.7 $\pm$  2.4   \tablenotemark{r}    \\

{[S II]} $\lambda$6731             &   12.4 $\pm$ 0.5    \tablenotemark{b}  &  16.3 $\pm$ 0.4     \tablenotemark{b}  & 20.9 $\pm$ 0.5   \tablenotemark{b}   &  31.4 $\pm$ 1.0    \tablenotemark{b}   & 18.6 $\pm$ 2.8   \tablenotemark{b}    \\
\nodata            &   12.9 $\pm$ 0.5   \tablenotemark{r}   &  17.0 $\pm$ 0.4     \tablenotemark{r}   & 22.2 $\pm$ 0.6   \tablenotemark{r}   &  33.0 $\pm$ 1.2     \tablenotemark{r}   & 17.1 $\pm$ 1.8     \tablenotemark{r}  \\

{[S II]} $\lambda$6716/$\lambda$6731     &  1.42 $\pm$ 0.05    \tablenotemark{b}  & 1.49 $\pm$ 0.03     \tablenotemark{b}  &  1.51 $\pm$ 0.03     \tablenotemark{b}  & 1.44 $\pm$ 0.04   \tablenotemark{b}  &  1.70 $\pm$ 0.24   \tablenotemark{b}    \\
\nodata     &  1.44 $\pm$ 0.05    \tablenotemark{r}   & 1.50 $\pm$ 0.03   \tablenotemark{r}    &  1.48 $\pm$ 0.03    \tablenotemark{r}    & 1.44 $\pm$ 0.04   \tablenotemark{r}   &  1.38 $\pm$ 0.14   \tablenotemark{r}    \\

He I $\lambda$7065                &  2.0 $\pm$ 0.1    \tablenotemark{b}  & 1.8 $\pm$ 0.1   \tablenotemark{b}  &  1.9 $\pm$ 0.1   \tablenotemark{b}  &    2.2 $\pm$ 0.1    \tablenotemark{b}    &  $-$    \\
\nodata            &  2.3 $\pm$ 0.2   \tablenotemark{r}   & 2.0 $\pm$ 0.1   \tablenotemark{r}   &  2.2 $\pm$ 0.1   \tablenotemark{r}   &    $-$    &  $-$    \\

{[Ar III]} $\lambda$7136           & 7.4 $\pm$ 0.3   \tablenotemark{b}  & 6.8 $\pm$ 0.1   \tablenotemark{b}  &   6.5 $\pm$ 0.2    \tablenotemark{b}  &   6.8 $\pm$ 0.3    \tablenotemark{b}  & 8.5 $\pm$ 2.3     \tablenotemark{b}  \\
\nodata         & 8.0 $\pm$ 0.4  \tablenotemark{r}  & 7.3 $\pm$ 0.2   \tablenotemark{r}   &   7.5 $\pm$ 0.3   \tablenotemark{r}   &   6.5 $\pm$ 0.3   \tablenotemark{r}   & 8.1 $\pm$ 1.7   \tablenotemark{r}     \\

\enddata
%% Text for table notes should follow after the \enddata but before
%% the \end{deluxetable}. Make sure there is at least one \tablenotemark
%% in the table for each \tablenotetext.
\tablecomments{Fluxes are given on a scale where F(H$\beta$)$=$100. The values obtained from the blue and red spectra are indicated with the letters $b$ and $r$ respectively.}
\end{deluxetable}

\begin{deluxetable}{lccccccc}
\tabletypesize{\scriptsize}
%rotate
\tablecaption{Derived properties in NGC~1705 for regions s5 $-$ s8-1}
%q\tablewidth{0pt}
\tablehead{
\colhead{Property} & \colhead{Obs} & \colhead{s5} &  \colhead{s6-1}  &  \colhead{s6-2}  &  \colhead{s7-1} &  \colhead{s7-2} &  \colhead{s8-1}  
}
\startdata
$E(B-V)$ & b & 0.01 $\pm$ 0.11   & 0.00 $\pm$ 0.02  & 0.02 $\pm$ 0.02   & 0.10 $\pm$ 0.01  & 0.16 $\pm$ 0.01   & 0.08 $\pm$ 0.01   \\ 
 & r & 0.24 $\pm$ 0.06 & 0.00 $\pm$ 0.01 & 0.00 $\pm$ 0.02 & 0.09 $\pm$ 0.01 & 0.16 $\pm$ 0.01 & 0.04 $\pm$ 0.01 \\ 
$n_e$ [$cm^{-3}$] \tablenotemark{a} & b & $-$  & $-$  & $<32$  & $<10$  & $-$  & $27^{+49}_{-27}$ \\ 
& r & $-$ & $-$  & $-$  & $<6$  & $<1$    & $20^{+40}_{-20}$ \\ 

$T_e (O^{+2})$ [K] & b & $-$ & $15300^{+560}_{-490}$ & $14030^{+900}_{-700}$   & $12950^{+370}_{-320}$ & $12230^{+580}_{-470}$   & $13080^{+480}_{-420}$  \\ 

\hline
$(N^+/H^+)\times 10^6$ & b &  $-$ &   $1.85_{- 0.07}^{+0.07}$ &   $1.78_{-0.13}^{+ 0.13}$  & $1.26_{-0.09}^{+ 0.09}$  & $1.28_{-0.11}^{+ 0.11}$  & $0.95_{-0.11}^{+ 0.11}$ \\
&  r & $-$ &  $1.69 _{-0.06}^{+ 0.06}$   & $1.67_{-0.12}^{+0.12}$  & $1.23_{-0.05}^{+ 0.06}$  & $1.39_{-0.12}^{+0.12}$   & $1.02_{-0.06}^{+ 0.06}$   \\
$12+\log(N/H)$ & b & $-$ & $6.54 \pm 0.02$ &  $6.73 \pm 0.03$ &  $6.53 \pm 0.03$ & $6.63 \pm 0.04$  & $6.63 \pm 0.05$  \\
& r & $-$ & $6.49 \pm 0.02$ &  $6.66 \pm 0.03$ &  $6.54 \pm 0.02$ & $6.62 \pm 0.04$  & $6.61 \pm 0.03$  \\

\hline
$(O^0/H^+)\times 10^6$ & b & $-$ & $-$ &  $-$ & $1.89_{-0.12}^{+0.13}$    & $-$  & $1.43_{-0.11}^{+ 0.12}$   \\
& r &  $-$  &  $-$  &  $-$  &  $-$  &  $-$  &  $-$  \\

$(O^+/H^+)\times 10^5$  & b & $-$ & $2.95_{-0.17}^{+0.17}$    &  $2.44_{-0.28}^{+0.30}$   & $3.21_{-0.22}^{+0.24}$    & $3.37_{-0.40}^{+0.50}$    & $1.94_{-0.17}^{+0.18}$ \\

$(O^{+2}/H^+)\times 10^5$ & b & $-$ & $2.62_{-0.21}^{+0.21}$   &  $4.90_{-0.74}^{+0.72}$  &  $5.21_{-0.38}^{+0.37}$  &  $7.95_{-0.99}^{+1.05}$  & $6.87_{-0.65}^{+0.62}$ \\   
& r & $-$ & $2.58_{-0.21}^{+0.21}$   &  $4.79_{-0.73}^{+0.71}$   &  $5.12_{-0.38}^{+0.36}$   & $7.68_{-0.96}^{+1.01}$   & $6.68_{-0.63}^{+0.61}$ \\   

$12+\log(O/H)$\tablenotemark{b}  & b &  $-$ & $7.75 \pm 0.03$  &  $7.87 \pm 0.06$ &  $7.93 \pm 0.03$ & $8.05 \pm 0.05$  & $7.95 \pm 0.04$ \\

\hline

$(Ne^{+2}/H^+)\times 10^5$ & b & $-$ &  $0.83_{-0.08}^{+0.08}$ &   $1.24_{-0.21}^{+0.22}$ &  $1.05_{-0.09}^{+0.09}$ &   $2.27_{-0.33}^{+0.36}$  &  $1.90_{-0.21}^{+0.20}$  \\

$12+\log(Ne/H)$ & b  & $-$ & $7.01 \pm 0.04$ &  $7.16 \pm 0.08$ &  $7.09 \pm 0.04$ & $7.41 \pm 0.07$  & $7.32 \pm 0.05$ \\

\hline
$(S^+/H^+)\times 10^7$  & b & $-$ & $7.93_{-0.27}^{+0.27}$   &  $6.23_{-0.52}^{+0.52}$  &  $4.16_{-0.17}^{+0.18}$  &  $5.18_{-0.44}^{+0.44}$  &  $3.71_{-0.19}^{+0.20}$ \\    
& r & $-$ & $7.62_{-0.26}^{+0.26}$   &  $6.16_{-0.44}^{+0.44}$  &  $4.32_{-0.18}^{+0.19}$   &  $5.74_{-0.49}^{+0.49}$  &  $3.95_{-0.21}^{+0.21}$ \\   

$(S^{+2}/H^+)\times 10^6$ & b & $-$ & $-$ &  $-$  & $1.49_{-0.17}^{+0.20}$   & $-$   &   $1.80_{-0.26}^{+0.31}$ \\     
& r & $-$ & $-$ & $-$ & $1.69_{-0.20}^{+0.23}$   & $-$  & $-$ \\   

$12+\log(S/H)$ & b  & $-$ & $-$ &  $-$ &  $6.31 \pm 0.04$ & $-$  & $6.40 \pm 0.06$ \\
& r  & $-$ & $-$ &  $-$ &  $6.36 \pm 0.05$ & $-$  & $-$ \\

\hline
 $(Ar^{+2}/H^+)\times 10^7$ & b & $-$ & $-$&  $-$&  $3.90_{-0.26}^{+0.29}$ & $5.22_{-0.66}^{+0.66}$ &  $4.07_{-0.32}^{+0.38}$  \\
& r & $-$ & $2.73_{-0.15}^{+0.17}$ &   $3.29_{-0.39}^{+0.39}$  & $4.24_{-0.28}^{+0.31}$ &   $5.08_{-0.69}^{+0.67}$ &  $4.29_{-0.34}^{+0.40}$ \\ 
 
$12+\log(Ar/H)$ & b & $-$ & $-$ &  $-$ &  $5.62 \pm 0.03$ & $5.75 \pm 0.03$  & $5.66 \pm 0.04$ \\
& r & $-$ & $5.47 \pm 0.02$ &  $5.55 \pm 0.05$ &  $5.66 \pm 0.03$ & $5.74 \pm 0.06$  & $5.68 \pm 0.04$ \\

\enddata
\tablecomments{Ion abundances are in unit of ionized hydrogen.}
\tablenotetext{a}{`$-$' indicates that a solution for the density could not be found within the $1 \sigma$ range of the [S II] line ratio. In those cases we assumed $n_e=30 \ cm^{-3}$ for the temperature calculation. } 
\tablenotetext{b}{Direct $T_e$ method oxygen abundance. The total abundance was derived from the sum of $O^0 + O^+ + O^{+2}$ for regions s71, s81, s82, s83, and s84, and neglecting the contribution from 
  $O^0$ in all the other regions.}  
\end{deluxetable}

\begin{deluxetable}{lccccccc}
\tabletypesize{\scriptsize}
%rotate
\tablecaption{Derived properties in NGC~1705 for regions s8-2 $-$ s10}
%q\tablewidth{0pt}
\tablehead{
\colhead{Property} & \colhead{Obs} & \colhead{s8-2} &  \colhead{s8-3} &  \colhead{s8-4} &  \colhead{s9} &  \colhead{s10} 
}
\startdata
$E(B-V)$ &  b & 0.04 $\pm$ 0.01& 0.05 $\pm$ 0.01 & 0.07 $\pm$ 0.01 & 0.04 $\pm$ 0.01  & 0.02 $\pm$ 0.03  \\
 &  r & 0.01 $\pm$ 0.01 & 0.01 $\pm$ 0.01  & 0.02 $\pm$ 0.01 & 0.01 $\pm$ 0.01 & 0.01 $\pm$ 0.02  \\

$n_e$ [$cm^{-3}$] \tablenotemark{a}   & b & $14^{+39}_{-14}$ & $-$  & $-$  & $<22$  & $-$ \\
 & r & $<35$ & $-$  & $-$  & $<26$  & $40^{+140}_{-40}$   \\

$T_e (O^{+2})$ [K] &  b & $13420^{+380}_{-340}$ & $12990^{+240}_{-210}$   & $13290^{+200}_{-190}$  & $14470^{+350}_{-320}$   & $14940^{+1720}_{-1160}$  \\ 
\hline
$(N^+/H^+)\times 10^6$  &   b & $1.04_{- 0.10}^{+ 0.10}$   &    $1.31_{- 0.09}^{+ 0.09}$   &    $1.92_{-0.09}^{+0.09}$    &  $2.65_{-0.16}^{+0.16}$ & $1.59_{-0.21}^{+0.21}$  \\
& r &  $1.17_{-0.05}^{+ 0.05}$    &  $1.60_{-0.05}^{+0.05}$   &  $2.21_{-0.08}^{+0.08}$  &  $2.73_{-0.13}^{+ 0.13}$   & $2.11_{-0.28}^{+0.28}$  \\
$12+\log(N/H)$ & b & $6.64 \pm 0.04$ &  $6.60 \pm 0.03$ &  $6.69 \pm 0.02$ & $6.75 \pm 0.03$  & $6.58 \pm 0.06$ \\
& r & $6.65 \pm 0.02$ &  $6.65 \pm 0.01$ &  $6.71 \pm 0.01$ & $6.73 \pm 0.02$  & $6.68 \pm 0.06$ \\

\hline
$(O^0/H^+)\times 10^6$ &  b & $1.50_{-0.24}^{+0.24}$   &  $2.98_{-0.12}^{+0.13}$   &  $3.80_{-0.12}^{+0.13}$  & $-$    & $-$   \\
 &  r & $-$  &  $-$  &  $-$  &  $-$  &  $-$   \\

$(O^+/H^+)\times 10^5$ &b&  $1.96_{-0.13}^{+0.13}$    &   $3.06_{-0.14}^{+0.15}$    & $3.48_{-0.12}^{+0.13}$   &    $3.22_{-0.14}^{+0.16}$    &    $2.93_{-0.54}^{+0.54}$ \\

$(O^{+2}/H^+)\times 10^5$  & b & $6.34_{-0.46}^{+0.45}$   &  $6.02_{-0.29}^{+0.29}$ &  $4.91_{-0.19}^{+0.20}$  &  $3.52_{-0.22}^{+ 0.22}$  & $4.05_{-0.90}^{+ 0.98}$ \\     
& r & $ 6.16_{-0.45}^{+0.44}$  &  $5.82_{-0.28}^{+ 0.28}$   & $4.80_{-0.19}^{+0.20}$  &  $3.34_{-0.21}^{+0.21}$   &  $4.21_{-0.94}^{+1.02}$ \\    

 $12+\log(O/H)$\tablenotemark{b}  & b &  $7.93 \pm 0.03$ & $7.97 \pm 0.02$  &  $7.94 \pm 0.02$ &  $7.83 \pm 0.02$ & $7.84 \pm 0.08$   \\

\hline
$(Ne^{+2}/H^+)\times 10^5$ & b &   $1.66_{-0.14}^{+0.14}$ &  $1.57_{-0.09}^{+0.09}$ &  $1.44_{-0.07}^{+0.07}$ &   $1.18_{-0.08}^{+0.09}$ &   $0.97_{-0.18}^{+0.28}$      \\

$12+\log(Ne/H)$ & b  & $7.27 \pm 0.04$ &  $7.26 \pm 0.02$ &  $7.24 \pm 0.02$ & $7.16 \pm 0.03$  & $7.06 \pm 0.12$ \\

\hline
$(S^+/H^+)\times 10^7$ & b &    $4.10_{-0.16}^{+0.15}$   &  $5.85_{-0.15}^{+0.16}$  &  $7.33_{-0.17}^{+ 0.17}$   & $9.55_{-0.30}^{+0.30}$   &  $6.03_{-0.88}^{+0.88}$ \\   
& r & $4.31_{-0.18}^{+0.18}$  &  $6.14_{-0.16}^{+0.17}$  &  $7.71_{-0.20}^{+0.20}$   &  $10.00_{-0.27}^{+0.29}$  &  $4.91_{-0.51}^{+0.51}$ \\

 $(S^{+2}/H^+)\times 10^6$  & b & $1.27_{-0.17}^{+0.17}$   &   $1.27_{-0.20}^{+0.20}$   &   $1.50_{-0.09}^{+0.10}$   &  $-$ &  $-$ \\ 
& r & $-$  & $-$ &  $-$  &  $-$   & $-$  \\ 

$12+\log(S/H)$ & b  & $6.29 \pm 0.05$  & $6.31 \pm 0.05$ &  $6.38 \pm 0.02$ & $-$  & $-$ \\
 & r  & $-$  & $-$ &  $-$ & $-$  & $-$ \\

\hline
 
$(Ar^{+2}/H^+)\times 10^7$  & b &  $3.89_{-0.24}^{+0.23}$  & $3.86_{-0.17}^{+0.20}$ &  $3.49_{-0.11}^{+0.12}$ & $3.04_{-0.14}^{+0.15}$ &  $3.57_{-0.97}^{+0.97}$ \\  
& r & $4.21_{-0.26}^{+0.25}$  & $4.15_{-0.18}^{+0.21}$ &  $4.03_{-0.14}^{+0.16}$   &  $2.91_{-0.13}^{+0.14}$ &   $3.40_{-0.71}^{+0.71}$ \\

$12+\log(Ar/H)$ & b & $5.64 \pm 0.02$ &  $5.62 \pm 0.02$ &  $5.57 \pm 0.01$ & $5.51 \pm 0.02$  & $5.58 \pm 0.12$ \\
& r & $5.67 \pm 0.02$ &  $5.65 \pm 0.02$ &  $5.64 \pm 0.02$ & $5.50 \pm 0.02$  & $5.56 \pm 0.10$ \\

\enddata
\tablecomments{Ion abundances are in unit of ionized hydrogen.}
\tablenotetext{a}{`$-$' indicates that a solution for the density could not be found within the $1 \sigma$ range of the [S II] line ratio. In those cases we assumed $n_e=30 \ cm^{-3}$ for the temperature calculation. } 
\tablenotetext{b}{Direct $T_e$ method oxygen abundance. The total abundance was derived from the sum of $O^0 + O^+ + O^{+2}$ for regions s71, s81, s82, s83, and s84, and neglecting the contribution from 
  $O^0$ in all the other regions.}  
\end{deluxetable}

\begin{deluxetable}{lccc}
\tablewidth{0pc} 
%rotate
\tablecaption{Comparison of our average abundances with literature values}
%q\tablewidth{0pt}
\tablehead{
\colhead{} &  \colhead{This work} &  \colhead{This work} & \colhead{Lee et al. 2004\tablenotemark{c} }  \\
\colhead{} &  \colhead{(total\tablenotemark{a})} &   \colhead{central (s7$+$s8 \tablenotemark{b} )} &  \colhead{}   \\
}
\startdata
$12 + \log(N/H)$      & $6.63 \pm 0.07$  & $6.62 \pm 0.05$ & $6.51 \pm 0.06$ \\
$12 + \log(O/H)$      & $7.91 \pm 0.08$  & $7.96 \pm 0.04$& $8.22 \pm 0.05$ \\
$12 + \log(Ne/H)$   & $7.19 \pm 0.12$  & $7.26 \pm 0.09$&  $7.81 \pm 0.07$ \\
$12 + \log(Ar/H)$    & $5.62 \pm 0.06$  & $5.65 \pm 0.05$& $5.93 \pm 0.05$ \\
\hline 
$\log(N/O)$   &  $-1.27 \pm 0.11$  &  $-1.34 \pm 0.06$  & $-1.69 \pm 0.10$  \\
$\log(Ne/O)$ &  $-0.71 \pm 0.06$  & $-0.70 \pm 0.07$  & $-0.41 \pm 0.05$    \\
$\log(Ar/O)$  & $-2.31 \pm 0.03$   &  $-2.32 \pm 0.03$  & $-2.30 \pm 0.07$    \\

%\tablenotetext{b}{}
\enddata
\tablenotetext{a}{Average abundances and standard deviations derived averaging regions s6 to s10}
\tablenotetext{b}{Average abundances and standard deviations derived averaging regions s7-1, s7-2, s8-1, s8-2, s8-3, s8-4}
\tablenotetext{c}{Average abundances and standard deviations derived from the five regions in \cite{lee04} (A3, B3, B4, B6 and C6) with direct temperature estimates from 
the [O III]$\lambda$4363 line. 
The poor [N II]$\lambda$6583 detection in H II region C6 was not included in the average.}
\end{deluxetable}

\end{document}